\documentclass[twocolumn,floatfix,amssymb,amsmath,secnumarabic,nofootinbib, nobalancelastpage,longbibliography]{revtex4-2}    %Two column layout

\usepackage[pdftex]{graphicx} % Include figure files
\usepackage{dcolumn}  % Align table columns on decimal point
\usepackage{bm}       % bold math
\usepackage[usenames,dvipsnames]{color}														
\definecolor{URLCOL}{rgb}{0,0.52,0.83} %external link color
\definecolor{LINKCOL}{rgb}{0.05,0.5,0} %internal link color
\definecolor{CITECOL}{rgb}{0.25,0,0.48} %link to bibliography
\usepackage{epstopdf}
\usepackage[pdftex,bookmarks,breaklinks,bookmarksopen,bookmarksnumbered,colorlinks,linkcolor=LINKCOL,linktocpage,citecolor=CITECOL,urlcolor=URLCOL,pdfpagemode=UseOutline,pdftex]{hyperref}
\usepackage{datetime}

\def\preprintlink{ \href{http://dft.uci.edu + PAPER REF}{title of paper} }
\def\preprinttext{~}

\usepackage{fancyhdr}
\makeatletter
\fancypagestyle{titlepage}
{
	\lhead{\textsc{\preprinttext\ ~ \href{http://dft.uci.edu/publications.php}{~}}}
	\chead{}
	\rhead{\preprintlink}
	\lfoot{}
}
\makeatother
\def\preprintlink{ 
	\href{http://dft.uci.edu}%INSERT the URL of your paper
        {%v\versionnumber\ - 
~}%INSERT the reference of your paper
	}

\usepackage{graphicx}
\usepackage{amsmath}
\usepackage{physics} %PO
\usepackage{mathrsfs} %PO
\usepackage{subfigure} %PO
\usepackage{dcolumn} %PO
\usepackage{tabularx}%PO
\usepackage{mathtools}%PO
\usepackage{bm}
\usepackage{threeparttable}
\usepackage[normalem]{ulem}
\usepackage{array,multirow}
\usepackage{appendix}
\usepackage{epstopdf}
\usepackage{booktabs}
\usepackage{natbib}
\usepackage{feynmp}
\usepackage{threeparttable}
\usepackage{CJKutf8}
\definecolor{TITLECOL}{rgb}{0.1,0.2,0.7} %title color
\definecolor{PCOL}{rgb}{0.5,0.06,0.01} %title color
\definecolor{CHAPCOL}{rgb}{0,0.48,0} %chapter color
\definecolor{SECOL}{rgb}{0.1,0.2,0.7} %sec color
\definecolor{CONTENTSCOL}{rgb}{0.1,0.2,0.7} %can choose the table of contents title to have same color as sec
\definecolor{SSECOL}{rgb}{0.25,0,0.48} %ssection color
\definecolor{SSSECOL}{rgb}{0.2,0.08,0.53} %subsubsection color  0.2,0.08,0.53
\definecolor{SHDCOL}{rgb}{0.4,0,0} % heading of section color
\definecolor{ITMCOL}{rgb}{0.4,0,0} % bulletted text of item color
\definecolor{EXCOL}{rgb}{0,0.47,0.01} %color of exercises
\definecolor{DEFCOL}{rgb}{0,0.42,0.01} %color in definitions file for definition headings

\def\coloredtitle#1{\title{\textcolor{TITLECOL}{#1}}} %title color
\definecolor{URLCOL}{rgb}{0,0.17,0.43} %external link color
\definecolor{LINKCOL}{rgb}{0.05,0.4,0} %internal link color
\definecolor{CITECOL}{rgb}{0.35,0,0.48} %link to bibliography

\definecolor{ngreen}{rgb}{0,0.48,0}

\def\sec#1{\section{\textcolor{SECOL}{#1}}}

\def\sectable#1{
\addcontentsline{toc}{subsection}{~~Table: \textcolor{SSECOL}{#1}}
\begin{table*}[h]
\caption{\bf \textcolor{SSECOL}{#1}}
}

\def\bea{\begin{eqnarray}}
\def\eea{\end{eqnarray}}
\def\ben{\begin{equation}}
\def\een{\end{equation}}
\def\benu{\begin{enumerate}}
\def\enu{\end{enumerate}}

\def\bei{\begin{itemize}}
\def\eei{\end{itemize}}
\def\beit{\begin{itemize}}
\def\eit{\end{itemize}}
\def\benu{\begin{enumerate}}
\def\enu{\end{enumerate}}

\def\n{n}

\def\sss{\scriptscriptstyle\rm}

\def\g{_\gamma}

\def\1var{(\bx_1...\bx)}

\def\half{\frac{1}{2}}

\def\br{{\bf r}}

\def\bx{{x}}

\def\bj{{\bf j}}

\def\x{_{\sss X}}
\def\c{_{\sss C}}
\def\s{_{\sss S}}

\def\LDA{^{\rm LDA}}

\def\TF{^{\rm TF}}

\def\W{^{\rm W}}
\def\NL{^{\rm NL}} %PO

\def\unif{^{\rm unif}}

\def\sph_int{ {\int d^3 r}}

\definecolor{SPECOL}{rgb}{0,0.47,0.01}
\definecolor{QUOCOL}{rgb}{0,0,0.2}
\definecolor{SHDCOLb}{rgb}{0.69,0.4,0.1}%{0.01,0.4,0} % heading of section color

\definecolor{SPEQ}{rgb}{0.01,0.4,0.05} %

\definecolor{SPEQv}{rgb}{0.45,0.05,0.45} %

\definecolor{SPEQb}{rgb}{0.01,0.1,0.65} %

\definecolor{SPEQr}{rgb}{0.57,0.05,0.1} %

\def\sec#1{\section{\textcolor{SECOL}{#1}}}

\def\bay{\begin{array}}
\def\eay{\end{array}}
\def\bit{\begin{itemize}}
\def\beit{\begin{itemize}}
\def\eit{\end{itemize}}

\def\ln{\text{ln} }
\def\tanh{\text{tanh} }

\def\floor{\text{floor} }

\def\dd{~ \rotatebox{320}{\hspace{-5pt}\vbox to 5 pt {\hspace{-5pt} \hbox to 5pt {$\cdots$}}}\!\! }
\def\th{$\frac{1}{3}$}%{\vbox to 15pt {\vspace{1pt} \hbox to 4pt {\!\!\! -2}}}

\begin{document}

 % % % % % % % % % % % % % % % % % % % % % % % % % % % % % % % % % 
%%%%%%%%%%%%%%%%%%%TITLE AND TABLE OF CONTENTS%%%%%%%%%%%%%%%%%%%%
% % % % % % % % % % % % % % % % % % % % % % % % % % % % % % % % % 

\sf %san serif
\coloredtitle{Semiclassics: The hidden theory behind the success of
DFT}
\author{\color{CITECOL} Pavel Okun, Kieron Burke}
\affiliation{Department of Chemistry,
	University of California, Irvine, CA 92697,  USA}
\affiliation{Departments of Physics and Astronomy and of Chemistry}
\date{\today}
\begin{abstract}
It is argued that the success of DFT can be understood in terms of a semiclassical expansion around a very specific limit.  This limit was identified long ago by Lieb and Simon for the total electronic energy of a system.  This is a universal limit of all (non-relativistic) electronic structure: atoms, molecules, and solids.  In the simple case of neutral atoms, this limit corresponds to an expansion of the total energy in powers of $Z^{-1/3}$.  For the total energy, Thomas-Fermi theory becomes relatively exact in the limit.  The limit can also be studied for much simpler model systems, including non-interacting fermions in a one-dimensional well, where the WKB approximation applies for individual eigenvalues and eigenfunctions.  Summation techniques lead to energies and densities that are functionals of the potential.  We consider several examples in one dimension (fermions in a box, in a harmonic well, in a linear half-well, and in the Pöschl-Teller well.  The effects of higher dimension are also illustrated with the three-dimensional harmonic well and the Bohr atom, non-interacting fermions in a Coulomb well.  Modern density functional calculations use the Kohn-Sham scheme almost exclusively.  The same semiclassical limit can be studied for the Kohn-Sham kinetic energy, for the exchange energy, and for the correlation energy.  For all three, the local density approximation appears to become relatively exact in this limit.  Recent work, both analytic and numerical, explores how this limit is approached, in an effort to deduce the leading corrections to the local approximation.  A simple scheme, using the Euler-Maclaurin summation formula, is the result of many different attempts at this problem.  In very simple cases, the correction formulas are much more accurate than standard density functionals.  Several functionals are already in widespread use in both chemistry and materials that incorporate these limits, and prospects for the future are discussed.
\end{abstract}

%\listoftables

% % % % % % % % % % % % % % % % % % % % % % % % % % % % % % % % % 
%%%%%%%%%%%%%%%%%%%%%%%%%%%%%%%MAIN%%%%%%%%%%%%%%%%%%%%%%%%%%%%%%%
% % % % % % % % % % % % % % % % % % % % % % % % % % % % % % % % % 
\maketitle
\def\floor#1{{\lfloor}#1{\rfloor}}
\def\sm#1{{\langle}#1{\rangle}}
\def\dis{_{disc}}
\newcommand{\Z}{\mathbb{Z}}
\newcommand{\R}{\mathbb{R}}
\def\w{^{(0)}}
\def\w{^{\rm WKB}}
\def\II{^{\rm II}}
\def\hd#1{\noindent{\bf\textcolor{red} {#1:}}}
\def\hb#1{\noindent{\bf\textcolor{blue} {#1:}}}
\def\eps{\epsilon}
\def\ew{\epsilon\w}
\def\ej{\epsilon_j}
\def\upet{^{(\eta)}}
\def\ejeta{\ej\upet}
\def\tjeta{\tj\upet}
\def\bej{{\bar \epsilon}_j}
\def\ewj{\epsilon\w_j}
\def\tj{t_j}
\def\vj{v_j}
\def\F{_{\sss F}}
\def\xt{x_{\sss T}}
\def\sc{^{\rm sc}}
\def\al{\alpha}
\def\ae{\al_e}
\def\bj{\bar j}
\def\bz{\bar\zeta}
\def\eq#1{Eq.\, (\ref{#1})}
\def\cN{{\cal N}}

\graphicspath{{./F/}}
\def\Lam{\Lambda}
\def\lam{\lambda}
\def\G{\Gamma}
\def\g{\gamma}
\def\eps{\epsilon}
\def\om{\omega}
\def\D{\Delta}
\def\d{\delta}
\def\r{\rho}
\def\a{\alpha}
\def\th{\theta}
\def\TH{\Theta}
\def\z{\zeta}
%\def\s{\sigma}

%PO's commands
\def\mbf{\mathbf}
\def\ra{\rightarrow}
\newcommand{\act}{z}
\DeclarePairedDelimiter{\cl}{\lceil}{\rceil}
\def\Ai{\mathrm{Ai}}
\def\ss{\scriptscriptstyle}
\def\cw{\textcolor{white}}
\def\cb{\textcolor{blue}}
\def\cm{\textcolor{magenta}}

\tableofcontents

\sec{Introduction}
\label{sec:Intro}
We begin with a very basic question:  {\em Why does DFT work at all?}  After all, the electronic structure problem requires solving a fermionic quantum many-body problem.  There are dozens if not hundreds of physics books in many fields explaining just how difficult this is \cite{SO96,N09}.  Yet modern approximations to the exchange-correlation (XC) energy, which can often be written as simple formulas on the back of an envelope, yield usefully accurate results in an astonishing variety of situations \cite{ED11}.  It is likely that at least 50,000 scientific papers will appear next year reporting results of such calculations.  How can this even be the case?

This book chapter describes a 15-year search for the underlying cause of such success, assuming it is not accidental.   A key piece of evidence was provided much earlier, when Lieb and Simon rigorously proved a result that had been intuited since at least the late 1940s \cite{FMT49}: the precursor of Kohn-Sham (KS) DFT, Thomas-Fermi (TF) theory, becomes relatively exact for the total electronic energy of a system in a very specific limit \cite{LS73,LS77}.  We call this the LS theorem.  This limit involves simultaneously scaling both the potential and the number of electrons, $N$, in a very specific way.  This is a universal limit of all electronic matter:  Atoms, molecules, and solids.

In some ways, the simplest interacting 3D many-electron problem is that of atoms and ions.   There is a long and interesting history of physics and mathematical exploration \cite{S80,S81,ES85,E88} of the expansion of the energy of neutral atoms as a function of $Z$, the nuclear charge:
\begin{equation}
E(Z) = - c_0\, Z^{7/3} + \half Z^2 - c_2\, Z^{5/3} +....,
\label{EZasy}
\end{equation}
where $c_0 \approx 0.768745$ and $c_2 \approx 0.269900$ are fundamental constants that can be easily calculated to arbitrary accuracy \cite{E88,ES85,ELCB08,S80,S81,PD77}, as we discuss in Sec. \ref{sec:TF}.  A simple calculation using TF theory (the local density approximation for the kinetic energy and the Hartree approximation for the electron-electron repulsion) yields
\begin{equation}
E\TF(Z) = - c_0\, Z^{7/3},
\end{equation}
consistent with the LS theorem.  The leading correction is called the Scott correction \cite{S52}, and can be deduced by considering electron orbitals near the nucleus.  Schwinger and Englert showed that $c_2$ is given exactly by evaluating the local density approximation (LDA) for exchange plus the second-order gradient expansion for the kinetic energy on the TF density \cite{D30,S81}.  All discussion in this chapter is for the pure non-relativistic limit.

\begin{figure}[!htb]
\includegraphics[width=0.9\columnwidth]{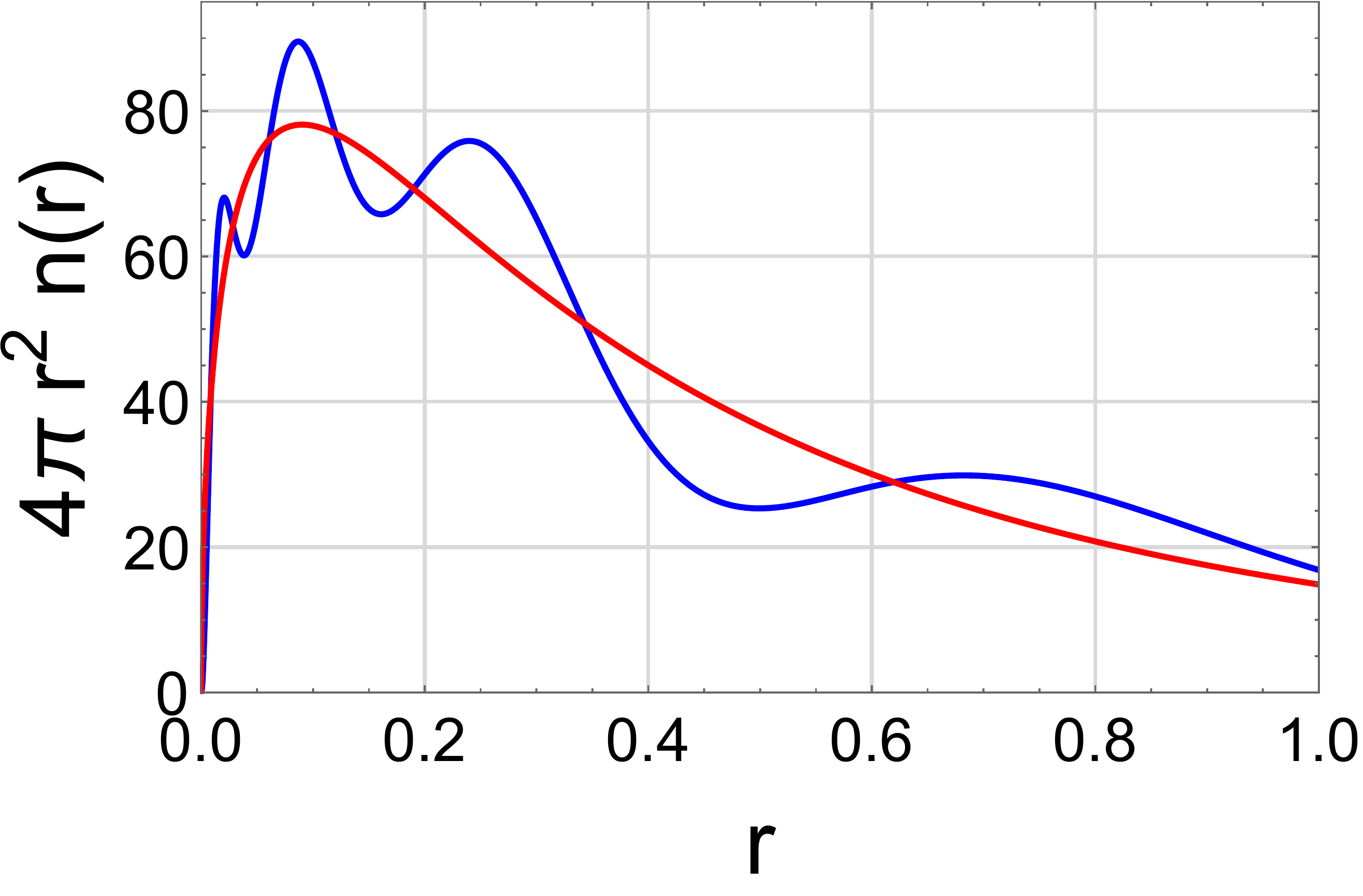}
\caption{Accurate radial density of Xe (blue) and its TF approximation (red), in atomic units.}
\label{fig:DenXe}
\end{figure}

Moreover, in a weak sense, as the limit is approached, the density approaches that of TF theory, and the error in any smooth integral over the density vanishes in the LS limit.  Fig. \ref{fig:DenXe} shows an accurate radial density of Xe and its TF approximation.  Despite behaving very differently for both small and large $r$, and missing shell structure, integrals over the TF density become relatively exact as $Z \to \infty$.

\begin{table}
$\begin{array}{|c|c|r|r|r|r|r|r|}
\hline
\multicolumn{3}{|c|}{} & \multicolumn{4}{c|}{\text{Error}}\\
\hline
\text{Atom} & \text{Z} & \multicolumn{1}{c|}{\text{Exact}} & \multicolumn{1}{c|}{\text{HF}} & \multicolumn{1}{c|}{\text{TF}} & \multicolumn{1}{c|}{1^{\text{st}}\text{ corr.}} & \multicolumn{1}{c|}{2^{\text{nd}} \text{ corr.}} \\
\hline
\text{H}  &  1 &     -0.500       & 0.000      &    -0.269      &   0.231      & -0.039      \\
\text{He} &  2 &     -2.904       & 0.042      &    -0.970      &   1.030      &  0.173      \\
\text{Ne} & 10 &   -128.937       & 0.390      &   -36.684      &  13.316      &  0.788      \\
\text{Ar} & 18 &   -527.539       & 0.722      &  -125.218      &  36.782      &  3.415      \\
\text{Kr} & 36 &  -2753.94\cw{0}  & 1.89\cw{0} &  -535.75\cw{0} & 112.25\cw{0} &  6.32\cw{0} \\
\text{Xe} & 54 &  -7235.23\cw{0}  & 3.09\cw{0} & -1237.72\cw{0} & 220.28\cw{0} & 12.06\cw{0} \\
\text{Rn} & 86 & -21872.5\cw{00}  & 5.8\cw{00} & -3223.9\cw{00} & 474.1\cw{00} & 21.8\cw{00} \\
\hline
\end{array}$
\caption{Accurate energies and errors for noble gas atoms and hydrogen where the TF results and the first and second corrections refer to Eq. (\ref{EZasy}).}
\label{tab:AE}
\end{table}
The LS limit is a jumping-off point for understanding approximations in DFT \cite{LS73,LS77}.  Most modern approximations to the exchange-correlation (XC) energy of KS-DFT begin with a generalized gradient approximation (GGA), a functional whose energy density depends on both the density and its gradient.  This idea was first suggested by Ma and Brueckner \cite{MB68} for the correlation energy of atoms.  They showed that a naive gradient expansion approximation (i.e., just using the gradient expansion for a slowly varying gas) fails miserably, but can be made much better by considering a more general functional of the gradient (hence the name GGA).  In various ways, many modern GGAs can be traced back to this initial work.

The reason the LS limit is the organizing principle behind the success of DFT is that it explains why local density approximations work as well (or as poorly) as they do:  They use that form
of the functional that is relatively exact in that limit, ensuring their relative error vanishes in that limit,
and producing a reasonable approximation even when the system is
highly inhomogeneous.  We will see that, in the simple case of non-interacting fermions
in one dimension, this local approximation is directly derivable from WKB theory, whose
semiclassical eigenvalues are often very accurate.
Because this limit applies to all matter, such approximations 'work' for all systems, including both molecules and materials.  One can then ask the question:  If the local approximation is the dominant term as the limit is approached, can we derive the leading correction to this limit?  And if so, how accurate would an approximation be that incorporates such corrections? Do modern GGA's accurately account for such corrections?   We will see that the answers are tantalizing.  In some extremely simple cases, the inclusion of just one or two more terms yields accuracy beyond the wildest dreams of any modern DFT calculation \cite{HET15,TSB16}.  Again, for simple cases, we can sometimes deduce many terms and achieve ridiculous levels of accuracy.  On the other hand, the difficulties in deriving such corrections for realistic systems are daunting.  But with guidance from simple systems, they can possibly be teased out numerically.  In any event, understanding this little-explored connection should put the art of DFT approximation on a surer footing.

\begin{figure}[!htb]
\includegraphics[width=0.9\columnwidth]{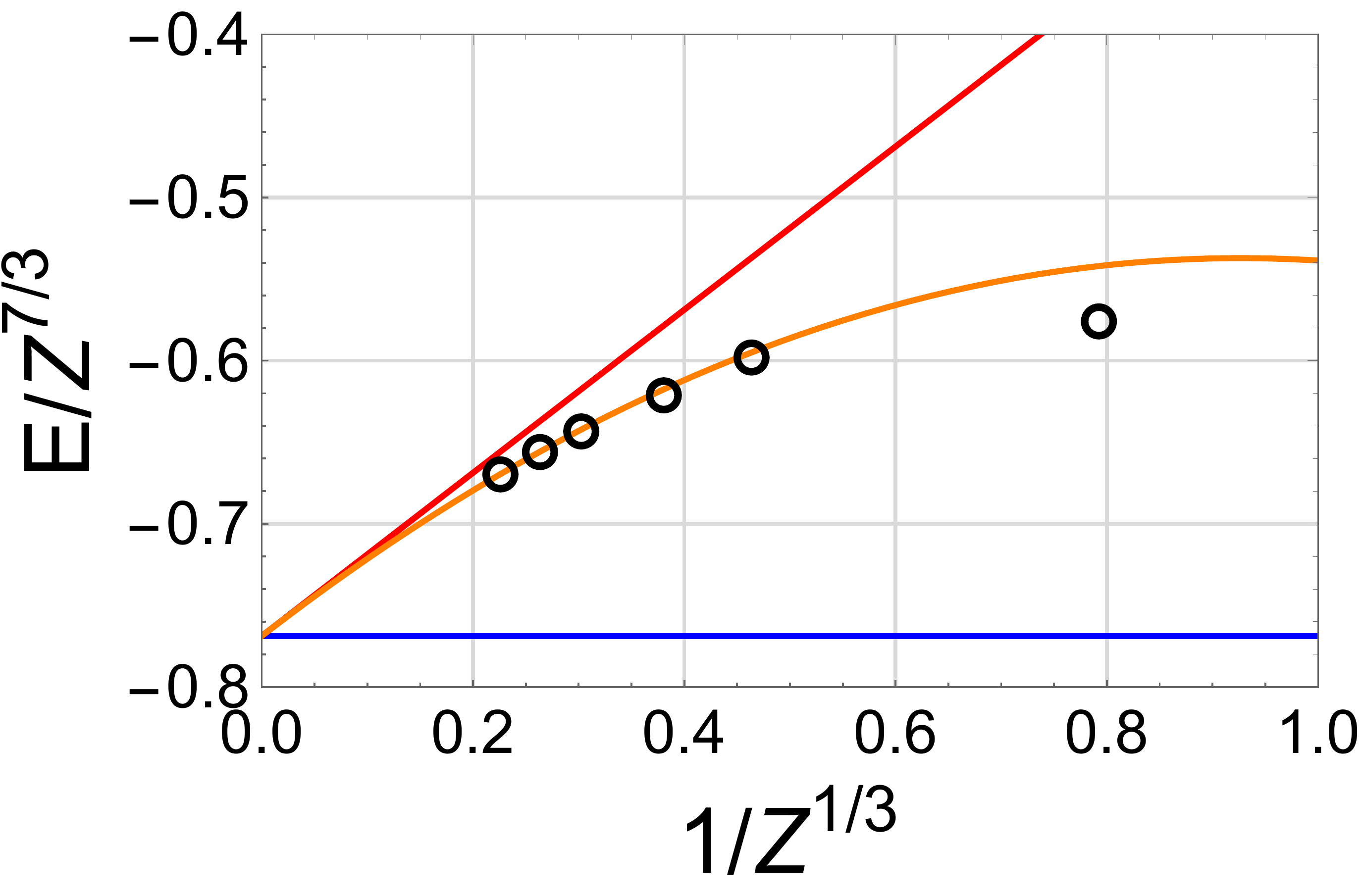}
\caption{The exact noble gas energies (black circles) compared with the expansion in Eq. (\ref{EZasy}): TF (blue), with first order correction (red), second order (orange).}
\label{fig:AtmEn}
\end{figure}

To better appreciate the power of such expressions, Table \ref{tab:AE} gives a list of total energies and the errors in several approximations to them for noble gas atoms, while Fig. \ref{fig:AtmEn} plots these energies, choosing variables consistent with the nature of the known asymptotic expansion.  There are many interesting points.  First, although very crude, the relative error of TF decreases as $Z$ becomes large, consistent with the LS theorem.  Next, we note that the expansion is in inverse powers of $Z^{1/3}$, so that the small parameter never even reaches below 0.2, even for Rn (Z = 86).  This makes numerical extrapolation quite difficult and imprecise.  We see that addition of each order of the expansion yields ever more accurate results.  The first three terms alone yield accuracies comparable to (but worse than) those of Hartree-Fock (HF) or a KS-DFT calculation using LDA.  One also sees that one could  very crudely  determine the coefficients in the expansion from such curves by fitting.  Here, the LS theorem is very important, as it is infinitely easier to perform a highly precise TF calculation than to perform the extrapolation from numerical results for individual atoms.  We are lucky to have all three terms to arbitrary accuracy, from solving the TF equation for atoms.

This book chapter describes an odyssey through various fields of theoretical physics, trying to find answers to these questions.  It is naturally divided into two sections.  The first describes studies of the non-interacting kinetic energy, mostly in 1D.  In 1D, the semiclassical expansions of eigenvalues which form the starting point of these studies are particularly simple, and many powerful tricks have been developed over the years, such as the WKB expansion \cite{GS18,BO99,W26,K26,B26,D32}.  Thus explicit derivations yield explicit answers, albeit with considerable work and ingenuity in some cases. On the other hand, very few chemical and materials problems can be solved just by knowing the 1D kinetic energy.   Thus the second half attempts to take insight from the first half, and apply it to make progress on the exchange-correlation energy for realistic systems.

Many of these questions were already asked more than 50 years ago.  In the sixties, before the advent of widespread and economical computing, there were many attempts to perform electronic structure calculations using semiclassical methods \cite{M68,M68b,M70}.  Ironically, early in 1965, Kohn and Sham showed how to calculate accurate approximate densities semiclassically, by performing contour integrals of the WKB Green's function, and showing how both shell structure and evanescence could be accurately found this way \cite{KSb65}.  Of course, later that year, they also published the rather more famous Kohn-Sham equations \cite{KS65}, pointing out their exactness in a legendary note added in proof, and simultaneously inventing the modern LDA approximation.  Their scheme proved successful beyond their wildest dreams \cite{Z14}, and Walter Kohn shared the Nobel prize in chemistry some 30 years later.

The aim of this chapter is to convince the reader that exactness in the LS limit is the least understood and possibly most fundamental reason why KS-DFT has been so successful.   Most modern functionals reduce to the uniform gas results in the limit of constant density, and so can recover the exact result in this limit.   Most GGAs appear to capture the leading corrections, at least for molecular systems, either by imposing relevant exact conditions or fitting to Coulomb-interacting systems.  Many of the successes and failures of standard DFT functionals can be understood from this viewpoint, as discussed throughout this chapter and in the key references.

But, more tantalizing than this, the most recent work shows that the leading corrections are sensitive to global boundary conditions \cite{PSHP86} that distinguish molecules from solids, and bulk from surfaces.  In simple model problems in 1D, when the right corrections accounting for these differences are included, tremendous improvements in accuracy are possible, suggesting that even KS-DFT calculations might attain much higher accuracy than is presently achieved.  

We use atomic units throughout, so that all energies are in Hartrees and all distances in Bohr radii.  We treat only the non-relativistic limit in the Born-Oppenheimer approximation.  We do not consider external magnetic fields \cite{VRG90} and give most results in terms of pure density functionals instead of spin-density functionals.  No statements should be considered mathematically rigorous.

This review is organized as follows.  In Sec. \ref{sec:TF} we review Thomas-Fermi theory for atoms, which is the starting point of our semiclassical expansion for the energy.  In Sec. \ref{sec:ill}, we illustrate semiclassical limits on four simple model systems in 1D, while in Sec. \ref{sec:LSScale}, we make the meaning of ''semiclassical" precise by introducing a scaling that produces an expansion around the semiclassical limit.  Secs. \ref{sec:box} and \ref{sec:RTP} describe semiclassical corrections to the TF density for 1D systems.  The former with box boundary conditions while the latter has open boundaries.  As all the expressions given so far are functionals of the potential, in Sec. \ref{sec:PFT} we briefly explain potential functional theory as an alternative to density functional theory.  In Sec. \ref{sec:GEA} we define the gradient expansion for slowly-varying gases.

In the rest of the review, we work away from simple models, all the way to practically useful XC approximations for use in modern KS-DFT codes.  To see some effects of degeneracy, in Sec. \ref{sec:3D}, we apply the
semiclassical formalism to two non-interacting systems in 3D: The Bohr atom and the 3D harmonic oscillator.
In Secs. \ref{sec:X} and \ref{sec:C} we describe the relevance of these ideas to exchange and
correlation functionals respectively.  Most importantly we argue that both exchange and
correlation become local, like the kinetic energy, in the semiclassical limit.  
In Sec. \ref{sec:IE} we show that TF theory becomes exact not just for energies, but
also for ionization energies in the semiclassical limit, for our model systems.
We review work showing that the periodicity of the periodic table remains significant, even as
$Z\to\infty$.  However, averaging over rows, numerical evidence suggests that (extended)-TF theory
yields  the  correct average in the large-$Z$ limit, at least for exchange.
In Sec. \ref{sec:PracFunc} we describe the relevance of this work to constructing useful functionals.
In Sec. \ref{sec:sums} we discuss a different approach to the semiclassical limit.  
Instead of focusing on finding corrections to the density (which may not improve the energy)
as in Secs. \ref{sec:box} and \ref{sec:RTP}, we focus on directly
finding approximations to the sums of eigenvalues (occupied energy levels).  
Finally, in Sec. \ref{sec:Conc}, we wrap everything up, connecting the results from the different sections.

\sec{Basics}
\label{sec:TF}
Thomas-Fermi theory was created around 1927 \cite{T27,F28} and Thomas does not mention the Schr\"odinger equation, perhaps because he had not yet heard of it.  In modern terms, he approximates the universal part of the energy functional \cite{B07,HK64} as
\begin{equation}
F\TF[\n] = T\TF[\n] + U[\n],
\end{equation}
where $T\TF[\n]$ is the local density approximation for the (spin-unpolarized) kinetic energy of non-interacting electrons:
\begin{equation}
\label{TFT3D}
T\TF[n] = \frac{3}{10} (3\pi^2)^{2/3} \int d^3 r\, n^{5/3}(\mbf{r}),
\end{equation}
and $U[\n]$ is the classical electrostatic self-repulsion of the electronic density, now called the Hartree energy:
\begin{equation}
U[\n] = \frac{1}{2} \int d^3 r \int d^3 r'\, \frac{n(\mbf{r}) n(\mbf{r}')}{|\mbf{r}-\mbf{r}'|}.
\end{equation}
For any $F[n]$, we can minimize the energy with respect to the density while holding the number of particles constant \cite{B07} yielding the Euler-Lagrange equation for the density:
\begin{equation}
\label{ELEq}
\frac{\d F}{\d n(\br)} + v(\br) = \mu.
\end{equation}
For TF theory, $\mu$ is the TF chemical potential, and for an atom or ion the external potential is $-Z/r$.  Because the functional derivative of $U$ is the Hartree potential, satisfying Poisson's equation, a second-order radial differential equation for the dimensionless potential results:
\begin{equation}
\Phi''(x)=\left[ \frac{\Phi^{3}(x)}{x} \right]^{1/2}_+,
\label{e343}
\end{equation}
where $x$ is a dimensionless coordinate, $x=Z^{1/3}r/b$, and $b=(1/2)(3\pi/4)^{2/3} \approx 0.885341$.  Here, we have used Lieb's notation \cite{L81}, where the subscript $+$ indicates that the function is set to zero unless its argument is positive.  For neutral atoms ($N=Z$), the unique solution has $\mu=0$ and \cite{S80}
\begin{equation}
\Phi(0) = 1\,, ~~~~\Phi'(0)=-B\,, ~~~~ B \approx 1.5880710226\,.
\label{eq:initcond}
\end{equation}
The density is then 
\begin{equation}
n\TF_{\ss Z}(r)=\frac{Z^{2}}{4\pi b^{3}}\left(\frac{\Phi}{x}\right)^{3/2}.
\label{bu1}
\end{equation}
While Eq. (\ref{e343}) can only be solved numerically, which is where the value of $B$ comes from, all its properties are perfectly well-defined.  Unlike the exact density, $n\TF(r)$ diverges as $1/r^{3/2}$ as $r\to 0$ and decays as $1/r^6$ for $r\to\infty$ \cite{LCPB09}.  Nonetheless, exact neutral densities weakly converge to this simple form as $Z$ grows, as seen in Fig. \ref{fig:NGDEN1}.

Inserting the TF density into the known contributions to the $Z$-expansion yields \cite{S80,S81}:
\begin{equation}
\label{EZasyCoeff}
c_0 = \frac{3B}{7b}, \qquad c_2 = \frac{44b}{9\pi^2} M_2,
\end{equation}
where \textcolor{Blue}{\cite{LCPB09}}
\begin{equation}
M_2 = \int_{0}^{\infty} dx\, \Phi^2(x) \approx 0.615434679.
\end{equation}
Thus the coefficients in Eq. (\ref{EZasy}) can be found by entirely elementary means, and yield errors only about 4 times larger than Hartree-Fock, but without any orbital calculation.  If we could achieve higher accuracy, and calculate them for molecules and solids, we might not need DFT at all.  A simpler version of our original question is:  How do we find HF-like accuracy without doing an electronic structure calculation?

\sec{Illustrations}
\label{sec:ill}
One of the simplest versions of this question can be deduced from the opening chapter of the ABC of DFT \cite{B07}.  It considers same-spin non-interacting fermions in a one-dimensional infinite well of width $L$, occupying the lowest $N$ levels.  This is a DFT analog of everyone's first quantum problem.   The individual eigenvalues are $\pi^2 j^2/(2L^2)$, with $j=1,2,...$  We find the total energy by summing over occupied eigenvalues:
\begin{equation}
\label{PIBE}
E_{\ss L}(N) = \frac{\pi^2 N^3}{6L^2} \left( 1 + \frac{3}{2 N}+\frac{1}{2 N^2} \right).
\end{equation}
The TF approximation, for same-spin fermions in a 1D potential $v(x)$, is (analogous to Eq. (\ref{TFT3D}))
\begin{equation}
\label{TF1d}
T\TF[\n] = \frac{\pi^2}{6} \int_{-\infty}^\infty dx\, \n^3(x).
\end{equation}
Here the one dimensional analog of Eq. (\ref{ELEq}) yields
\begin{equation}
\n\TF(x) = \frac{p\F(x)}{\pi},
\end{equation}
where
\begin{equation}
p\F(x) = \big(2[\mu-v(x)]\big)^{1/2}_+,
\end{equation}
is the classical momentum at energy $\mu$, which is found by normalizing the density:
\begin{equation}
\int_{-\infty}^{\infty} dx\, n(x) = N.
\end{equation} 
In general 
\begin{equation}
E\TF(N) = \int_{0}^{N} dN' \mu(N'), \qquad \mu =\frac{dE\TF}{dN},
\end{equation}
since $E\TF(0) = 0$.  For our flat box problem, $v=0$, the results are trivial:
\begin{equation}
\mu_{\ss L} = \frac{\pi^2 N^2}{2 L^2},~~~~\n\TF_{\ss L}(x) = \frac{N}{L},
\end{equation}
and
\begin{equation}
\label{PIBTF}
E\TF_{\ss L}(N) = \frac{\pi^2 N^3}{6L^2},
\end{equation}
consistent with the semiclassical limit, i.e., the relative error of TF vanishes as $N\to\infty$.  Nor is this an artifact of the potential being constant inside the box.  The TF statement is true for essentially any reasonable $v(x)$ for which the limit exists.  For these 1D non-interacting fermions, the local density approximation to the kinetic energy is a universal approximation, in the sense that it is a universal limit for all such problems, and relative errors must always vanish as the limit is approached.  Moreover, even for cases not close to this limit, the use of the functional form can yield surprisingly accurate results.

\begin{figure}[!htb]
\includegraphics[width=0.9\columnwidth]{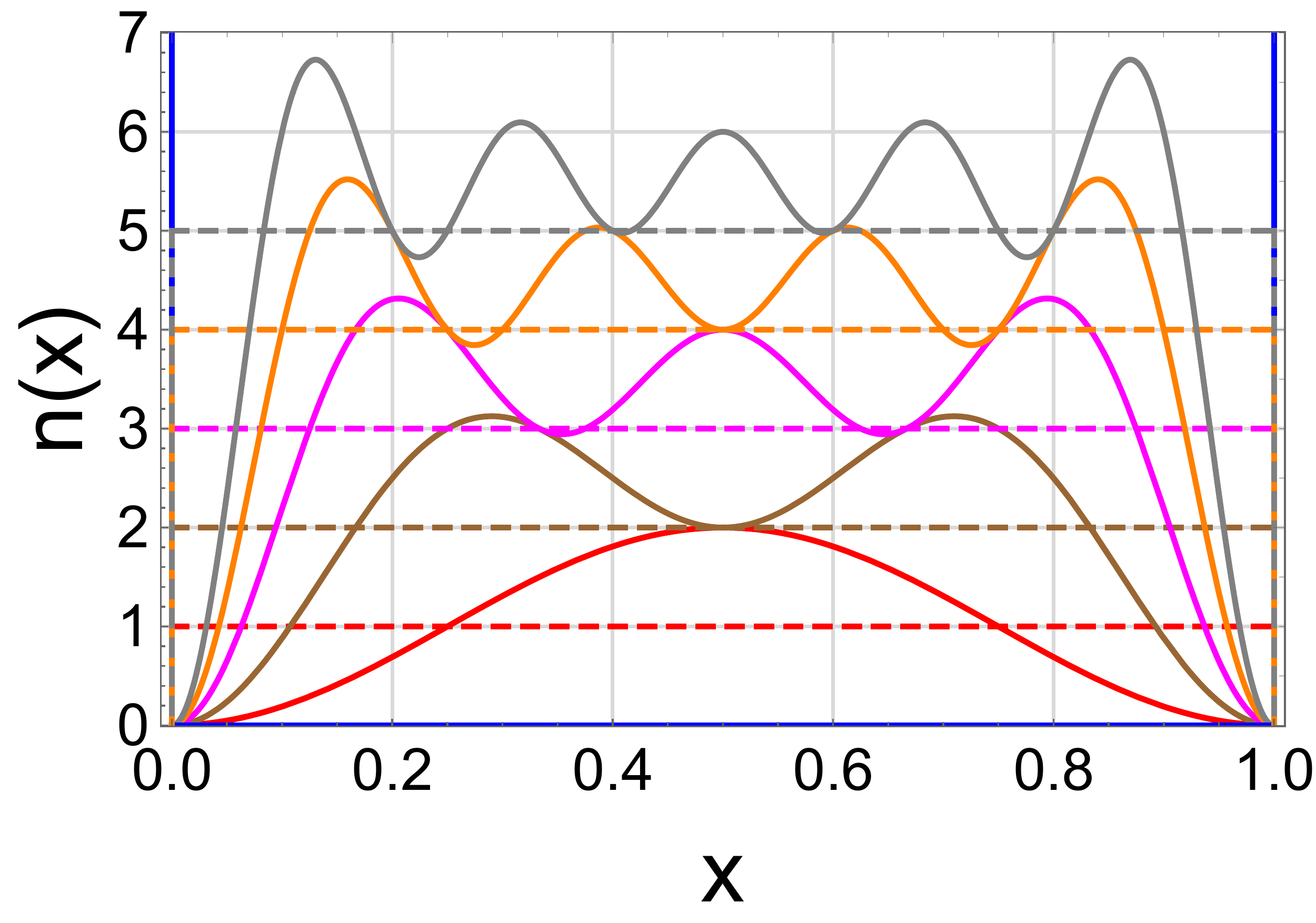}
\caption{Exact (solid) and TF (dashed) densities for the flat box, from $N = 1$ (red) to $N = 5$ (gray).}
\label{fig:PIB}
\end{figure}
In Fig. \ref{fig:PIB}, we plot the exact and TF densities for up to 5 particles in a flat box.  The inadequacies of our 1D TF approximation mimic those of the TF approximation for real atoms, Fig. \ref{fig:DenXe}.  The TF density fails to satisfy the boundary conditions and misses the quantum oscillations.  On the other hand, as $N$ grows, the TF density errors in the interior are of order 1, while at the edge they are of order $N$, but the edge region shrinks to within $O(1/N)$ of the walls.  The semiclassical limit requires that any integrals over smoothly varying functions of the density, such as $\int n^2$, will also become relatively exact in TF theory as $N\to\infty$.

\begin{figure}[!htb]
\includegraphics[width=0.9\columnwidth]{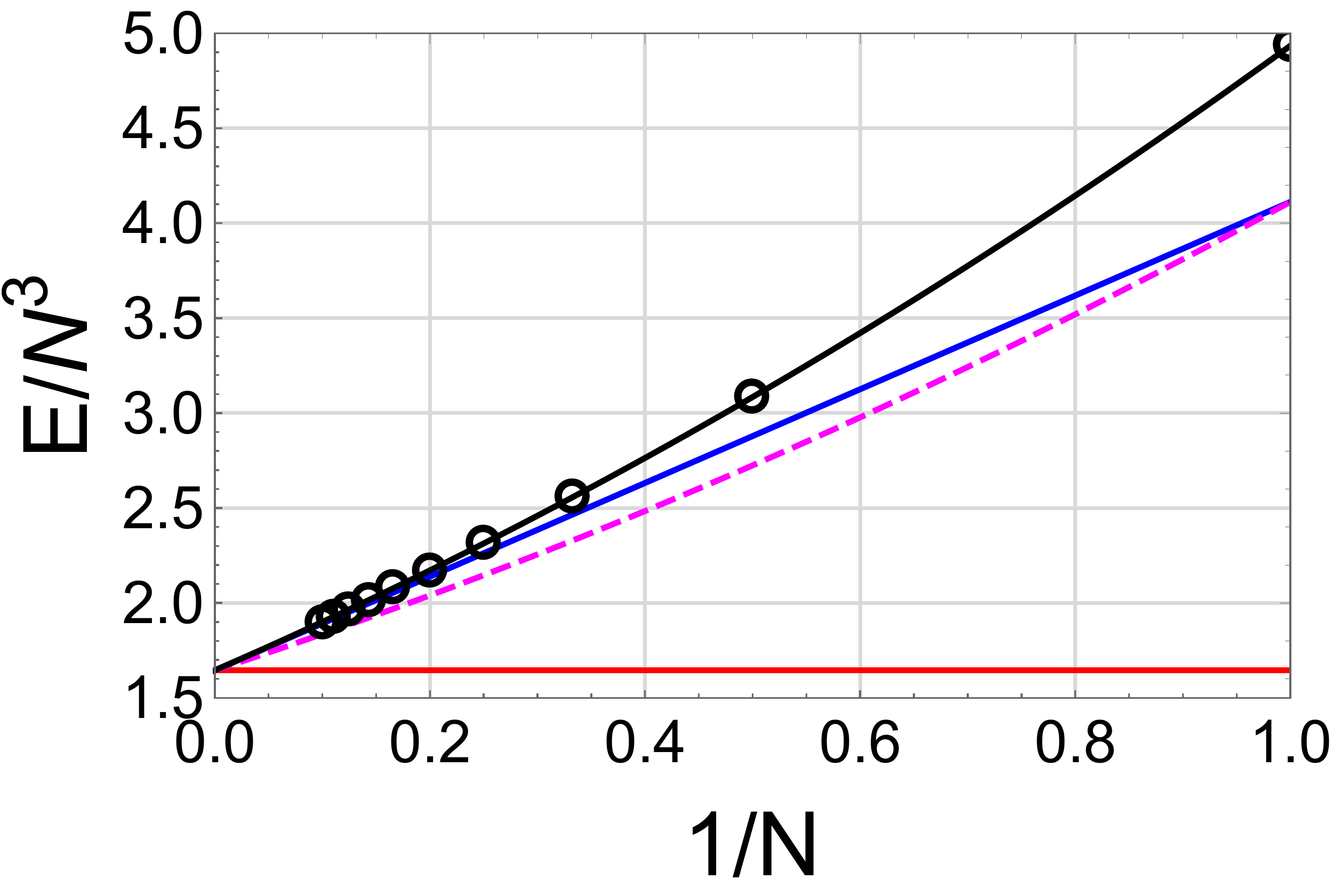}
\caption{Flat box energy with $L = 1$: exact (black circles), TF (red), 1st (blue) and 2nd (black) corrections from Eq. (\ref{PIBE}), and TF on exact density from Eq. (\ref{TTFn}) (magenta, dashed).}
\label{fig:PIBTComp}
\end{figure}
To illustrate how local approximations can be more accurate than they have a right to be, consider instead applying the TF approximation to the exact density, which can be found analytically with some effort:
\begin{equation}
\label{PIBDen}
\n_{\ss L}(x) = \frac{\bar{N}}{L}
 - \frac{\sin(2 \bar{N} \pi x/L)}{2 L \sin(\pi x/L)}, \qquad \bar{N} = N + \half.
\end{equation}
This yields:
\begin{equation}
T\TF[\n_{\ss L}] = \frac{\pi^2 N^3}{6L^2} \left( 1 + \frac{9}{8 N} + \frac{3}{8 N^2} \right).
\label{TTFn}
\end{equation}
In Fig. \ref{fig:PIBTComp} we compare the above expression to the TF energy in Eq. (\ref{PIBTF}) and the exact result in Eq. (\ref{PIBE}).  Fig. \ref{fig:PIBTComp} shows that, when evaluated on the exact density instead of the self-consistent one, the TF approximation is much more accurate.  Eq. (\ref{TTFn}) contains all three terms, and each is a good approximation to its exact counterpart.  [In modern DFT language, this illustrates that the TF approximation is dominated by density-driven errors \cite{VSKS19}, and qualitatively inaccurate results, such as Teller's non-binding theorem \cite{T62}, might not occur with better densities].  Note, however, from the Figure that simply including the leading correction to TF theory is everywhere more accurate than using the TF functional on the exact density.

\begin{figure}[!htb]
\includegraphics[width=0.9\columnwidth]{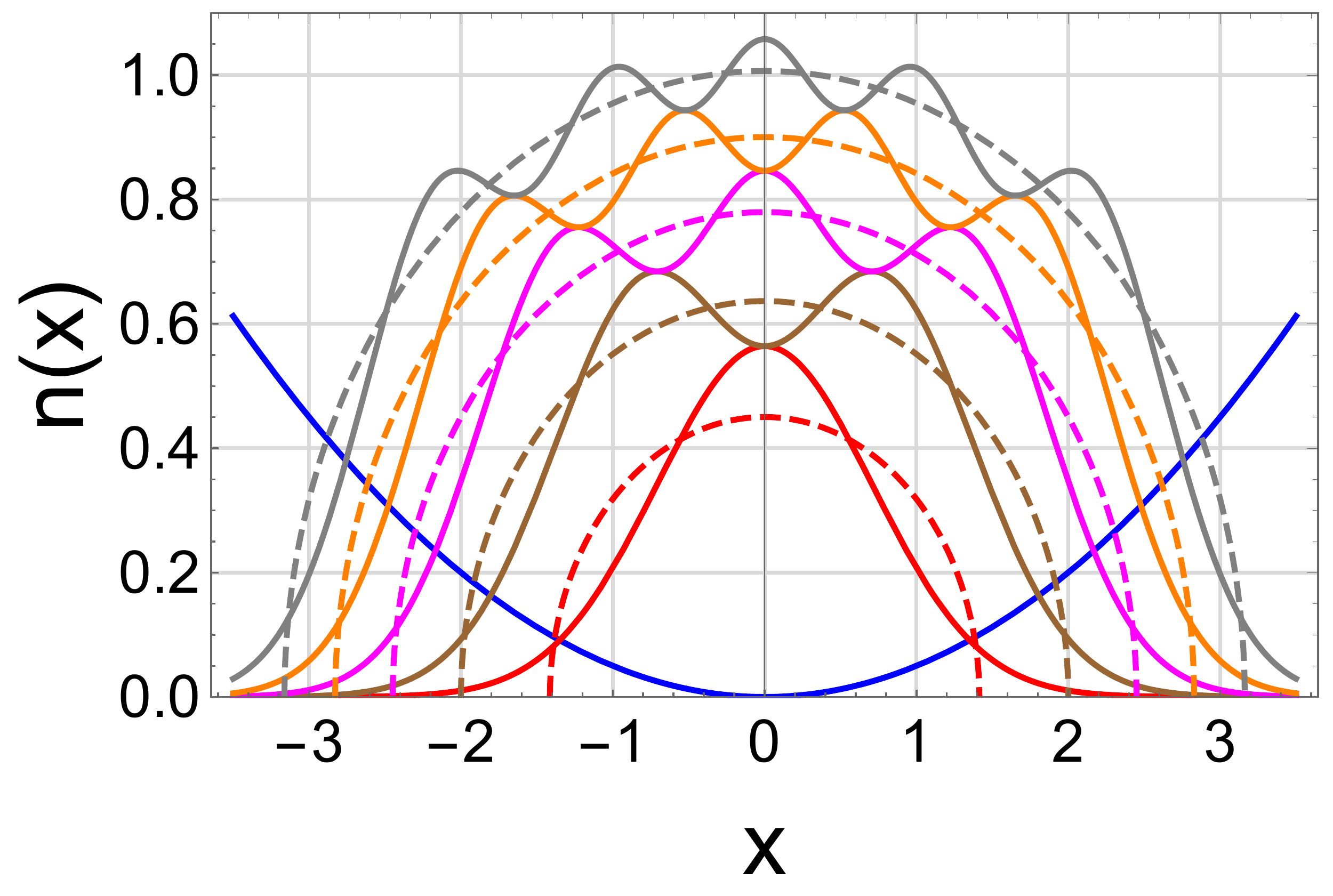}
\caption{Same as Fig. \ref{fig:PIB} but for the harmonic oscillator.  The blue curve is $v(x)/10$.}
\label{fig:HO}
\end{figure}
To make clear that there is nothing special about the flat box, we repeat this exercise for a harmonic oscillator with potential $v(x) = \omega^2 x^2/2$.  Here, the eigenvalues are $\omega (j+\half)$, where $j=0,1,2...$, so 
\begin{equation}
E_\om(N) = \om\, \frac{N^2}{2}.
\label{ENHO}
\end{equation}
In this case, instead of solving the TF equations, we could invoke the semiclassical limit, to deduce that TF must yield the exact answer for all $N$ for this problem, because there is only one term in the energy expression.   To see this explicitly, the TF density is
\begin{equation}
\label{HOTFDen}
n\TF_\om(x) = \frac{[2\mu_\om - (\om x)^2]^{1/2}_+}{\pi}, \qquad \mu_\om = \om\, N.
\end{equation}
Insert this density into Eq. (\ref{TF1d}) to find the exact answer of Eq. (\ref{ENHO}).  We plot both exact and approximate densities in Fig. \ref{fig:HO}.  The overall behavior and deficiencies of the TF density are similar to those of the flat box.  The semiclassical limit still guarantees vanishing relative error for both energies and expectation values.  Some details are slightly different from the flat box case.  Here the TF density has finite measure, whereas the exact density does not.  In the bulk region, the exact density oscillates around the TF density, whereas in the box, the exact density is almost everywhere above the TF density in the interior.  For future use, we also define Fermi turning points, $\pm x\F$, as the positions at which the TF density vanishes i.e. $v(x\F) = \mu$.  Beyond that, one sees the exact density decaying as a Gaussian.  We note the overall similarities to Fig. \ref{fig:DenXe}, which shows the same comparison for the Xe atom (5 filled shells).

We also note a slightly perplexing question.  For the HO the TF functional acting on the TF density yields the exact answer.  But the TF functional applied to the exact density yields a different (and therefore worse) answer, unlike the box example.  From a DFT perspective, surely this is a case of the right answer for the wrong reasons.  Or one might say this is the most extreme cancellation of errors ever.  Yet, almost all semiclassical approaches, such as WKB eigenvalues, yield the exact answer for the harmonic oscillator, so there is no surprise here for the semiclassics community \cite{G90,BB18,C14}.  We discuss this further in Sec. \ref{sec:RTP}.

So far, so simple.   But the crucial question is this:  If TF theory becomes exact in the semiclassical limit, do we have a procedure for calculating the next correction for the same set of reasonable $v(x)$, and can we design a density functional that will capture this correction?   The answer to the first question is yes, the second, no, so far.  In answering the first question, we will see (Sec. \ref{sec:sums}) that the correction is sensitive to the details of the boundary conditions.   Such a correction must yield exactly the $Z^2/2$ term in Eq. (\ref{EZasy}). 

\begin{figure}[!htb]
\includegraphics[width=0.9\columnwidth]{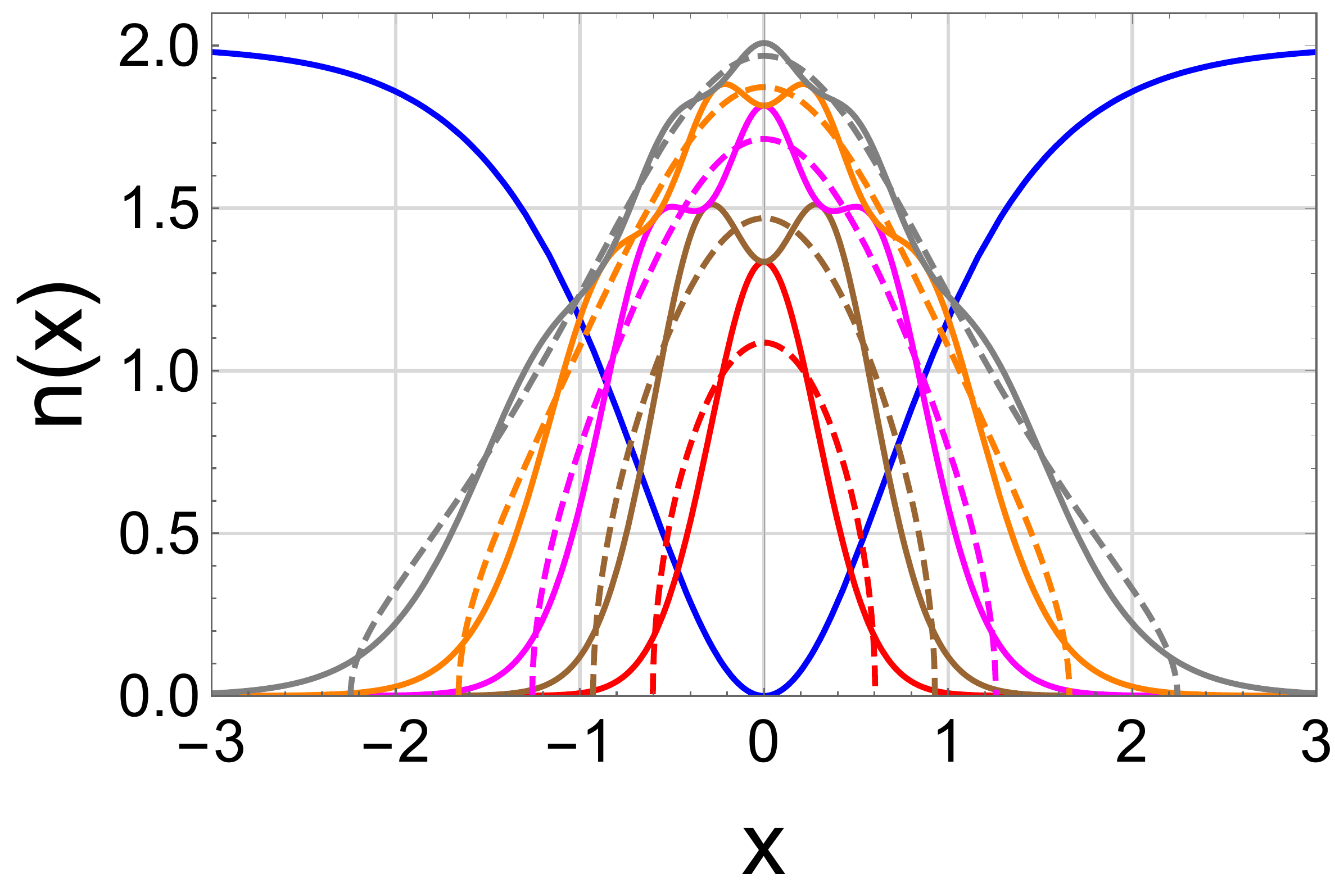}
\caption{Same as Fig. \ref{fig:HO} but for the PT well with $D = 20$.  The blue curve is $v(x)/10$.}
\label{fig:PTWDen}
\end{figure}
Our next example is the Pöschl-Teller (PT) well \cite{RM32,PT33} with potential,  
\begin{equation}
v_{\ss D}(x) = D\, \tanh^2 x.
\end{equation}
Because $v \ra D$ as $|x| \ra \infty$, this well (unlike the box or harmonic oscillator) binds only a finite number of states.  Even so, as the semiclassical limit is approached, the spacing between energy levels decreases and the number of states grows.  The eigenvalues are
\begin{equation}
\eps_{{\ss D},j} = D - \frac{1}{2} (\lam - j)^2, \qquad j=0,1,...,j_{\max},
\end{equation}
where
\begin{equation}
\lam = \sqrt{2D + \frac{1}{4}} - \half, \qquad D = \frac{\lam(\lam + 1)}{2},
\end{equation}
and $j_{\max} = \floor{\lam}$, the highest integer $\leq \lam$.  The exact solution is given in Ref. \cite{LL77}.  The semiclassical expansion is an expansion around large $D$, keeping $j$ proportional to $\lambda$, yielding
\begin{equation}
\label{PTWKB}
\eps_{{\ss D},j} = \eps^{(0)}(z_j) + \D\eps^{(2)}(z_j) + \D\eps^{(4)}(z_j) + ...,
\end{equation}
where $z_j = j + 1/2$ and
\begin{align}
\begin{split}
\eps^{(0)}(z) &= \sqrt{2 D} z - z^2/2,\\
\D \eps^{(2)}(z) &= \frac{1}{8} \left( \frac{z}{\sqrt{2D}} - 1 \right),\\
\D \eps^{(4)}(z) &= - \frac{z}{256 \sqrt{2} D^{3/2}}.\\
\end{split}
\end{align}
Unlike our previous examples, this expansion does not end at any finite order.  However, it is convergent unless $D$ is very small.  The exact energy of $N \leq \lam + 1$ occupied orbitals is
\begin{equation}
E_{\ss D}(N) = [6 N \lam - (2N - 1)(N-1)] \frac{N}{12}.
\end{equation}
The TF density is
\begin{equation}
\label{PTTFDen}
n\TF_{\ss D}(x) = \frac{[2(\mu_{{\ss D}} - D\, \tanh^2 x)]^{1/2}_+}{\pi}, \qquad \mu_{\ss D} = \left(\sqrt{2D} - \frac{N}{2} \right) N,
\end{equation}
and the TF energy is
\begin{equation}
E\TF_{\ss D}(N) = \left(\sqrt{\frac{D}{2}} - \frac{N}{6}\right)N^2,
\end{equation}
and the LS limit is $N \ra \infty$, with $N/\sqrt{D}$ fixed.

\begin{figure}[!htb]
\includegraphics[width=0.9\columnwidth]{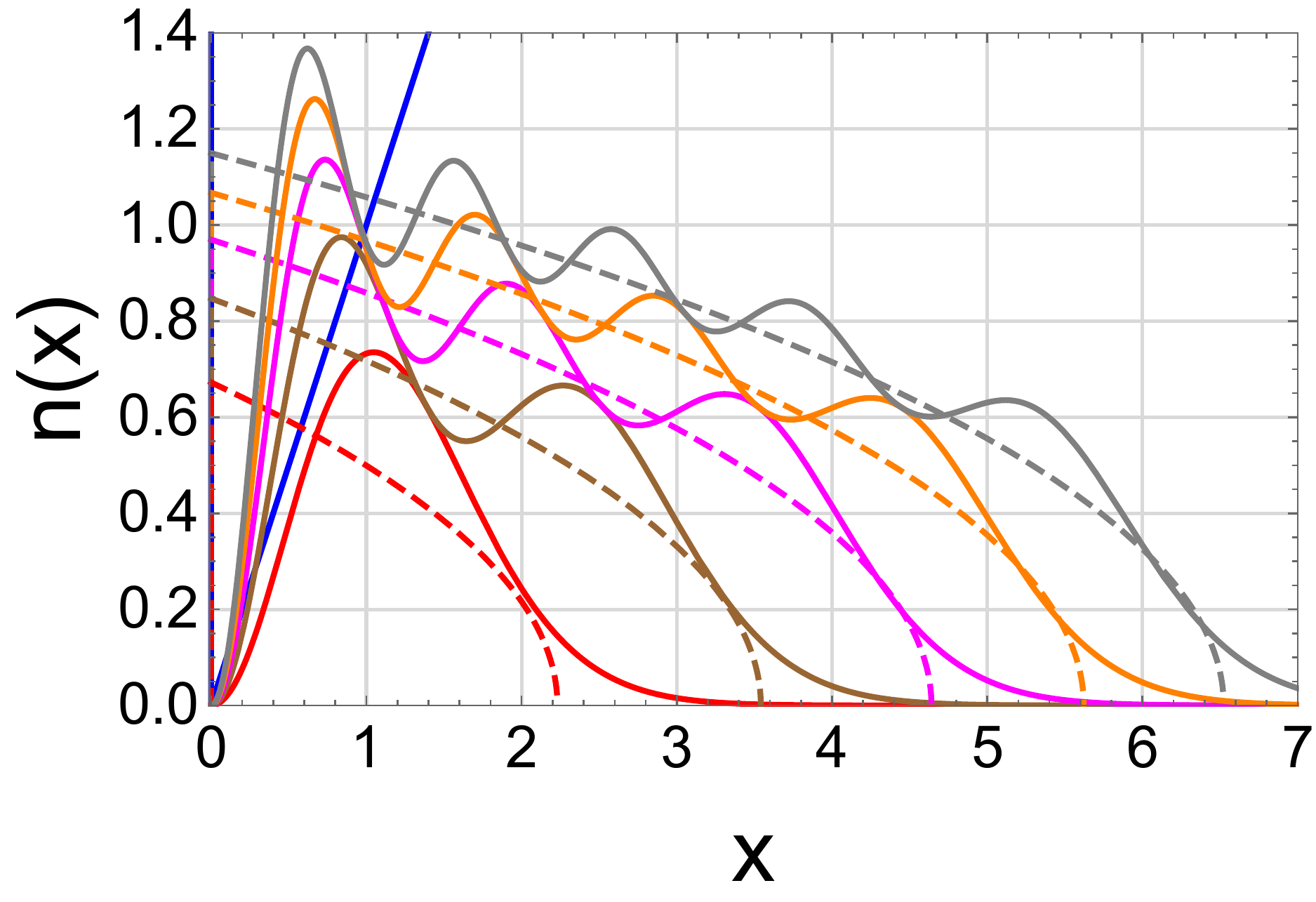}
\caption{Same as Fig. \ref{fig:PIB} but for the LHW with $F=1$.}
\label{fig:LHWDen}
\end{figure}
Our last model system is the most prototypical, because the semiclassical expansion is purely asymptotic, i.e., it never converges.  We will discuss this system in detail in Sec. \ref{sec:sums}.  The flat box has two hard walls (where the slope of the potential is infinite), while the harmonic oscillator has two real turning points where the slope is finite.  Now we consider the linear half well (LHW), defined only for $x \geq 0$, with potential
\begin{equation}
v(x)=Fx,
\end{equation}
which has one hard wall (at $x = 0$) and one real turning point.  The exact LHW solution is written in terms of the Airy function (Sec. 9 of Ref. \cite{DLMF}).  The unnormalized orbitals are
\begin{equation}
\label{LHWOrbitals}
\phi_n(x) = \Ai[(2F)^{1/3}x - a_n], \qquad n = 0,1,2,3,...
\end{equation}
where the $a_n$ are the negative of the zeroes of the Airy function: $\Ai(-a_n) = 0$.  The eigenvalues are
\begin{equation}
\label{LHWEigenval}
\eps_{{\ss F},n} = \left(\frac{F^2}{2}\right)^{1/3} a_n.
\end{equation}
There is no closed form expression, but the zeroes have a well-known asymptotic expansion (Sec. 9.9 of \cite{DLMF}) which corresponds to the semiclassical expansion: 
\begin{equation}
\label{LHWWKB}
\eps_{{\ss F},n} = F^{2/3} \sum_{m = 0}^{\infty} d_m z_n^{2/3-2m}, 
\end{equation}
where $z_n = 3 \pi (n + 3/4)$ and
\begin{equation}
d_0 = \half, \qquad d_1 = \frac{5}{24}, \qquad d_2 = -\frac{10}{9}.
\end{equation}
The exact density and energy are given by sums over Eqs. (\ref{LHWOrbitals}) and (\ref{LHWEigenval}).  The TF density is
\begin{equation}
\label{LHWTF}
n\TF_{\ss F}(x) = \frac{[2(\mu_{{\ss F}} - F x)]^{1/2}_+}{\pi}, \qquad \mu_{\ss F} = \frac{(3 \pi F N)^{2/3}}{2},
\end{equation}
The asymptotic expansion of the energy for large $N$ is
\begin{equation}
\label{LHWSU}
E_{\ss F}(N) = \frac{(3 \pi F)^{2/3}}{2} N^{5/3} \left(\frac{3}{5} + \frac{1}{4N} - \frac{g_1}{N^2}  + \frac{g_3}{4N^3}  + ...\right),
\end{equation}
where $g_j = 36^{-1}[1/4 + 5(j\pi^2)^{-1}]$ and the leading term is the TF energy.

\sec{Scaling}
\label{sec:LSScale}
In this section, we discuss the LS limit about which we are interested in expanding.
The most physical definition is to change the external potential {\em while simultaneously} scaling the particle number:
\begin{equation}
v_\zeta(\br) = \zeta^{4/d}\, v(\zeta^{1/d} \br),~~~~~~N\to\zeta N,
\label{vzeta}
\end{equation}
where $d$ is the dimension, and $\z$ varies continuously from 1 to $\infty$.  We call this $\z$-scaling.  For molecules with nuclear positions ${\bf R}_\alpha$ and charges $Z_\a$, under this scaling, $Z_\a\to \z Z_\a$ and ${\bf R}_\a \to \z^{-1/3} {\bf R}_\a$ \cite{L81}.  Thus, for an atom, $\zeta$ simply scales both $Z$ and $N$.  Thus our Eq. (\ref{EZasy}) is simply an expansion in $\z$, made explicit by replacing $Z$ with $\z Z$ everywhere.  The Lieb-Simon statement was proven only for Coulomb repelling electrons in Coulomb attracting potentials for 3D.  But this is in many ways the most difficult case and likely the result applies to any `reasonable' one-body potential.  We apply our $\z$-scaling to any reasonable one-body potential and interaction, and generally expect TF to become relatively exact in the limit of large $\z$.  We use this scaling even in non-interacting cases, where other choices satisfy the same criteria, in order to provide a unified treatment of both interacting  and non-interacting cases.

We can also apply $\z$-scaling to our 1D non-interacting problems, where the results are trivially related to changing $N$.  For the box,
\begin{equation}
T_{{\ss L},\z}(N) = T_{{\ss L}/\z}(\z N) = \frac{\pi^2 \z^5 N^3}{6L^2} \left(1+ \frac{3}{2 \z N} + \frac{1}{2(\z N)^2}\right),
\end{equation}
while for the oscillator,
\begin{equation}
E_{\om,\z}(N) = E_{\z^3 \om}(\z N)= \frac{\z^5 \omega\, N^2}{2}.
\end{equation}
The density is slightly more complicated, as the position coordinate must also be scaled.  But for our simple particle in a box, $v=0$, we have:
\begin{equation}
\label{PIBnvz}
\n_{\ss L}[v_\z] (x) = \n_{{\ss L}/\z,\z N}(x) = \frac{\zeta}{L}\left[\bar{N}_\z - \frac{\sin(2 \bar{N}_\z \pi \z x/L)}{2\sin(\pi \z x/L)}\right],
\end{equation}
with $\bar{N}_\z = \z N + 1/2$.  This density integrates to $\z N$ electrons and has width $L/\z$.  Note that this makes no sense unless $\z N$ is also an integer, to ensure the density still vanishes at $x=L/\z$.  For practical and aesthetic reasons, we will often plot a renormalized version of the exact density on the scaled potential, 
\begin{equation}
\label{TFNIZeta}
\tilde \n_\z(\br) = \z^{-2}\n[v_\z](\br/\z^{1/d}),
\end{equation}
so that it spans the original space and integrates to $N$, the number of particles in the original density, for all values of $\zeta$.  Then $\tilde{n}_\z(x)$ converges (weakly) to a fixed limit as $\z \to \infty$, namely the TF density with $N$ particles.  For example, for the flat box, the TF density of the $\z$-scaled problem is $\zeta^2 N/L$, and $\tilde\n_\z(x)=N/L$, for any $\zeta$.  The TF density of any system is invariant under the scaling of Eq. (\ref{TFNIZeta}). 

For real electronic systems, in what sense is this a semiclassical limit?  For large $\z$, even in the scaled coordinate system, the particle number is growing, and the local Fermi wavelength is shrinking as $\z^{1/d}$.  Hence any finite smooth potential varies ever more slowly.  In Coulombic systems, there will always be a region around the nucleus where this assumption fails \cite{S52}.  In general, slow variation in the potential is equivalent to a semiclassical approximation. Thus, each orbital in a KS calculation approaches its semiclassical approximation, and the contribution from any given orbital becomes vanishingly small.  However, the kinetic energy contains two derivatives and the sum over $N$ orbitals becomes large, so that the entire term scales as $\z^{7/3}$.  Likewise, the contribution from any pair of orbitals to the electron-electron repulsion becomes very small, but in such a way that the double sum over pairs remains significant and scales the same way as the kinetic energy, and the total tends to the Hartree energy.  Thus this is also a (very specific) mean-field limit.  

The very fact that the three distinct energy contributions scale the same way under $\z$-scaling to yield a non-trivial TF problem (i.e. the electrons do not become non-interacting, the system does not become unbound, etc.) is what defines the choices behind $\z$-scaling, and ensures that local density approximations yield useful results in precisely this limit.  It also links the importance of local approximations to the fact that both potential operators $\hat{V}$ and $\hat{V}_{ee}$ are diagonal in coordinate space.  But it is a very difficult limit to treat carefully, as it involves many subtleties, many of which are not yet understood.  For example, naively one would expect that since orbital pairs interact ever more weakly as the limit is approached, one could use perturbation theory, and weak interaction expressions.  But even when the density in most of an atom is large, the valence region remains finite and even contains regions of low density.   Since the valence region is vital to ionization and chemical reactions, a weak-interaction treatment will never be accurate in this limit.

\begin{figure}[!htb]
\includegraphics[width=0.9\columnwidth]{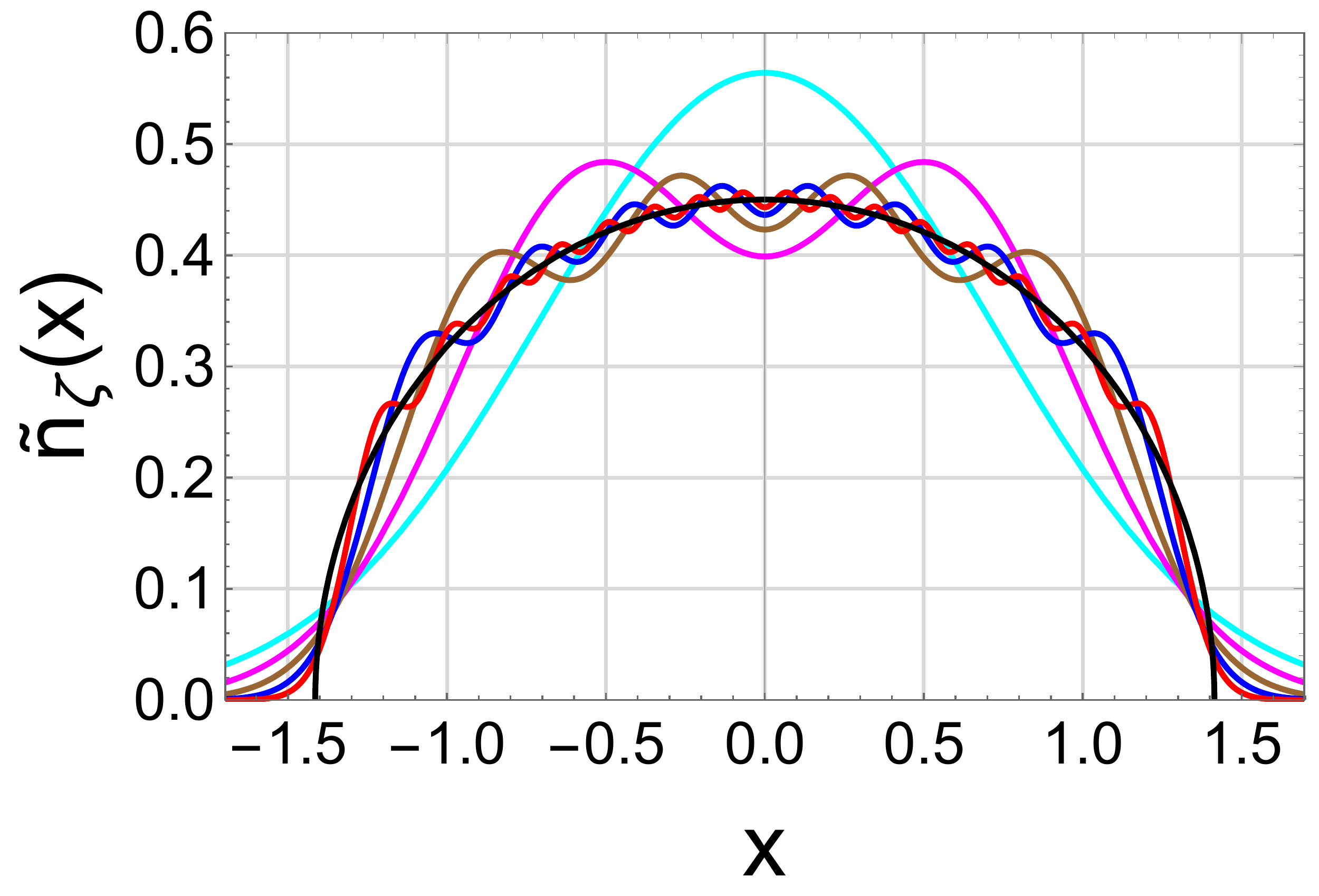}
\caption{Exact and TF harmonic oscillator ($\om = 1$) densities, $\z$-scaled according to Eq. (\ref{TFNIZeta}) with $v(x) = x^2/2$ and $N = 1$: TF (black), $\z = $ 1 (cyan), 2 (magenta), 4 (brown), 8 (blue), and 16 (red).}
\label{fig:SHO}
\end{figure}
\begin{figure}[!htb]
\includegraphics[width=0.8\columnwidth]{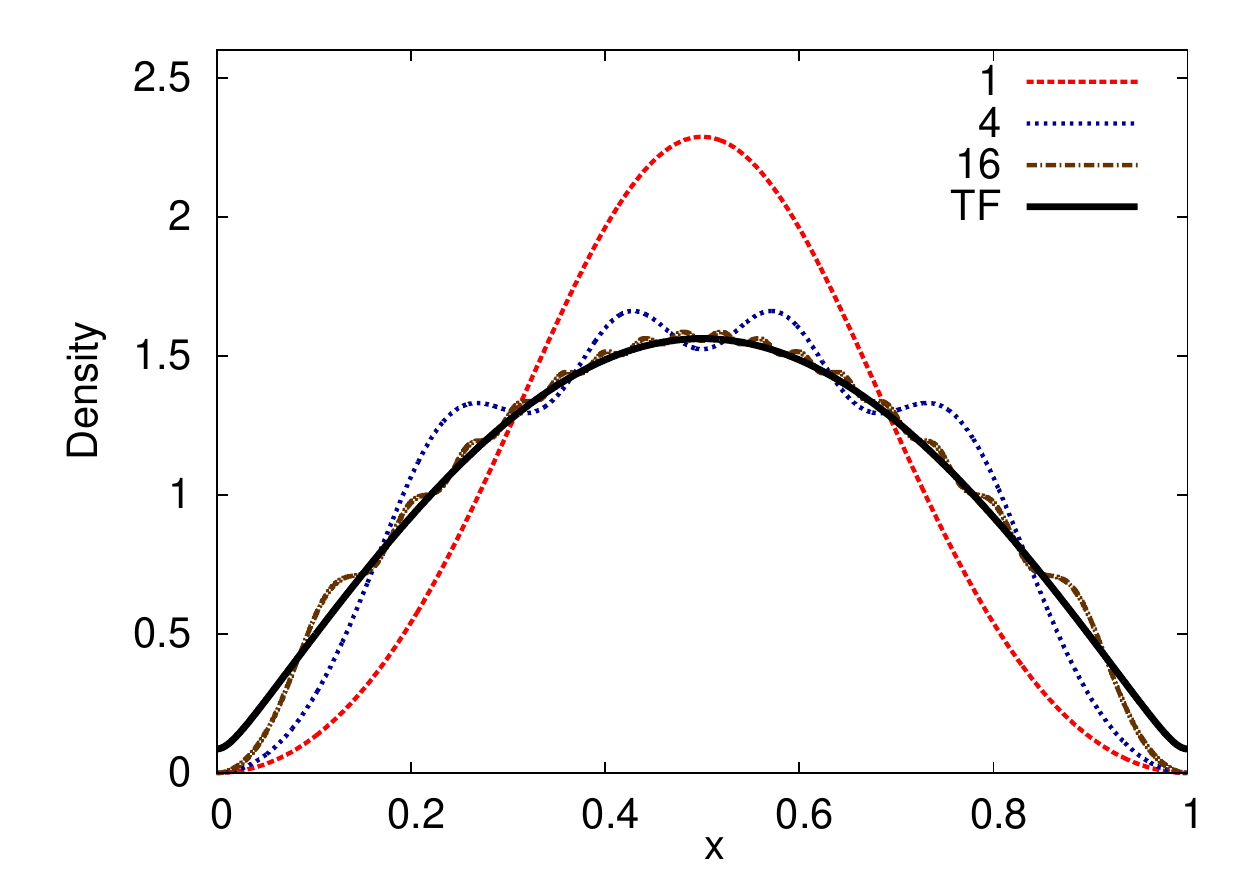}
\caption{TF and $\z$-scaled densities for $v(x) = 12 \cos^2(\pi x)$, box boundary conditions, and $\z = 1,4,16$.  Reproduced from Ref. \cite{CLEB10}.}
\label{fig:CLEB10FIG2}
\end{figure}

We can see this effect in Fig. \ref{fig:SHO}.  As $\z$ grows, the separation between adjacent peaks in the exact density, proportional to the Fermi wavelength, shrinks as does the amplitude of the oscillations.  In Fig. \ref{fig:CLEB10FIG2}, we show the renormalized potential-scaled density approaching the TF limit for a box ($0 \leq x \leq 1$), with potential $v(x) = D\, \cos^2(\pi x)$.  Now the distance between peaks in this scaled density is shrinking as $1/N$, and their amplitude is shrinking in the same manner.  The TF density is
\begin{equation}
\label{TFCosBox}
n_{\ss D}\TF(x) = \frac{\big(2[\mu - D \cos^2(\pi x)]\big)^{1/2}_+}{\pi},
\end{equation}
yielding
\begin{equation}
N = \frac{2^{3/2} \sqrt{\mu}}{\pi^2} \mathcal{E}_+\left(\frac{D}{\mu} \right),
\end{equation}
where
\begin{equation}
\mathcal{E}_\mp(x) = \int_{0}^{\pi/2} d\phi\, (1 - x \sin^2\phi)^{\mp 1/2},
\end{equation}
are the complete elliptic integrals of the first and second kind (Sec. 19.2(ii) of Ref. \cite{DLMF}).  The TF energy for this potential is
\begin{equation}
E_{\ss D}\TF = \frac{2^{3/2}\sqrt{\mu}}{(3\pi)^2} \left[ (4 D + \mu) \mathcal{E}_+\left( \frac{D}{\mu} \right) + 2 (\mu - D) \mathcal{E}_-\left( \frac{D}{\mu} \right)\right].
\end{equation}
We plot $\tilde\n\TF_{\z,D}(x)$, with $N = 1$ and $D = 12$, in Fig. \ref{fig:CLEB10FIG2}.

An alternative approach to this limit, and a more useful one in some circumstances, is to use the chemical potential \cite{B20b}.  For the exact case for non-interacting fermions, we define the energy and particle number as a function of $\mu$:
\begin{eqnarray}
N(\mu)= \sum_j \Theta (\mu-\eps_j),\nonumber\\
G(\mu) = \sum_j (\eps_j-\mu) \Theta (\mu-\eps_j),\\
E(\mu)=G(\mu)+\mu\, N(\mu).\nonumber
\end{eqnarray}
This is just the zero-temperature limit of the Fermi distribution.  Then simply scale
\begin{equation}
\tilde v_\zeta(\br) = \zeta^{4/d}\, \tilde v(\zeta^{1/d} \br),
\label{vmuzeta}
\end{equation}
where $\tilde v(x) = v(x)-\mu$.  This approaches the semiclassical limit also, but along a slightly different path, so the leading corrections can be different.

For the purpose of density functional theory, it is often useful to define the conjugate scaling for the density:
\begin{equation}
\n_\z (\br) = \z^2 \n(\z^{1/d} \br).
\label{denscale}
\end{equation}  
Note that this combines the usual coordinate scaling of DFT \cite{LP85} with scaling the particle number, i.e., $N_\zeta = \zeta N$.  So it is not the same as either coordinate scaling, or scaling the potential with a fixed particle number.  For the flat box,
\begin{equation}
\label{PIBnz}
\n_{{\ss L},\z}(x)= \frac{\z^2}{L} \left[{\bar N}_\z-\frac{\sin(2 {\bar N}_\z \pi \z x/L)}
{2\sin(\pi \z x/L)}\right],
\end{equation}
where $\bar{N}_\z = N \z + 1/2$.  Of course, just as in coordinate scaling, scaling the potential and the density are two different operations, and the density of the ground-state wavefunction for the scaled potential does not match the scaled density.  This means that in general
\begin{equation}
\n[v_\z](x) \neq \n_\z(x),
\end{equation}
despite both integrating to $\z N$, both being legitimate densities for the $\z$-scaled problem, and being the same for $\z=1$.  Careful examination (or plotting) of the densities from Eqs. (\ref{PIBnvz}) and (\ref{PIBnz}) shows their important difference, even for a particle in a box.  However, it is a simple exercise to show that these scalings are equivalent within TF theory, where, for our box problem,
\begin{equation}
\n\TF_{\ss L}[v_\z](x) = \n\TF_{{\ss L},\z} (x) = \z^2 \frac{N}{L},~~~~0 \leq x \leq L/\z.
\end{equation}

\begin{figure}[!htb]
\includegraphics[width=0.8\columnwidth]{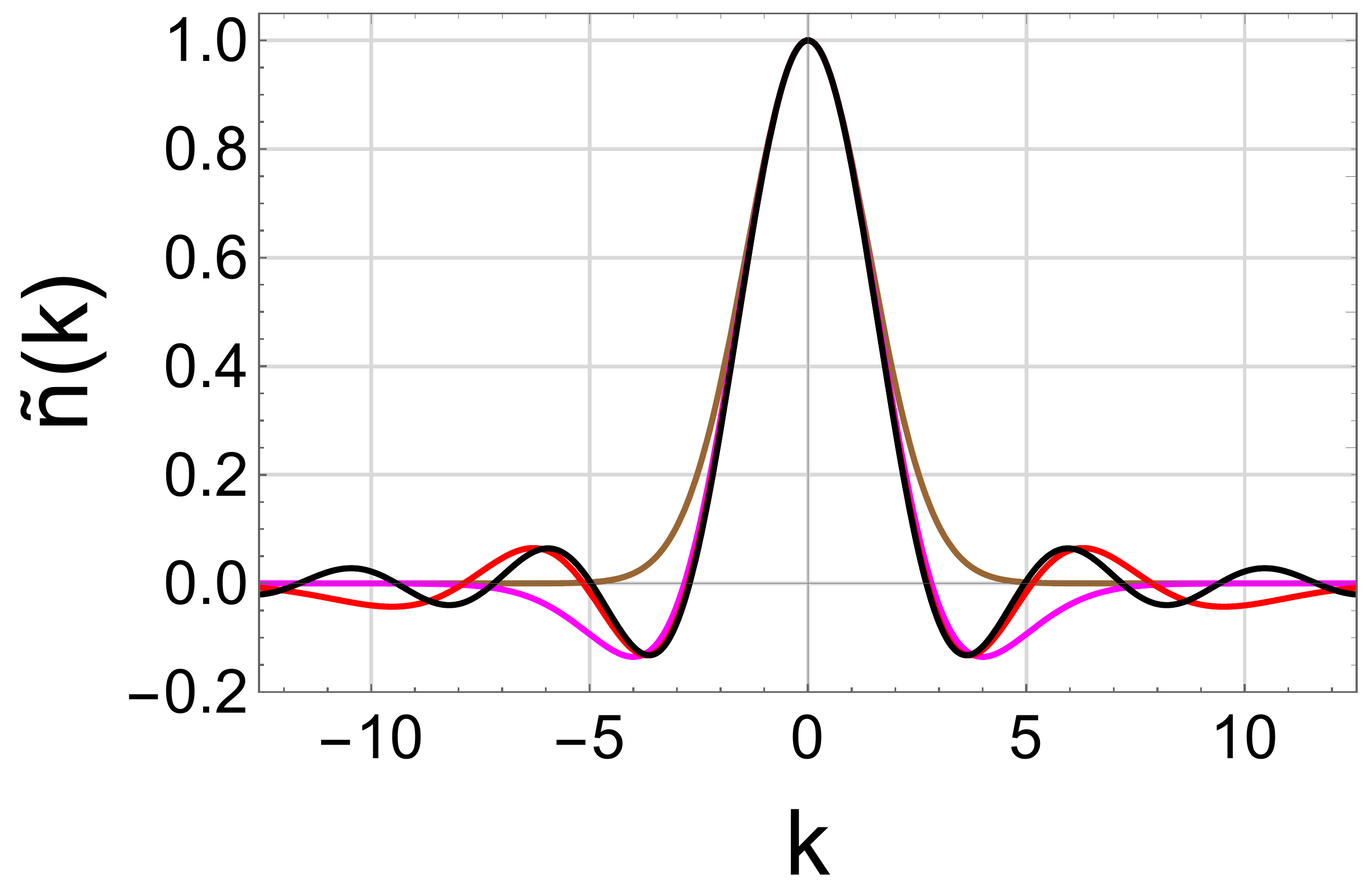}
\caption{Fourier transforms of scaled harmonic oscillator densities: TF (black), $N = 1$ (brown), $N=2$ (magenta), $N=4$ (red).}
\label{fig:FTHO}
\end{figure}

We now discuss the Fourier transform of the density for the harmonic oscillator:
\begin{equation}
\label{FTDef}
n(k) = \int_{-\infty}^{\infty} dx\, n(x) \cos(k x).
\end{equation}
For $\om = 1$ Eq. (\ref{FTDef}) yields
\begin{equation}
\label{FTHO}
\tilde{n}\TF(k) = \frac{\sqrt{2} J_1(\sqrt{2} k)}{k},
\end{equation}
where $J_m(x)$ is the Bessel function of the first kind (Sec. 10.2(ii) of Ref. \cite{DLMF}).  We plot this and the Fourier transforms of the exact $\z$-scaled densities of Eq. (\ref{TFNIZeta}) in Fig. \ref{fig:FTHO}.  At $k = 0$ all the densities have the same values of $\tilde{n}(k)$ (normalization) and $\tilde{n}''(k)$ (kinetic energy).  Fig. \ref{fig:FTHO} clearly shows the increasing accuracy of TF theory as $N$ grows.  For a given value of $|k|$, the exact and TF densities approach each other as $N$ increases.  Such an exercise is reminiscent of how imposing cut-offs in momentum space for XC holes produced the prototype of modern GGAs \cite{LP77,LM81}.

\sec{Box boundaries}
\label{sec:box}

In order to improve upon the TF approximation, we first show its relation to the famous WKB \cite{GS18,BO99} approximation for eigenstates.  Consider a box from $x = 0$ to $x = 1$ with $\mu > v(x)$ everywhere inside the box.  Things are relatively simple since the only turning points at the Fermi level are hard wall turning points.  The (leading order) WKB approximation to the wavefunction is
\begin{equation}
\label{PsiWKB}
\phi(\eps,x)=\frac{\sin\th(\eps,x)}{\sqrt{p(\eps,x)}},
\end{equation}
where $p(\eps,x)$ is the classical momentum at energy $\eps$ and 
\begin{equation}
\label{PhaseWKB}
\theta(\eps,x) = \int_{0}^{x}dx'\ p(\eps,x'),
\end{equation}
is the phase accumulated from the left wall.  We denote $\theta(\eps) = \theta(\eps,L)$ as the phase across the entire well.  The eigenvalues are determined by the requirement that the wavefunction vanish at the right wall, yielding the WKB eigenvalue condition for this problem:
\begin{equation}
\label{BoxQuantCond}
\theta(\eps) = j\pi,~~~j=1,2,3,...
\end{equation}

The trick is now to find the semiclassical approximation to the sum of the squares of the eigenstates, i.e., the density, which is {\em not} a simple sum of the WKB eigenfunctions squared, but the semiclassical approximation to this sum \cite{CLEB10,ELCB08}.  As $N$ grows, $\phi^2$ from Eq. (\ref{PsiWKB}) contains a term which oscillates ever more rapidly (with $2\theta\F$, where F denotes evaluation at the TF Fermi energy $\mu$, i.e $p\F(x) = p(\mu,x)$) which must exactly cancel the TF density at the boundaries.

While there is a long history of derivations of this kind of formula (see next section), a key condition of a correct solution is that it be a uniform approximation in all space, i.e., it must capture the leading correction to the TF density everywhere in space.  In Ref. \cite{ELCB08}, the first paper in which such a formula was found (because the  problem is simplified by the box boundary conditions), this was achieved by clever contour tricks using the semiclassical Green's function.  Later \cite{CLEB10}, it was simplified using the Euler-Maclaurin summation formula.  Either derivation yields a beautifully simple formula for the density, purely in terms of classical quantities.  Write
\begin{equation}
\label{BoxTime}
\tau(\eps,x) = \frac{\partial }{\partial \eps} \theta(\eps,x) = 
\int_0^x \frac{dx'}{p(\eps,x')},
\end{equation}
as the classical time required for a particle with energy $\eps$ to reach $x$, starting from the left wall.  Then $\tau(\eps,L)$ is the time to reach the opposite wall at that energy.  Here we use the subscript $F$ as shorthand for evaluation at $\mu$, the TF chemical potential, but for $N + 1/2$ particles \cite{ELCB08}.  This semiclassical density is then simply
\begin{equation}
\n\sc(x)
= \frac{p\F(x)}{\pi}-\frac{\om\F \sin{[2\theta\F(x)]}}{2 \pi\, p\F(x) \sin \a\F(x)},
\label{ns}
\end{equation}
where $\a\F(x) = \pi \tau\F(x)/\tau\F(L)$ and $\om\F = \pi/\tau\F(L)$.  An analogous formula, using the same ingredients, can be derived for the kinetic energy density, and so yields corrections to the TF approximation for the kinetic energy.

There are many remarkable features of these formulas.  It turns out that many researchers had sought such formulas over decades in several different fields \cite{A61,SZCB62,P63,P64,E88,ELCB08,KSb65,G66,BZ73,LY73,LL75}.  It was only by using box boundaries to avoid turning points that it was possible to perform the derivation with elementary techniques.  Perhaps the most remarkable feature is that the approximation to the fully quantum density contains only classical quantities.  No differential equation need be solved to evaluate it.   Moreover, these quantities are evaluated at only one energy, the TF chemical potential.  Thus, all properties are determined by the (semiclassical) highest occupied orbital.   This is reminiscent of Fermi liquid theory \cite{P12}. Only knowledge in the vicinity of the Fermi energy is relevant to the result.

We also note that these are approximate densities in terms of the potential, which is not the way DFT usually operates.  In DFT, one usually starts from an energy functional and uses the Euler equation, Eq. (\ref{ELEq}), to find the density as a functional of the potential.  (Recently there has been interest and progress in finding XC potentials as direct functionals of the density \cite{TB09}, the reverse of what is accomplished here).  In the next section, we will discuss what that means at a formal level.

Analyzing Eq. (\ref{ns}) as a functional of the potential, we see that the TF term (the first one) is a local functional of the potential, but using the (globally determined) TF chemical potential.  The leading correction is  also a local functional of the accumulated phases, which themselves are highly non-local functionals of the potential (their spatial derivatives are local).  Thus, the correction is not a higher-order correction in the sense of the gradient expansion (no derivatives of $v(x)$ occur), but is a phase-accumulated term arising from the boundary.  It would be absent if we had used periodic boundary conditions.  If the derivation were continued to one more order, one would expect to find the leading gradient correction to TF, plus a phase-dependent term that depends on gradients of the potential.

Naively, there is no such thing as a semiclassical expansion of the density, because the form of the expansion itself varies with $x$.  If we expand semiclassically (equivalent to powers of $1/N$), the TF term is order $N$ everywhere, whereas the correction has precisely those features mentioned above needed to achieve correct boundary conditions, and to smoothly vary from fully canceling the TF term at the boundaries to becoming a relatively small correction in the interior.  In this sense, it is a uniform approximation to the density:  Its relative error vanishes for all $x$, despite the different nature of the different regions.

\begin{figure}[!htb]
\includegraphics[width=0.9\columnwidth]{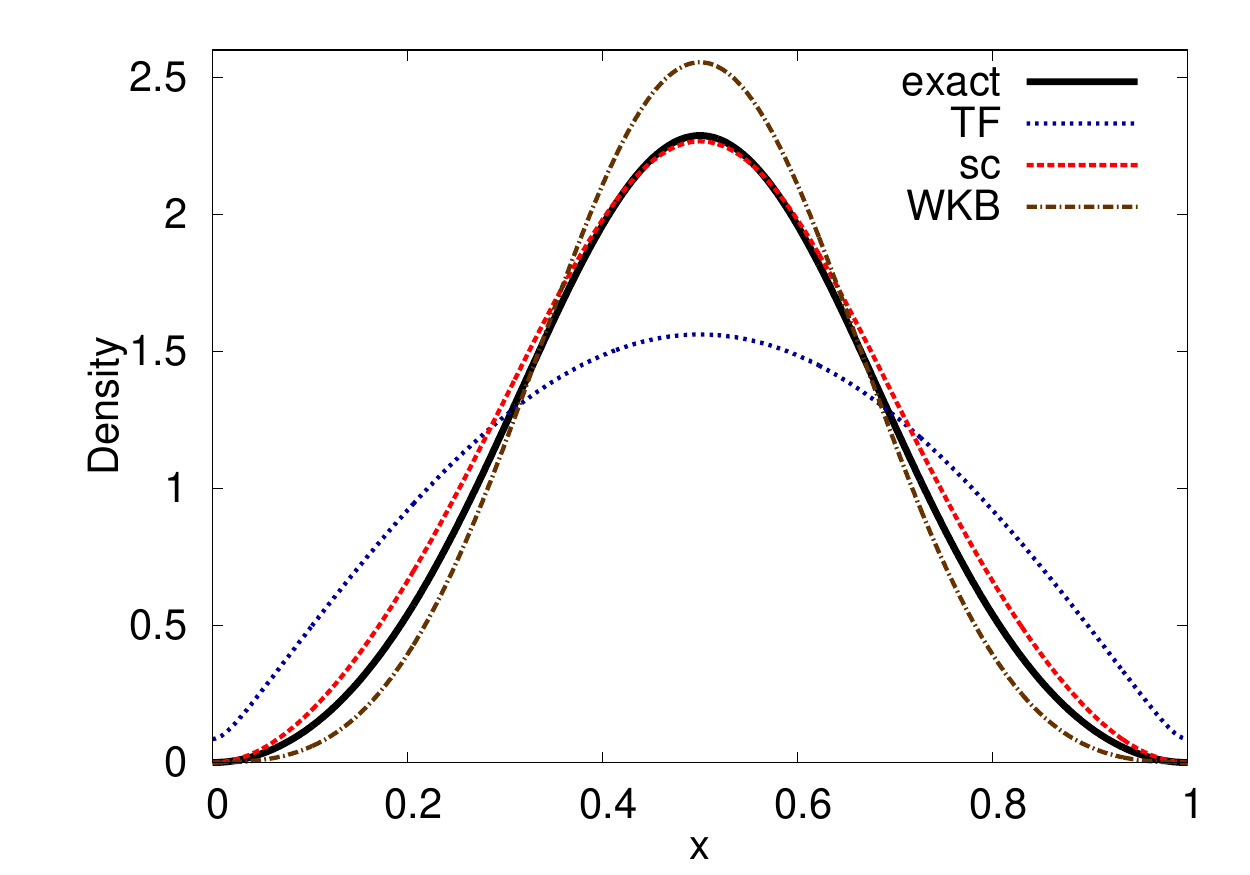}
\caption{Density of one particle in the single-dip potential of Fig. \ref{fig:CLEB10FIG2}: exact, TF, Eq. (\ref{ns}), and its WKB approximation (square of the WKB orbital in Eq. (\ref{PsiWKB})).  Reproduced from Ref. \cite{CLEB10}.}
\label{fig:PostTFBoxN1}
\end{figure}

Lastly, we note that, for $v(x)=0$, all phases become linear in $x$, and the semiclassical formula is {\em exact}.  In Fig. \ref{fig:PIB}, Eq. (\ref{ns}) has no error.  Moreover, it has been tested for many different non-constant potentials \cite{ELCB08,CLEB10}.  By construction, Eq. (\ref{ns}) becomes very accurate as $N$ increases, so almost all tests are done for $N=1$, the most difficult case.  For {\em any} $N > 1$, the difference between the exact density and Eq. (\ref{ns}) are indistinguishable to the eye.  In Fig. \ref{fig:PostTFBoxN1}, we plot results for a single particle in a potential with one dip that is infinitely differentiable.  Clearly, the semiclassical approximation is a huge improvement over both the TF and WKB approximations.

\begin{figure}[!htb]
\includegraphics[width=0.9\columnwidth]{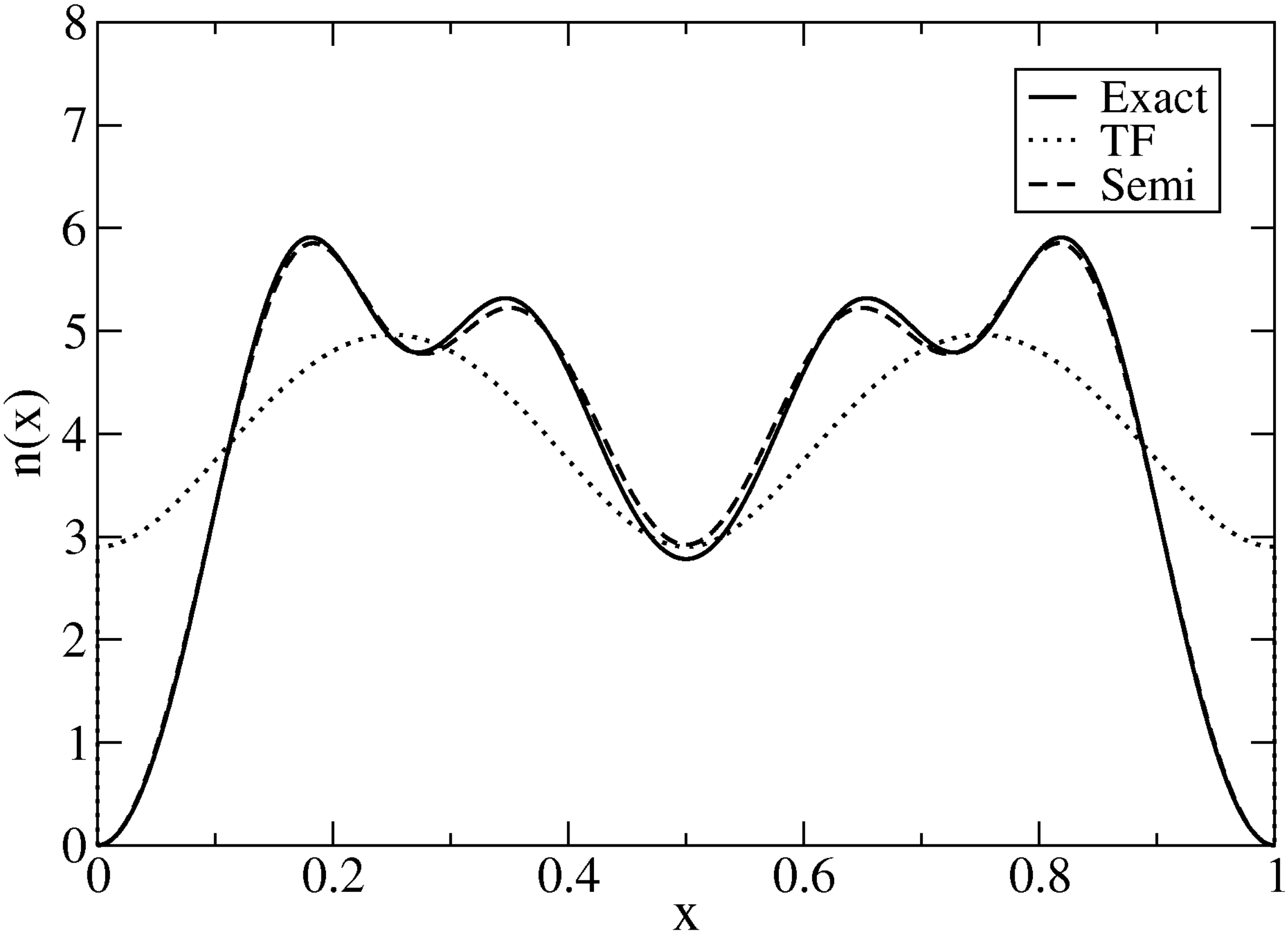}
\caption{Here $v(x) = - 80 \sin^2(2\pi x)$ and $N = 4$.  Reproduced from Ref. \cite{ELCB08}.}
\label{fig:PostTFBoxN4}
\end{figure}
An even more dramatic example is that of Fig. \ref{fig:PostTFBoxN4}.  In this case, the lowest two levels have negative energies and are almost degenerate, as the well has two dips.  By one measure, the reduction in error of the semiclassical density relative to TF is by a factor of 40 \cite{ELCB08}.  The one case where errors can be clearly seen is the one designed to make it fail: a potential whose TF chemical potential is only barely above the maximum of $v(x)$.   Even then, when almost all the density is decaying, the approximate density is not too bad (see Figs. 10 and 11 of Ref. \cite{CLEB10}).  

Thus, even from these simple beginnings, there was the suggestion that, if post-TF corrections could be derived, they might transform the entire nature of DFT development.  It holds out the possibility of DFT becoming a much higher accuracy theory, by involving ingredients that come out of such derivations, and look nothing like those in common use in DFT today.

\sec{Real turning points}
\label{sec:RTP}
The work described in Sec. \ref{sec:box} provided a definitive derivation of the leading correction to the local approximation to the kinetic energy, but only for the very special case of one dimension with box boundary conditions, and with the semiclassical chemical potential above $v(x)$ everywhere, thereby avoiding complications of real turning points (i.e., where the slope of the potential is finite when the classical momentum vanishes).  This was very important in showing, at least in one very simple case, that such formulas are possible to derive, contain only classical ingredients evaluated at the Fermi energy, and are more accurate than simple sums over WKB orbitals.

But atoms, molecules, and solids do not have box boundary conditions, so the next important step was to generalize to real turning points.  By real turning points, we mean those where the potential is finite at the turning point.   This is tricky.  In elementary discussions of WKB, such turning points are often treated by inelegant stitching formulas, using distinct generic WKB solutions in the traveling and tunneling regions, and approximating the potential linearly in the region of the turning point.  The domains of validity of these three regions overlap, so that this produces a spatially {\em uniform} approximation, i.e., its error vanishes in the limit no matter the value of $x$.  Langer found a more elegant solution to the turning point problem by showing \cite{L37} that an Airy function of an appropriate classical argument yields a uniform approximation \cite{B69,S05} in all three regions, no stitching required.  Thus uniform approximations to individual eigenstates can be found this way.

However, the trick is to find the density, by analyzing the sum of the squares of the orbitals.  One performs an asymptotic analysis of this form, yielding the TF result as the dominant contribution, and finding the leading correction that still is a uniform approximation.   This is a subtle problem in mathematical physics, and had been attempted over the decades in different disciplines (notably electronic structure and chemical and nuclear physics) as we discussed in Sec. \ref{sec:box}.  Previous attempts had often yielded partial results, but none produced the general uniform approximation that was needed.  This was achieved in Ref. \cite{RLCE15}, for both the density and a specific choice of kinetic energy density.  Despite being derived in the large-$N$ limit, these formulas are remarkably accurate for almost any system they can be applied to.

\begin{figure}[!htb]
\includegraphics[width=0.9\columnwidth]{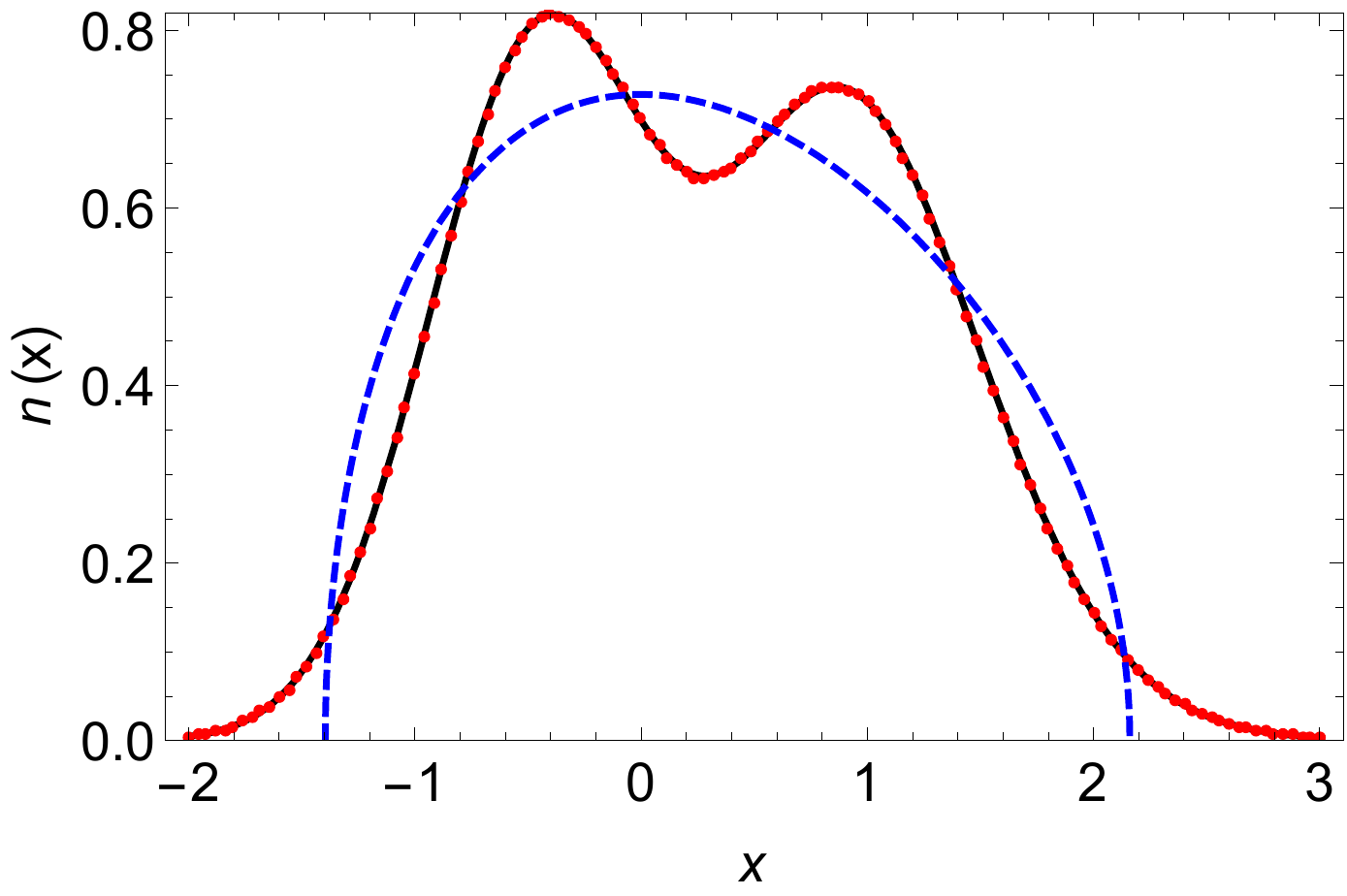}
\caption{Densities in a Morse potential with $N = 2$: exact (black solid), TF (blue dashed), and  semiclassical from Eq. (\ref{nsRTP}) (red dots).  Reproduced from Fig. 1 of Ref. \cite{RLCE15}.}
\label{fig:PostTFRTP}
\end{figure}

To write the semiclassical density, we first generalize the phase formula given in Eq. (\ref{PhaseWKB}) to
\begin{equation}
\label{RTPPhase}
\theta(\eps,x) = \int_{-x(\eps)}^{x} dx\, p(\eps,x).
\end{equation}
where $x(\eps)$ is the classical turning point at energy $\eps$ and, for simplicity we assume $v(x) = v(-x)$.  If $x < x(\eps)$, the phase becomes complex and we choose that branch on which the accompanying semiclassical wavefunction decays in space.  The box formula Eq. (\ref{PhaseWKB}) is just a special case of this, where the boundaries occur before true turning points appear.  The WKB energy condition becomes
\begin{equation}
\label{WKBQuantRule}
\theta(\eps)=\theta[\eps,x(\eps)]= \left(j+\half\right) \pi,~~~~~j=0,1,2...
\end{equation}
The $\half$ is called the Maslov index \cite{MF81} (and becomes 1 for two box boundaries, as in Eq. (\ref{BoxQuantCond})).  When this condition is satisfied, the right evanescent region is given by choosing the other branch.  Then $\tau(\eps,x)$ etc. can be defined analogously to Eq. (\ref{BoxTime}).  The density formula is \cite{RLCE15}
\begin{equation}
\label{nsRTP}
n\sc(x) = p_{\F}(x) B[\theta\F(x)] + q_{\F}(x) C[\theta\F(x)],
\end{equation}
where $B$ and $C$ are specific combinations of Airy functions.  Here F denotes TF quantities evaluated at $j = N - 1/2$ in Eq. (\ref{WKBQuantRule}), and $q\F(x)$ contains the quantum oscillations, analogous to those of the second term in Eq. (\ref{ns}).  In particular
\begin{align}
\begin{split}
B(\theta) &= \sqrt{z} \Ai^2(-z) + \frac{\Ai^{\prime 2}(-z)}{\sqrt{z}},\\
C(\theta) &= \Ai(-z) \Ai'(-z),\\
\end{split}
\end{align}
where $z=(3\theta/2)^{2/3}$, and
\begin{equation}
q_{\F}(x) = \frac{\om_{\F} }{\sin[\a\F(x)]\, p\F(x)} - \frac{p\F(x)}{3 \theta\F(x)},
\end{equation}
and $\a\F(x)$ is given after Eq. (\ref{ns}).  These formulas yield uniform approximations to the density. As the semiclassical limit is approached, the fractional error vanishes everywhere, i.e., in the classically allowed region, the evanescent region, and the vicinity of the turning points.  For more details see Refs. \cite{RLCE15,RB18}.  In Fig. \ref{fig:PostTFRTP} we show the remarkable accuracy of this approximation.  Note that even with $N = 2$, the post-TF density in Eq. (\ref{nsRTP}) is indistinguishable from the exact density.  And semiclassical approximations only improve as $N$ increases!

These results first appeared briefly in Ref. \cite{RLCE15}, while the full derivation appeared several years later \cite{RB18}.  Their derivation involves rather exquisite manipulations of products of Airy functions and their integrals \cite{VS10}.  The final result, Eq. (\ref{nsRTP}), reproduces all previous attempts \cite{A61,SZCB62,P63,P64,E88,ELCB08,KSb65,G66,BZ73,LY73,LL75}  under the specific circumstances in which they were derived.  The previous methods of Euler summation appeared too difficult to apply to this more complex situation, and instead the Poisson summation formula was used \cite{RLCE15,RB18}.

Next, the results were applied to about half a dozen special cases for which analytic formulas were available, and in which the leading correction could be explicitly checked \cite{RB17}.  Moreover, they were also applied numerically to several cases where no analytic example was available, showing great accuracy under almost all conditions.

Consider a potential that rises very rapidly outside a well-defined region, such as $0 < x < 1$.  As the limit of infinite rise is approached, the density will tend to 0 at the edges, and be dominated by its behavior in the traveling region.  Very likely, Eq. (\ref{nsRTP}) then reduces to Eq. (\ref{ns}), as suggested by Eq. (A.3) of Ref. \cite{RB18}.

But, in a surprise turn, the integrated energy from such approximations (which are very accurate pointwise), would sometimes be {\em less} accurate than the TF result \cite{RB17}.  This was finally traced to a very simple source.  The expansion of the density (and kinetic energy density) is in powers of $\zeta^{1/3}$ (here due to this being 1D), but the expansion of the energy is in powers of $\zeta$.  So the crucial question is what happens to integrated values in the semiclassical limit?  For example, exactly $(6 \pi \sqrt{3})^{-1} \approx 0.03$ of a particle leaks into the evanescent region beyond a turning point.  This can be seen in Figs. \ref{fig:HO}-\ref{fig:LHWDen} for larger $N$.  So while, pointwise, the corrections to the kinetic energy are highly accurate, their integrated effect over the entire system, becomes identically {\em zero} in the semiclassical limit, and likewise for the contributions to the potential energy due to the density expansion \cite{RB17}.  To find the leading correction to the energy from this expansion, one would need the leading three corrections, not just one!  This is very different from the box case, where the leading corrections in the density and kinetic energy density, when integrated, directly yielded the leading correction to the energy, being of the same order.

Moreover, in one dimension, TF theory for the harmonic oscillator, applied self-consistently, yields the exact energy, as we saw in Sec. \ref{sec:ill}.   Since most smooth potential wells can be approximated harmonically, this means TF theory is exact at the harmonic level, and only anharmonic contributions lead to corrections.  Thus the TF energy in 1D is typically highly accurate already.

\sec{Potential functionals}
\label{sec:PFT}

The astute reader will have noticed that Eqs. (\ref{ns}) and (\ref{nsRTP}) (and their analogs for the kinetic energy density) are functionals of the potential.  Moreover, unlike the circumstances of the gradient expansion (Sec. \ref{sec:GEA}), there is no obvious path to turn them into density functionals.  An alternative is to take them at face value, as functionals of the potential.  This is the potential functional theory (PFT) approach.  But then one can ask, in analogy to DFT, is there a corresponding variational principle, using the potential as the basic variable, and could it be applied to these formulas?  If so, would it yield even more accurate answers?

The answer is yes.  Yang, Ayers, and Wu \cite{YAW04} had in fact already explored such a theory for the KS scheme, demonstrating the duality between potentials and densities, with specific application to the optimized effective potential procedure \cite{GKG97}, used in the XC problem.  But we discuss it here at the more general level of Hohenberg-Kohn theory \cite{HK64}.  For simplicity, we present formulas for the non-interacting case, but they all apply to fully interacting arbitrary electronic systems \cite{CLEB11}.

Define the kinetic energy as a functional of the potential:
\begin{equation}
T[v] =  \langle \Psi_v | \hat T | \Psi_v \rangle,
\end{equation}
where $\Psi_v$ is the ground-state wavefunction of potential $v(x)$.  Similarly, define $\n[v](x)$ as the density of that wavefunction.  We do not use a mark to distinguish potential functionals from density functionals, but the argument dictates which functional it is.  They are simply related:
\begin{equation}
T[v]=T[\n[v]],~~~~T[\n]=T[v[\n]],
\end{equation}
where $v[\n]$ is the inverse of the $\n[v]$ map.  Then clearly we may write
\begin{equation}
\label{Ev1}
E[v] = T[v] + \int_{-\infty}^{\infty} dx\, \n[v](x)\, v(x).
\end{equation}
For any pair of potential functional approximations to $T$ and $\n(x)$, we get an approximation to $E$ for any potential.  We call this direct application of the potential functionals.  But what about a variational principle?  From the Ritz principle, clearly
\begin{equation}
\label{Ev2}
E[v] = \min_{v'} \left\{ T[v'] + \int_{-\infty}^{\infty} dx\, \n[v'](x)\, v(x)\right\},
\end{equation}
for the exact pair of functionals.  For any pair of approximations, one could perform this search and (possibly) improve an energy estimate.  But a drawback of PFT is the need for two different functionals.  Is there some consistency condition that these two functionals should satisfy?

The answer is yes, the stationary condition at the minimum, i.e., the Euler equation in PFT, relates the functional derivatives with respect to the potential.  If this relation is not satisfied, the solution to the minimization problem makes no sense, and so this should be a required condition.  Moreover, one can functionally integrate the Euler equation to derive a potential functional for $T[v]$ 
via a coupling constant integral over $\n[v](x)$.  (This constant mutiplies the one-body potential,
whereas usually such constants multiply the interaction potential.)
This is important, because then any approximation for $\n[v](x)$ uniquely determines an approximation for $T[v]$, and hence the energy.  The constructed kinetic energy functional is a functional of a functional, which we call an ffunctional \cite{CGB13}.  Moreover, variational compatibility of these two functionals is only guaranteed if the density-density response function derived from $\n[v](x)$ is symmetric under interchange of its arguments.   This general condition is compatible with an earlier result of Gross and Proetto \cite{GP09}, which showed variational compatibility if both the energy and density approximations had been derived from the same approximation to the Green's function.

\begin{figure}[!htb]
\includegraphics[width=0.9\columnwidth]{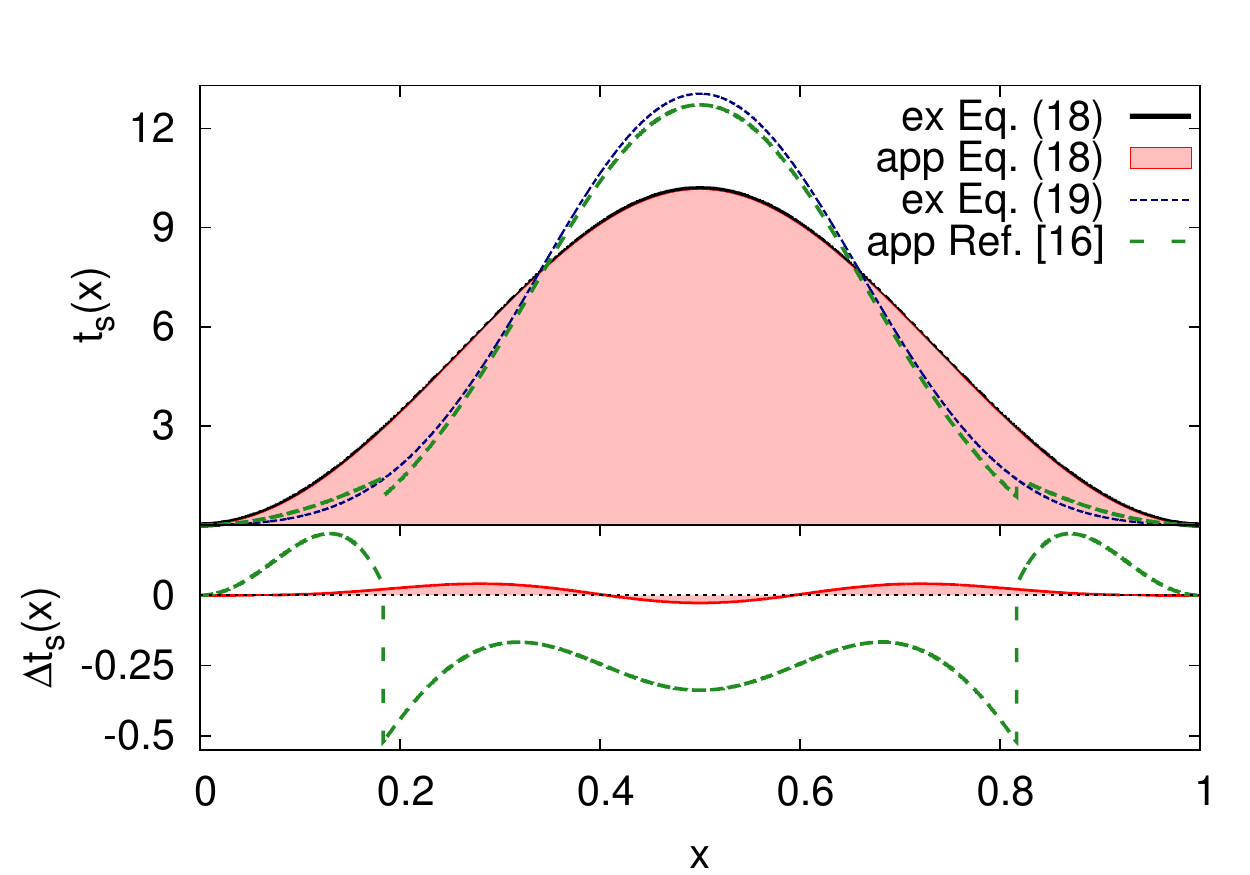}
\caption{Kinetic energy densities for one particle in a box with $v(x) = - 5 \sin^2(\pi x)$, and their errors (lower panel).  Reproduced from Ref. \cite{CLEB11}, to which the equations and references refer.  Ref. 16 in Ref. \cite{CLEB11} is Ref. \cite{CLEB10} and the approximation in green comes from Eq. (\ref{ns}).  How can there be two exact kinetic energy densities?  Because only the integrated kinetic energy is physically meaningful, so as long as both exact densities integrate to the exact value, they are correct.  This is the ambiguity of any energy density \cite{PRSB14}.}
\label{fig:PFTDEMO}
\end{figure}

The box semiclassical approximations (Eq. (\ref{ns}) for the density and its kinetic energy density analog) and the coupling-constant kinetic energy derived from Eq. (\ref{ns}) alone, via the procedure described above, were compared for simple cases.  The coupling-constant formula was found to be much more accurate than the original formulas, and to capture more terms exactly in the asymptotic expansion.  It was found that neither combination minimizes at the exact potential, as neither satisfies the symmetry condition.  Both are more accurate in direct evaluation than at their minima.  The failure of the symmetry condition was traced to the small normalization error in Eq. (\ref{ns}), which is only guaranteed to yield $N$ as $\z \ra \infty$.  In Fig. \ref{fig:PFTDEMO} we show how well PFT approximates the kinetic energy density.

The potential functionals described here are not the same as the Lieb potential functional \cite{L83} or the more general bifunctional of Englert \cite{E92}, which is simultaneously a functional of both the density and a potential, and was used to extract several quantum corrections to TF theory \cite{ES84,ES84b}.
	 
\sec{Gradient expansions}
\label{sec:GEA}
If one considers a slowly-varying potential, periodic throughout space, in which the chemical potential is above $v(x)$ everywhere, then the traditional gradient expansion of density functional theory corresponds to our $\z$-expansion.  Essentially, powers of $\hbar$ correspond to gradients of the density.  We illustrate this here in one dimension.

Begin with the density as a potential functional \cite{SP99}:
\begin{equation}
\label{SPnv}
\n[\tilde{v}](x) = \frac{p[\tilde{v}](x)}{\pi} \left[ 1 + \frac{\tilde{v}''(x)}{12 p^4[\tilde{v}](x)} + ...\right],
\end{equation}
where $p[\tilde{v}](x) = [-2\tilde{v}(x)]^{1/2}_+$ and $\mu$ is determined by normalization and contains corrections to its TF value.  Similarly, the kinetic energy density is
\begin{equation}
t[\tilde{v}](x) = \frac{p^3[\tilde{v}](x)}{2\pi} \left[ \frac{1}{3} + \frac{\tilde{v}''(x)}{4p^4[\tilde{v}](x)} + ...\right].
\end{equation}
Many higher orders can be easily generated \cite{SP99}.  Applying $\z$-scaling by simultaneously scaling $v(x)$ and $\mu$, as in Eq. (\ref{vmuzeta}), we find
\begin{equation}
\label{GEAScaleDenPot}
\n[\tilde{v}_\z](x) = \z^2\, \frac{p[v](\z x)}{\pi} \left( 1 + \frac{\tilde{v}''(\z x)}{12 \z^4 p^4[v](\z x)} + ...\right),
\end{equation}
and similarly for $t[v_\z](x)$.  Integrating up, and accounting for the scaling of the coordinate yields
\begin{equation}
T[v_\z] = \z^5\, T\TF[v] + \z^3\, \D T^{(2)}[v] + \z\, \D T^{(4)}[v]+...,
\end{equation}
where $\Delta T^{(j)}$ is the $j$th-order gradient correction \cite{SP99}.  Thus orders in the gradient expansion correspond to orders in $\z^2$.  To make the corresponding density functionals, we invert Eq. (\ref{SPnv}), power by power, yielding the potential in gradients of the density, and insert, to find:
\begin{equation}
T[\n_\z] = \z^5\, T\TF[\n] + \z^3\, \Delta T^{(2)}[\n] + \z\, \Delta T^{(4)}[\n]+...,
\end{equation}
where (in 1D) $\Delta T^{(2)}[\n] = - T\W[\n]/3$, and $T\W$ is the well-known
von Weisacker functional \cite{W35},
\begin{equation}
T\W[\n] = \frac{1}{8} \int_{-\infty}^{\infty} dx\, \frac{\n^{\prime 2}(x)}{\n(x)}.
\end{equation}
We note that the leading gradient correction in 1D is negative, which implies that it is not useful to find the densities of finite systems self-consistently with the GEA in 1D.  The three dimensional analog \cite{K57,SD85,M81,Y86,H73} is
\begin{equation}
T[\n_\z] = \z^{7/3}\, T\TF[\n] + \z^{5/3}\, \Delta T^{(2)}[\n] + \z\, \Delta T^{(4)}[\n]+...,
\end{equation}
and the individual contributions differ, e.g., $\Delta T^{(2)}[\n] =  T\W[\n]/9$.  In this section we have described the GEA in 1D and 3D.  The GEA in 2D is more subtle, and we refer interested readers to Ref. \cite{TLNM16}.

For any infinitely differentiable periodic $v(x)$, the gradient expansion is likely an asymptotic expansion, just as the WKB expansion is.  We will show later that modern methods of dealing with asymptotic expansions \cite{BH93} can produce tremendously accurate approximations.  For now, our point is simply that $\z$-scaling can be applied to all electronic systems, but in the special case of a slowly varying gas, an expansion in $\z$ coincides with the long known gradient expansion.  For every system, the leading order term is the local approximation, but the corrections, for finite systems, depend on the boundaries, and do not appear in the traditional gradient expansion.

\sec{Three dimensions}
\label{sec:3D}
A fun and instructive exercise that introduces the complications of degeneracy in 3D without the complications of interactions is to consider truly non-interacting electrons in an attractive Coulomb well.   All the orbitals are hydrogenic, as are their energies. This looks very different from real atoms, as there is no screening of the nuclear charge.  Almost all physical properties are very different, even the shell structure.  However, several analytic results are derivable, making it quite instructive.  Moreover, the (single-particle) potential does not change, so it is truly an example of potential scaling.  This is called a Bohr atom \cite{E88,HL95}.  

In this section, because of the analogy to real systems, we switch to doubly occupying the orbitals, so the lowest non-trivial case is $N=2$.  

\begin{figure}[!htb]
\includegraphics[width=0.9\columnwidth]{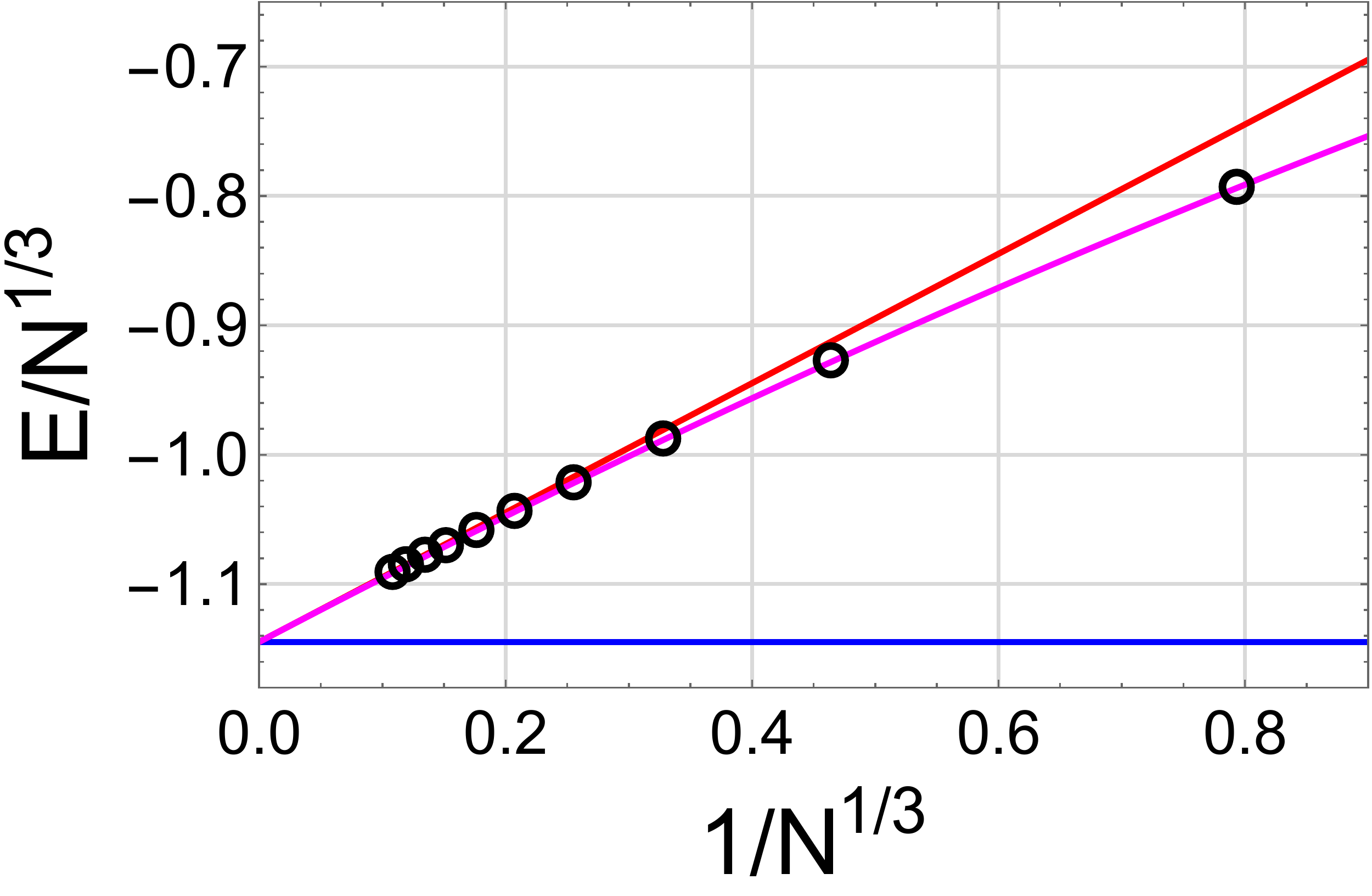}
\caption{The exact closed shell Bohr atom energies (black circles, Eq. (\ref{BAE})) and their approximation in Eq. (\ref{BAEA}), to zeroth order (TF, Blue), first order (red), and second order (magenta) in $N^{-1/3}$.}
\label{fig:BAE}
\end{figure}

First we give the expansion for the energy in powers of $N$.  We can find the kinetic energy from the virial theorem: $E(N) = -T(N)$.  The first shell holds 2 electrons with orbital energy 1/2, the next holds 8 with 1/8, the third holds 18 with  1/18, and so forth.  By elementary means, the exact expression, for closed shells only, is
\begin{equation}
\label{BAE}
E_Z(N) = -\frac{Z^2}{2} \left(A - 1 + \frac{1}{3A} \right),
\end{equation}
where $A^3 = 6 N [1 - \sqrt{1 - (972 N^2)^{-1}}]$.  Unlike for real atoms, we have an analytic exact expression for the energy.  The number of particles in the $s$-th closed shell is
\begin{equation}
N_s = \frac{1}{3} s (s+1) (2s+1),
\end{equation}
which is quite different from the usual Madelung rule \cite{M36} for filling shells in the periodic table.  The large-$N$ expansion of Eq. (\ref{BAE}) is
\begin{equation}
\label{BAEA}
E_Z(N) = -Z^2 \left[ \left(\frac{3N}{2}\right)^{1/3} - \half + \frac{1}{6(12N)^{1/3}} + ...\right]
\end{equation}
\cite{BCGP16}. For $Z=1$ and $N=2$, this yields 1.0000297, i.e., an error of only 30 microHartrees!  We show how well Eq. (\ref{BAEA}) approximates the exact result in Eq. (\ref{BAE}) in Table \ref{tab:BA}.  Eq. (\ref{BAEA}) is the analog of Eq. (\ref{EZasy}) in the introduction, with the electron-repulsion turned off.  The exact and approximate curves are shown in Fig. \ref{fig:BAE}, which is the analog of Fig. \ref{fig:AtmEn}.  Clearly, the first three terms are sufficient to yield extremely accurate energies for $N \geq 2$.  

\begin{table}
$\begin{array}{|c|c|c|c|c|r|}
\hline
\multicolumn{3}{|c|}{} & \multicolumn{3}{c|}{\text{Error}}\\
\hline
\text{Shells} & \text{N} & \text{Exact} & \text{TF} & 1^{\text{st}}\text{ corr. (mH)} & \multicolumn{1}{c|}{2^{\text{nd}}\text{ corr. }(\mu\text{H})} \\
\hline
1 & 2   & -1 & -0.44 & 58 & -29.68 \\
2 & 10  & -2 & -0.47 & 34 &  -2.09 \\
3 & 28  & -3 & -0.48 & 24 &  -0.38 \\
4 & 60  & -4 & -0.48 & 19 &  -0.11 \\
5 & 110 & -5 & -0.48 & 15 &  -0.03 \\
6 & 182 & -6 & -0.49 & 13 &  -0.02 \\
\hline
\end{array}$
\caption{Bohr atom energies with $Z=1$ and the errors of the asymptotic expansion in Eq. (\ref{BAEA}).}
\label{tab:BA}
\end{table} 

\begin{figure}[!htb]
\includegraphics[width=0.9\columnwidth]{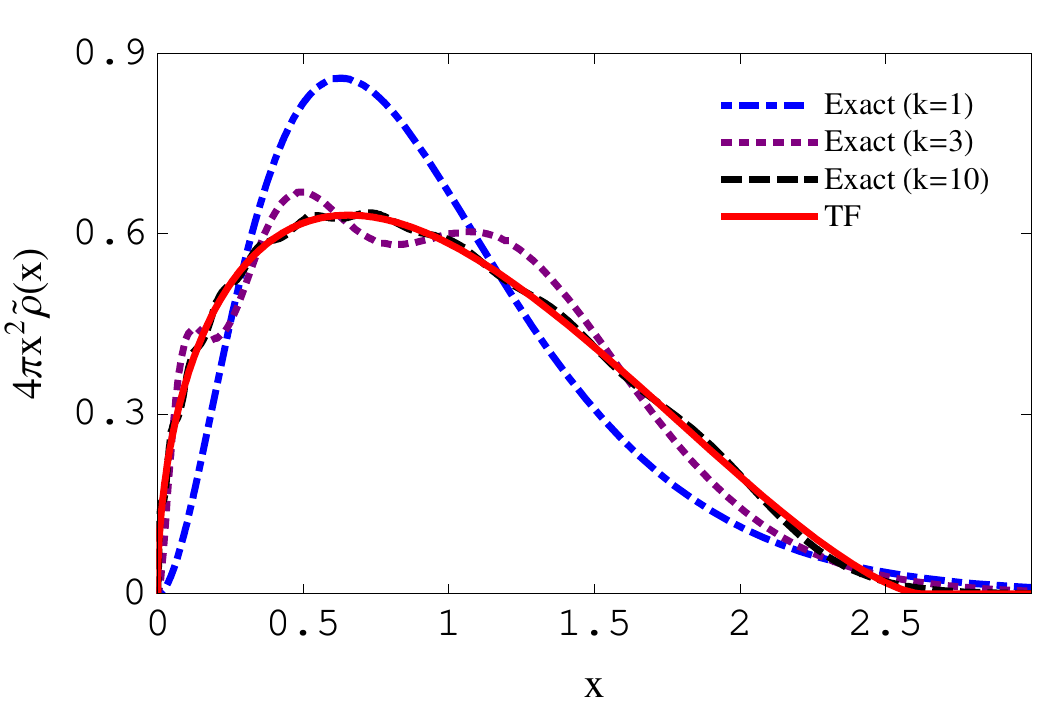}
\caption{Exact and TF radial scaled densities of closed shell Bohr atoms, where $k$ is the number of filled shells and $\tilde{\rho} = \tilde{n}$.  Compare with Fig. \ref{fig:DenXe}.  For details about the scaling, see Eqs. (\ref{BAScale}) and (\ref{BATFScale}).  Reproduced from Ref. \cite{SOLR11}.}
\label{fig:BAdens}
\end{figure}
The TF density is \cite{BCGP16},
\begin{equation}
\label{BATFDen}
n_{{\ss Z}}\TF(r) = \frac{(2 Z)^{3/2}}{3\pi^2} \left( \frac{1}{r} - \frac{1}{r_0} \right)^{3/2}_+,
\end{equation}
where
\begin{equation}
r_0 = \frac{(18 N^2)^{1/3}}{Z}, \qquad \mu_{{\ss Z}} = - \frac{Z}{r_0}.
\end{equation}
This is the solution of Eq. (\ref{e343}) without interaction (so that $\Phi''(x) = 0$).  Inserting Eq. (\ref{BATFDen}) in $T\TF[n]$ of Eq. (\ref{TFT3D}) yields the leading-order in Eq. (\ref{BAEA}).  The approach of the exact density to the TF density is shown in Fig. \ref{fig:BAdens}, where we choose $Z = 1$ and scale the densities by
\begin{equation}
\label{BAScale}
\tilde{n}(x) = N n(x), \qquad x = N^{-2/3} r,
\end{equation}
yielding
\begin{equation}
\label{BATFScale}
\tilde{n}\TF(x) = \frac{2^{3/2}}{3\pi^2} \left( \frac{1}{x} - \frac{1}{18^{1/3}} \right)^{3/2}_+.
\end{equation}
This is trivially related to the $\z$-scaling in Sec. \ref{sec:LSScale}.

\begin{figure}[!htb]
\includegraphics[width=0.9\columnwidth]{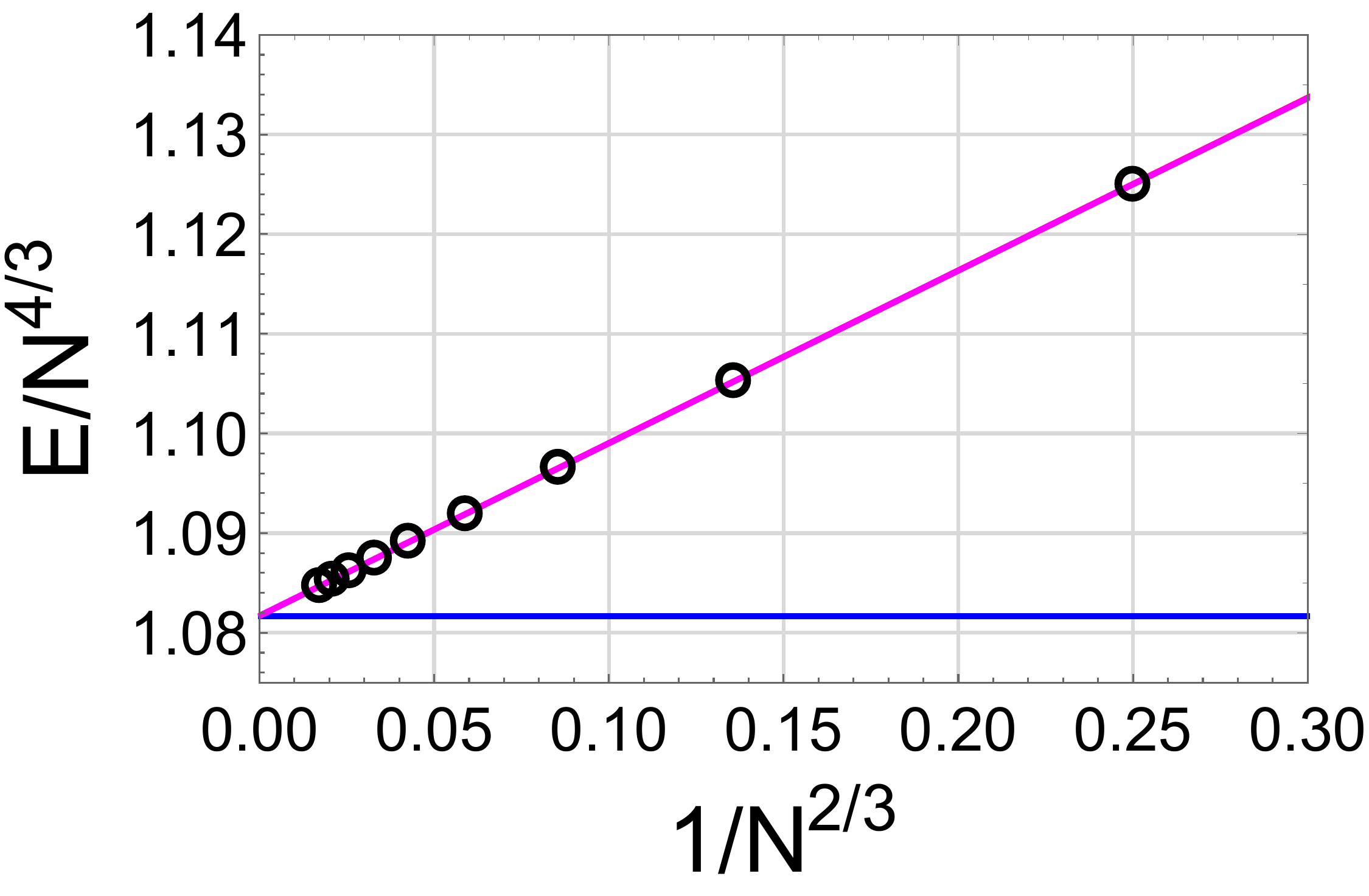}
\caption{Comparing the exact 3D harmonic oscillator closed shell energies (Eq. (\ref{3DHOCSEN}), black circles) with their asymptotic approximation (Eq. (\ref{3DHOENExp})): lowest order (TF, blue), first order (red), and second order (magenta).  We set $\om = 1$.}
\label{fig:3DHOE}
\end{figure}

\begin{figure}[!htb]
\subfigure{\includegraphics[width=0.9\columnwidth]{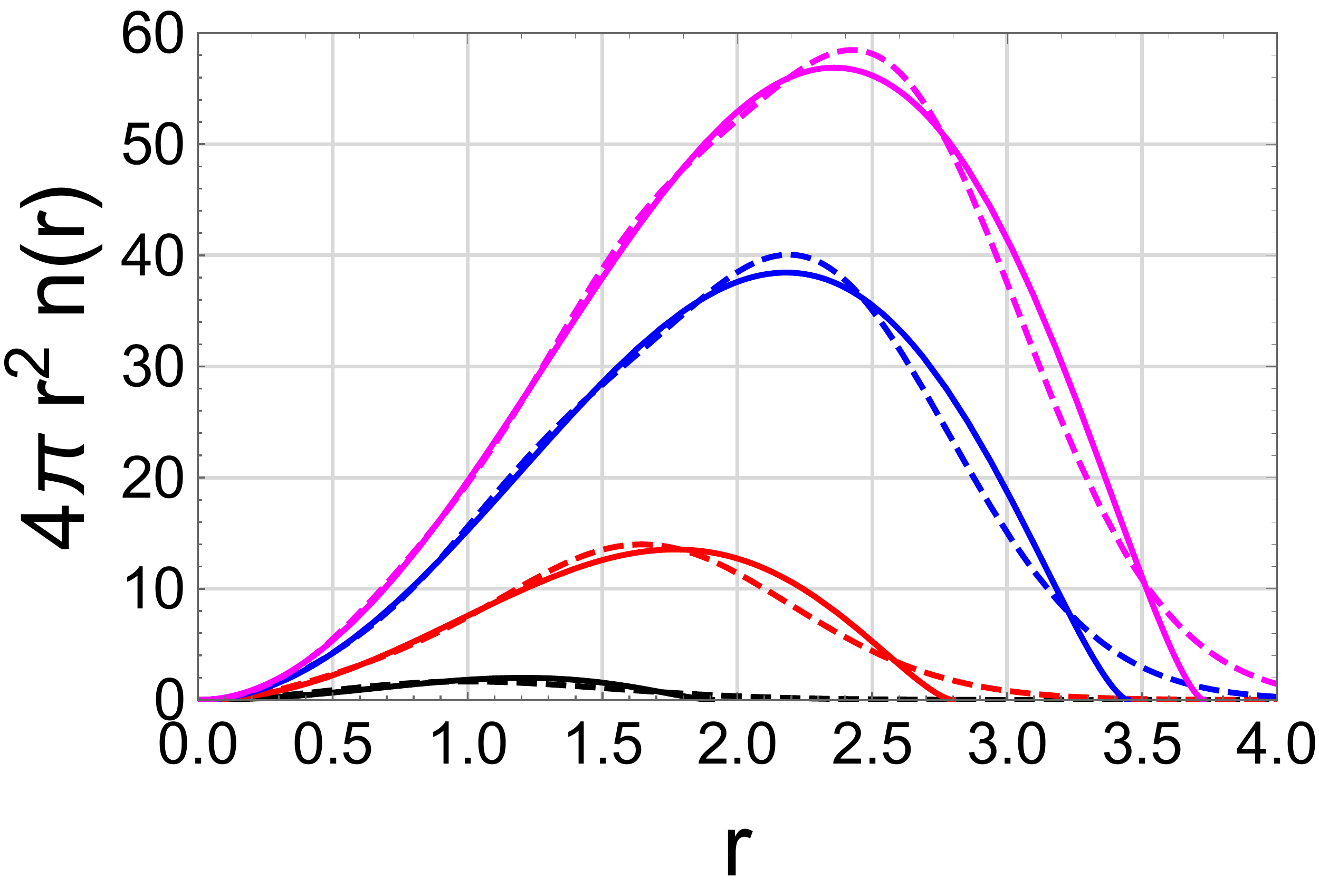}}
\caption{The 3D harmonic oscillator radial densities (dashed) and their TF approximations (solid) for $k$ full shells with $k = 1$ ($N = 2$, black), $k = 3$ ($N = 20$, red), $k = 5$ ($N = 70$, blue),  and $k = 6$ ($N = 112$, magenta).  We choose $\om = 1$.}
\label{fig:3DHO}
\end{figure}

\begin{figure}[!htb]
\subfigure{\includegraphics[width=0.9\columnwidth]{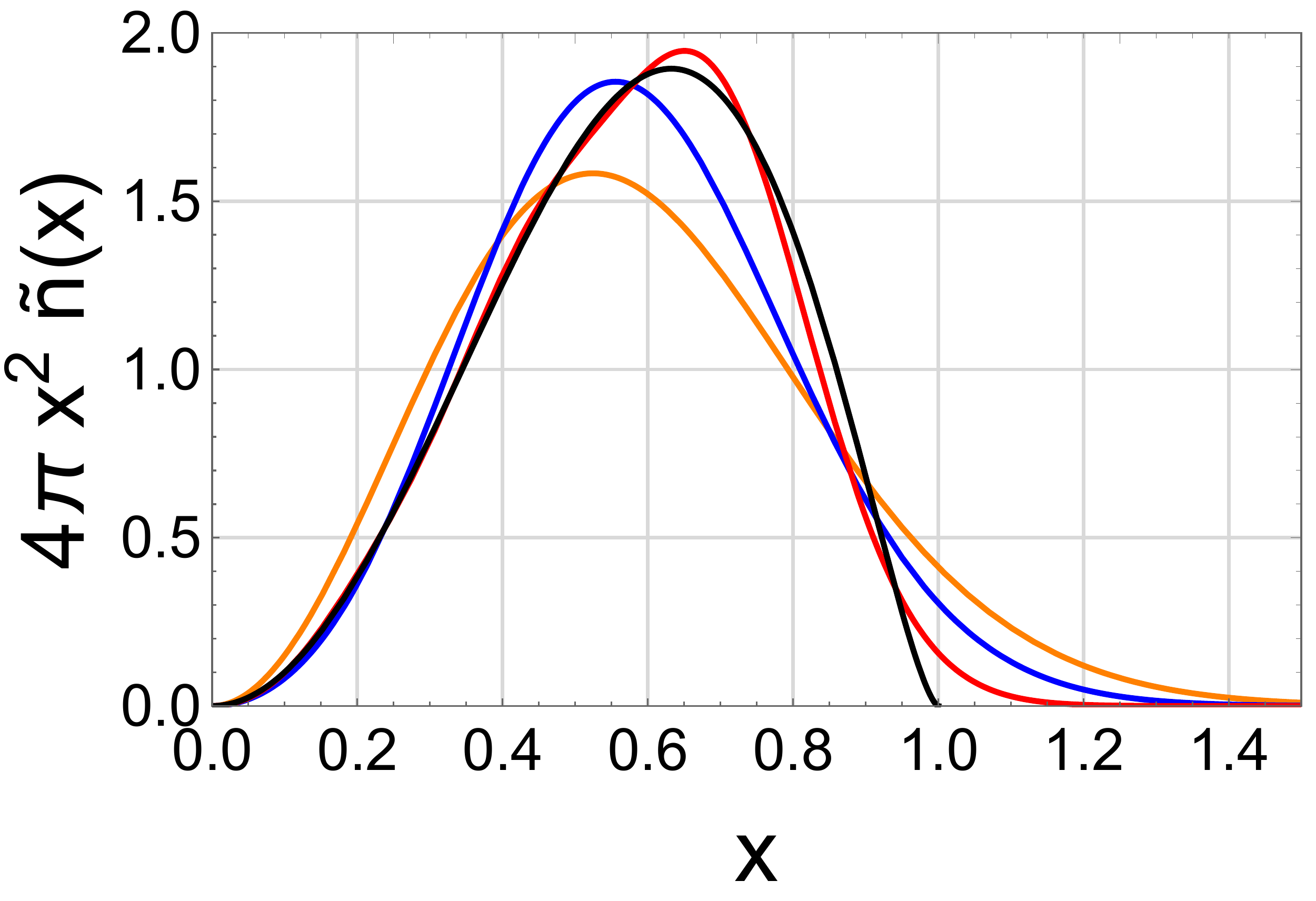}}
\caption{Same as Fig. \ref{fig:3DHO}, but scaled (Eqs. (\ref{3DHOScale}) and (\ref{3DHOScaleTF})) so that the TF density (black) is invariant with respect to $N$ with 1 (orange), 2 ($N = 8$, blue), and 6 (red) filled shells.}
\label{fig:3DHOS}
\end{figure}

\begin{figure}[!htb]
\includegraphics[width=0.9\columnwidth]{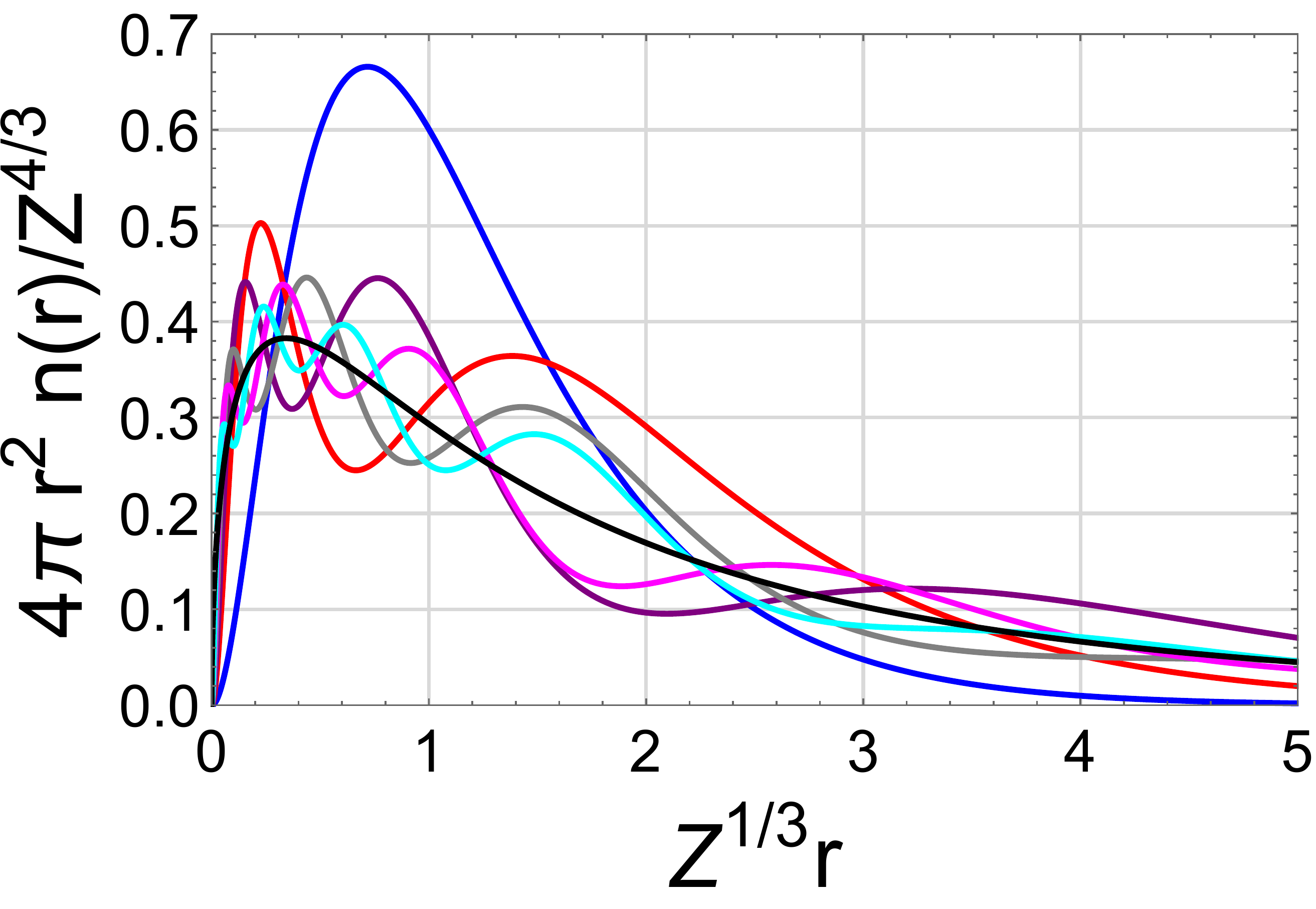}
\caption{Accurate $\z$-scaled radial noble gas densities: TF (black), He (blue), Ne (red), Ar (purple), Kr (gray), Xe (magenta), and Rn (cyan).}
\label{fig:NGDEN1}
\end{figure}

Another instructive example is the 3D harmonic oscillator, with $v(r)=\omega^2 r^2/2$.  Here, there is no Coulomb singularity and the potential is infinitely differentiable.  The number of particles in the $s$-th closed shell is now
\begin{equation}
\label{3DHOclosedshellN}
N_s = \frac{1}{3} s (s + 1) (s + 2).
\end{equation}
For closed shells the exact energy is
\begin{equation}
\label{3DHOCSEN}
E_\om(N) = \frac{\om}{2} \left( \frac{N^2}{12 \a} \right)^{1/3} \left[ 1 + 3 \left( \frac{3 N \a}{2} \right)^{2/3} \right],
\end{equation}
where $\a = 1 - \sqrt{1 - 4 (243 N^2)^{-1}}$.  Here, $T(N) = E(N)/2$ according to the virial theorem.  We expand Eq. (\ref{3DHOCSEN}) for large $N$:
\begin{equation}
\label{3DHOENExp}
E_\om(N) = \frac{(3N)^{4/3}}{4} \om \left[ 1 + \frac{1}{3(3N)^{2/3}} - \frac{1}{81(3N)^2} + ... \right],
\end{equation}
where the leading term is given by TF theory.  We show how well Eq. (\ref{3DHOENExp}) approximates the exact result in Eq. (\ref{3DHOCSEN}) in Table \ref{tab:3DHO} and Fig. \ref{fig:3DHOE}, where the last term makes an indistinguishable change.  The TF density is
\begin{equation}
n_\om\TF(r) = \frac{[2 \mu_\om - (\om r)^2]^{3/2}_+}{3\pi^2}, \qquad \mu_\om = \om (3N)^{1/3}.
\end{equation}
We plot the exact and TF densities for several filled shells in Fig. \ref{fig:3DHO}.  In Fig. \ref{fig:3DHOS} we take $\om = 1$ and use the scaling
\begin{equation}
\label{3DHOScale}
\tilde{n}(x) = 2 \sqrt{\frac{6}{N}} n(x), \qquad x = \frac{r}{\sqrt{2}(3N)^{1/6}},
\end{equation}
so that the exact scaled densities approach the TF density,
\begin{equation}
\label{3DHOScaleTF}
\tilde{n}\TF(x) = \frac{8}{\pi^2} (1-x^2)_+^{3/2},
\end{equation}
as $N \ra \infty$.

\begin{table}
\scalebox{0.9}{
$\begin{array}{|c|c|c|c|c|r|}
\hline
\multicolumn{3}{|c|}{} & \multicolumn{3}{c|}{\text{Error}}\\
\hline
\text{Filled Shells} & \text{N} & \text{Exact} & \text{TF} & 1^{\text{st}}\text{ corr. (mH)} & \multicolumn{1}{c|}{2^{\text{nd}}\text{ corr.}(\mu\text{H})} \\
\hline
1 & 2   & 3   & -0.3 & 0.84 & -93.2 \\
2 & 8   & 18  & -0.7 & 0.36 & -14.8 \\
3 & 20  & 60  & -1.3 & 0.20 &  -4.4 \\
4 & 40  & 150 & -2.0 & 0.13 &  -1.7 \\
5 & 70  & 315 & -2.9 & 0.09 &  -0.8 \\
6 & 112 & 588 & -4.0 & 0.06 &  -0.5 \\
\hline
\end{array}$}
\caption{Same as Table \ref{tab:BA} but for the 3D harmonic oscillator with $\om=1$.}
\label{tab:3DHO}
\end{table}

\sec{Atoms}
\label{sec:Atoms}
We now turn to the more difficult problems of interacting electrons.  This section provides a demonstration of how one can use insight from simple analytic formulas in more realistic situations.  The basic idea is to assume behaviors of density functionals inspired by the simpler cases, and then use extremely accurate numerical calculations to extract coefficients in those formulas.  This is most easily done for atoms and their ions, because of their spherical symmetry.

In the atomic case, the virial theorem implies that $T=-E$.  The exact kinetic energy is written as $T\s$, the KS kinetic energy, plus $T\c$ a correlation contribution of the same order as $E\c$, the total correlation energy.  Since $T\c$ is smaller than $Z^{5/3}$ \cite{BCGP16}, the asymptotic expansion for $T\s$ is the negative of that of Eq. (\ref{EZasy}),
\begin{equation}
T\s = c_0\, Z^{7/3} - \half Z^2 + c_2\, Z^{5/3} +....
\end{equation}
In Ref. \cite{LCPB09}, accurate KS calculations were performed for atoms up to $Z=92$, and the values for $T\s$ extracted.  These are essentially those of a HF calculation.  From these, numerical estimates were made of the constants in the asymptotic expansion, and were found to agree with the known theoretical values to within about 1\%.

Next the gradient expansion for the KS kinetic energy (in 3D) was applied, term by term, to the highly accurate densities, and the asymptotic expansion was extracted.  Remarkably, the TF approximation, applied to the accurate densities (this is the analog of putting the exact density into the TF functional, Eq. (\ref{TTFn})), yields a leading correction of $-0.66 Z^2$, only a 25\% overestimate of the Scott correction.  This was reduced to -0.54 and then -0.52 with the addition of the 2nd- and 4th-order terms.  Moreover, when the coefficients in the gradient expansion were scaled to ensure the exact asymptotic expansion was recovered, percentage errors for large atoms (beyond Ca) were of order 0.1\%, about 5 times smaller than the 2nd order gradient expansion and much better than fourth-order for small atoms \cite{LCPB09}.  Unfortunately, the improvement for molecular energies was far more modest.  Tests were also run on jellium surface energies, showing the modifications could worsen the results of the regular gradient expansion.  On the other hand, the modifications appeared to improve the curvature energies of jellium clusters.

A new and improved parameterization for the neutral atom TF density was also given, guaranteeing certain exact conditions, and ensuring various measures of error were extremely small (often of order $10^{-8}$).  Formulas for various measures of the local gradient and higher-order gradients were also given.  This parameterization is much more faithful to the numerical solution of the TF equation than the older ones of Latter \cite{L55} and Gross and Dreizler \cite{GD78}.  It would be useful to construct a parameterization as a function of the $N/Z < 1$ ratio.   The energies as a function of $N/Z$ were given by Tal and Levy \cite{TL80}, but no parameterization of the corresponding densities has been performed.  In the limit $N/Z \ra 0$, we recover the Bohr atom results of Sec. \ref{sec:3D}. 

\begin{figure}[!htb]
\includegraphics[width=0.9\columnwidth]{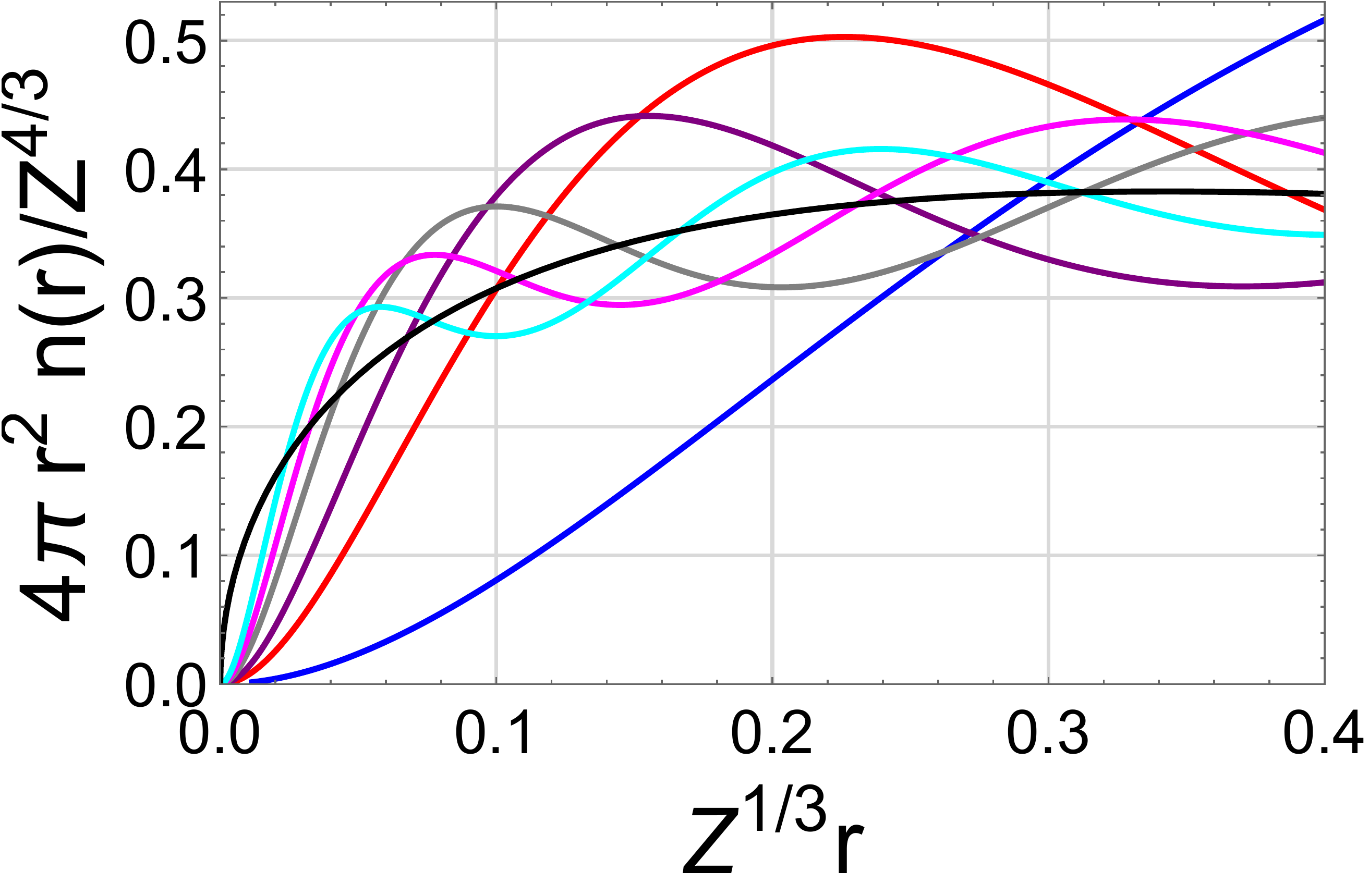}
\caption{Region near nucleus of Fig. \ref{fig:NGDEN1}.}
\label{fig:NGDEN2}
\end{figure}
The densities used in Ref. \cite{LCPB09} are plotted in Fig. \ref{fig:NGDEN1}, scaled appropriately via Eq. (\ref{TFNIZeta}) so that all tend to a single TF curve, Eq. (\ref{bu1}) with $Z=1$.  Fig. \ref{fig:NGDEN1} shows the weak approach of the density to its TF counterpart, as already seen in models (see Figs. \ref{fig:BAdens} and \ref{fig:3DHOS}).  In Fig. \ref{fig:NGDEN2}, we show how, while the TF density misbehaves at the nucleus (it diverges), the region of misbehavior shrinks as $Z$ grows.   To appreciate the subtleties of the density behavior in this limit, the reader is referred to Fig. 2 of Ref. \cite{L81}, which identifies 7 distinct regions, to be contrasted with the relatively simple 3 regions of the Bohr atom (Fig. \ref{fig:BAdens}) which are $ r \leq 1/Z$ (core), $1/Z \leq r \leq Z^{-1/3}$ (bulk), and $r \geq Z^{-1/3}$ (evanescent).  The study of the non-interacting kinetic energy functional has a long and diverse history (see e.g. Ref. \cite{DG90}), and Refs. \cite{CR17,CSK16,LMA14} make some recent contributions.

\sec{Exchange}
\label{sec:X}
In this section, we attempt to leap-frog from studies of the KS kinetic energy (often in 1D) to the more practical issue of the exchange-correlation energy in KS calculations.  The primary purpose of the earlier studies is to build understanding of the local density approximation and the semiclassical expansion for which it is the dominant term.  The overarching hypothesis is that the success of modern DFT is precisely because of the accuracy of this expansion.  Most modern approximations start from the generalized gradient approximation, often combined with a fraction of HF exchange \cite{B93,PEB96,SC20}.  Many of the successes and failures of semilocal DFT can be understood in terms of this hypothesis, including the differences between weakly and strongly correlated \cite{MG12} systems, and most failures of such approximations.

The first attempt at such an analysis appears in Ref. \cite{PCSB06}.   An important finding is the density scaling conjugate to the potential scaling of the Lieb-Simon work, as discussed in Sec. \ref{sec:LSScale}.  Regular coordinate scaling has led to many of the most fundamental exact conditions in DFT, and can often easily be applied to suggested approximations.  Simultaneous scaling of the particle number is much more difficult to analyze, but is crucial to our work.

A crucial question is whether our reasoning also applies to the XC energy used in KS calculations.  Does the local density approximation (LDA) become relatively exact in the limit of large $\zeta$?  If so, and it appears to be so, then LDA is a universal limit of all electronic structure, not just some approximation with some reference model, as it is often described.   More importantly, what are the next corrections?  One of the greatest improvements in the accuracy of KS-DFT calculations occurred going from LDA to generalized gradient approximations (GGA's) within the KS scheme.   These (at least in their early forms) began from the gradient expansion for slowly-varying gases.  But as we have already seen in our kinetic energy calculations in 1D, the gradient expansion does not apply for matter with evanescent regions (i.e., all atoms and molecules).  We shall see later that it may apply to more solids than expected.

Exchange typically dominates over correlation, at least for total energies.  For neutral atoms Ref. \cite{PCSB06} conjectured that the atomic exchange energy has the asymptotic expansion \cite{EB09}
\begin{equation}
E\x(Z) = - c\x\, Z^{5/3} + a\x\, Z +...,
\label{EXZasy}
\end{equation}
where $c\x = 9 c_2/11 \approx 0.220827$ and $c_2$ is given in Eq. (\ref{EZasyCoeff}).  As shown originally by Schwinger for atoms \cite{S81},  the LDA for exchange \cite{D30} (compare with the TF approximation for $T_s$ in Eq. (\ref{TFT3D})):
\begin{equation}
\label{LocX}
E\LDA\x[\n]= - \frac{3}{4} \left( \frac{3}{\pi} \right)^{1/3} \int d^3r\, n^{4/3}(\mathbf{r}),
\end{equation}
produces the dominant term exactly.  This has been proven by Conlon \cite{C83} for any system, but not in the strictest mathematical sense, as the Coulomb singularity is rounded off.  Estimates of $a\x$ were made numerically, from tables of exchange energies of atoms \cite{EB09,PRCV08}.

We define the non-local (NL) contribution to a functional by subtracting the LDA approximation from it.  Thus $E\NL\x = E\x - E\x\LDA$, and $E\NL\x/E\x \ra 0$ as $Z \ra \infty$.  If Eq. (\ref{EXZasy}) is correct $E\NL\x \ra \D a\x Z$ ($\D a\x = a\x - a\x\LDA)$ in the semiclassical limit.  It was found that the gradient expansion underestimates $\D a\x$ by just about a factor of 2.  

Refs. \cite{EB09,BCGP16} found that the two most commonly used GGA's for exchange, the Becke 88 GGA \cite{B88} and the exchange contribution to PBE \cite{PBE96}, both recover $\D a\x$ in Eq. (\ref{EXZasy}) very accurately.  This makes sense from a pragmatic viewpoint.  Their most significant improvement over LDA is in atomization energies, the difference in energy between a molecule and its constituent atoms (a.k.a the cohesive energy of a solid).  Thus, if these approximations were not highly accurate for atoms, they would be unlikely to yield accurate atomization energies.  

Another insight from these results is a reverse-engineering derivation of the single empirical parameter in B88 \cite{EB09}.  Assume the asymptotic correction for the exchange energy of atoms is precisely a factor of 2 larger than it's value in the gradient expansion, and then solve for the value of that parameter that reproduces this result.   This value is within about 10\% of the value chosen by Becke, based on fitting to the exchange energies of noble gas atoms \cite{B88} (and Becke even considered earlier fitting the large-$Z$ limit, before doing this \cite{B86}).  This work showed that Becke's insight was correct in two important ways:  (a) he recovered the appropriate coefficient with his procedure and (b) he did not wait 18 years for a more detailed derivation before publishing.  The B88 functional played a crucial role at a crucial time in the adoption of DFT within chemistry \cite{B14}.  The B3LYP functional, a global hybrid of GGA (B88) and Hartree-Fock exchange, is still the most used functional in chemistry today \cite{MH17}.

\begin{figure}[!htb]
\includegraphics[width=0.9\columnwidth]{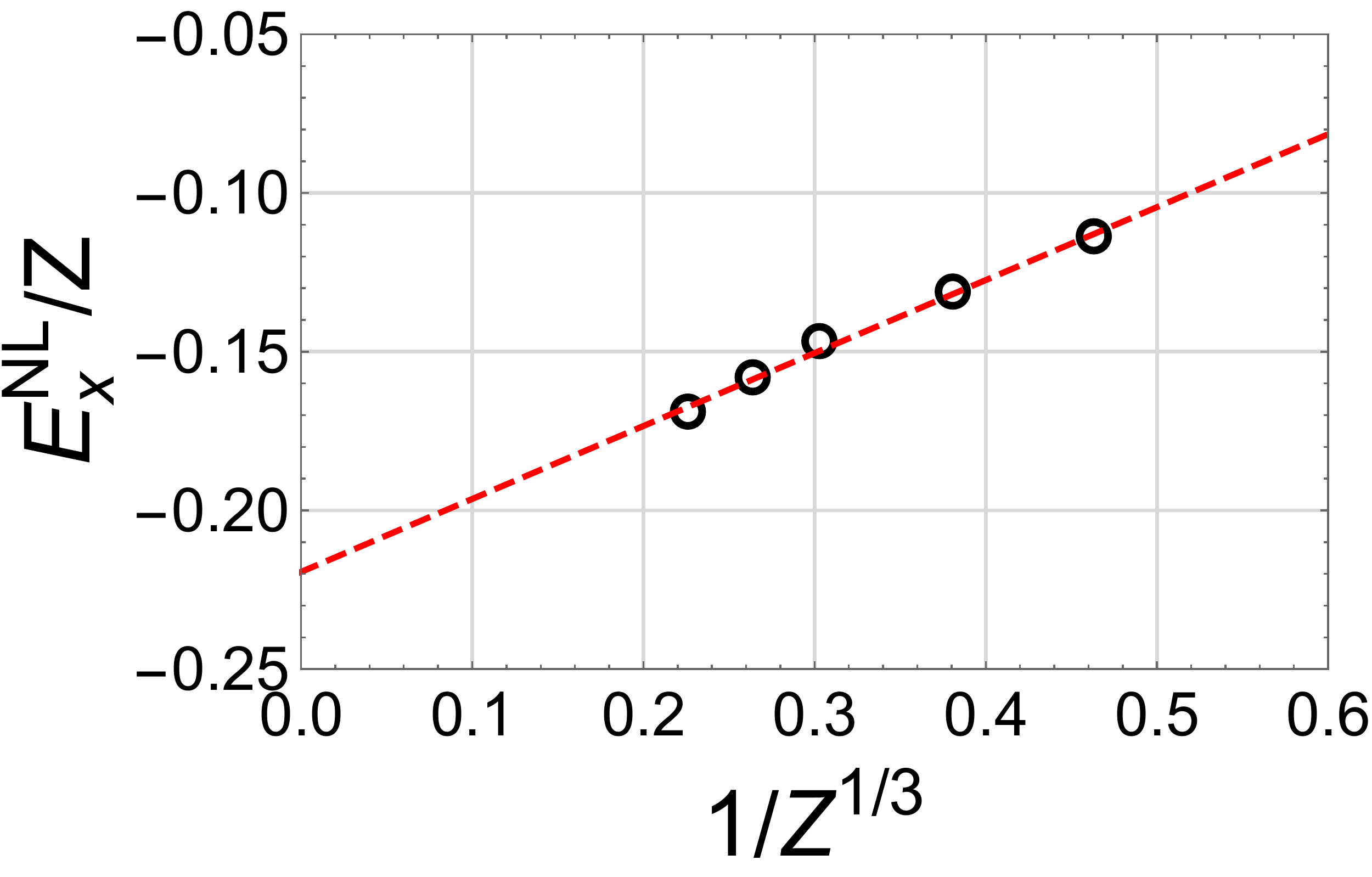}
\caption{Exact noble gas non-local exchange energies (not including He) and their linear extrapolation.}
\label{fig:DiffEx}
\end{figure}
\begin{figure}[!htb]
\includegraphics[width=0.9\columnwidth]{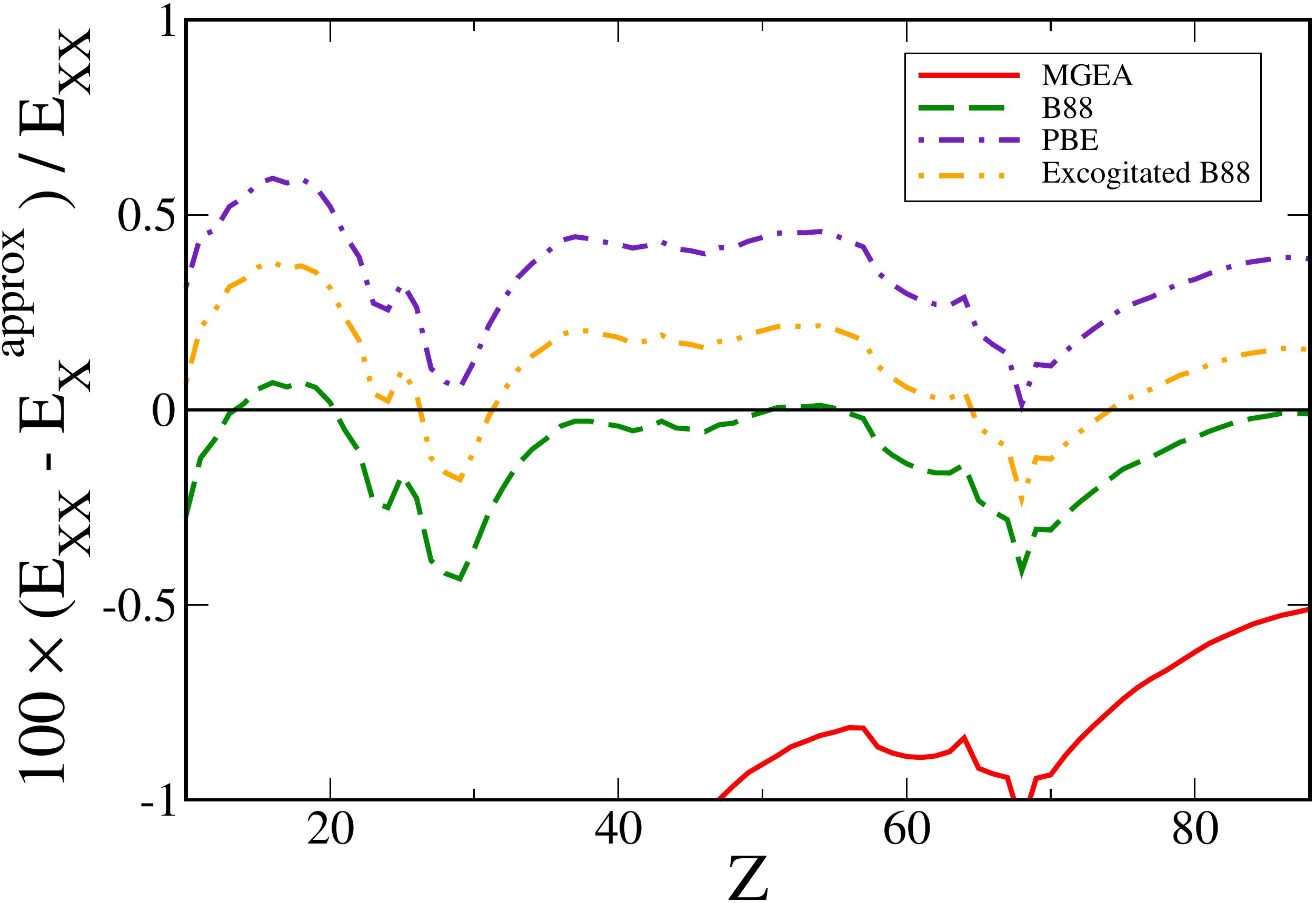}
\caption{Percentage error in various GGAs for exchange energies of atoms.  The details of the MGEA functional can be found in Ref. \cite{EB09}, from which this figure is reproduced.}
\label{fig:Exper1}
\end{figure}

Fig. \ref{fig:DiffEx} shows an extrapolation of $E\x\NL$ as a function of $Z$, to deduce the value of $\D a\x$, while Fig. \ref{fig:Exper1} shows the percentage error of various GGA approximations to the exchange energy with increasing $Z$.  All of these reduce to LDA when the density is uniform.  Excogitated B88 is B88 with the asymptotically correct parameter \cite{EB09}.

\begin{figure}[!htb]
\includegraphics[width=0.9\columnwidth]{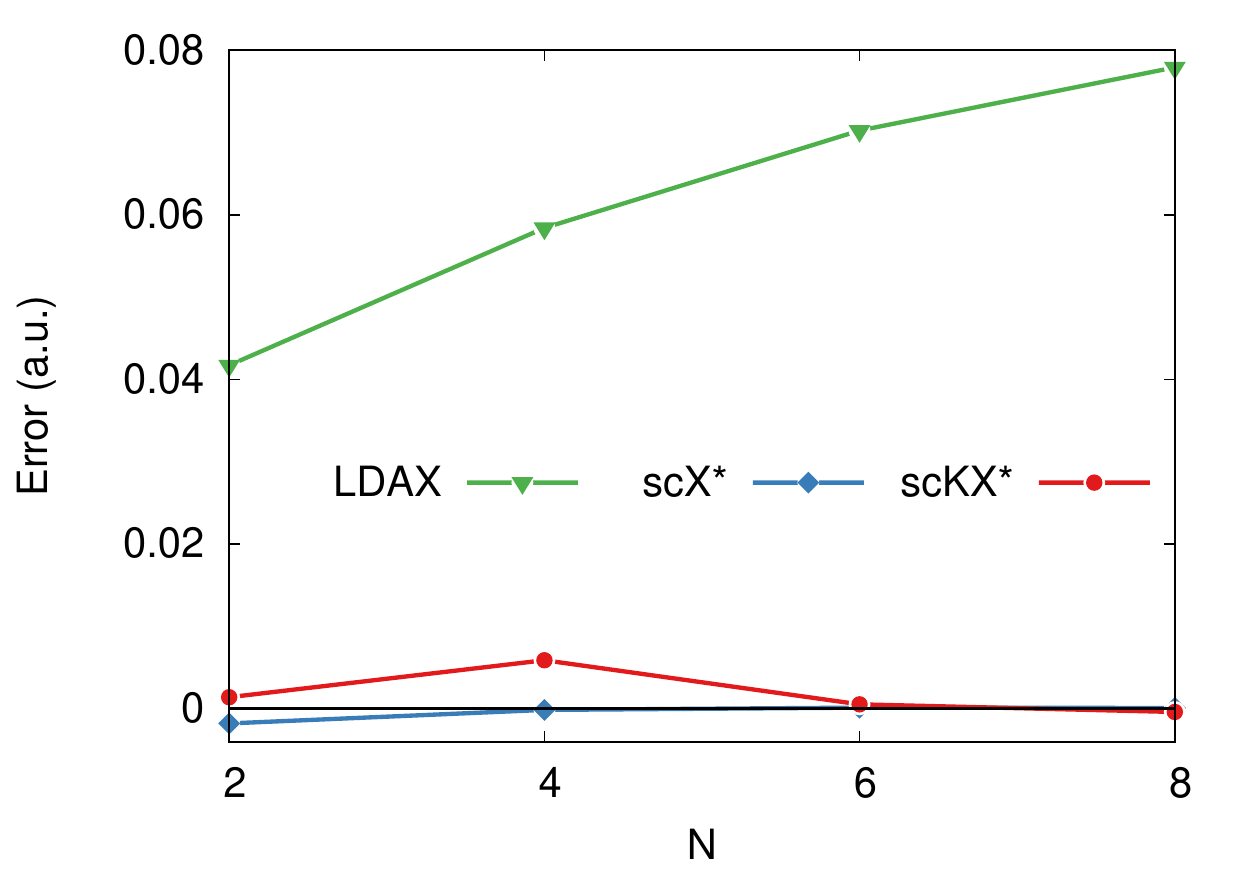}
\caption{Errors in a model 1D exchange calculation as a function of particle number.  Reproduced from Fig. 1 of Ref. \cite{ECPG15}.}
\label{fig:Exper2}
\end{figure}

We briefly return to 1D problems, where WKB-style approximations are straightforward.  A key question is: if we have good approximations for densities as functionals of the potential, can we generalize these to make approximations to the density matrix of non-interacting particles:
\begin{equation}
\n(x,x')=\sum_{j=0}^N \phi_j^*(x') \phi_j (x),
\end{equation}
as $E\x$ can be extracted from this object?  (The density is just the diagonal of the density matrix).  In fact, this is possible for the 1D box problem of Sec. \ref{sec:box}, where we found that the leading corrections to local potential approximations for the kinetic energy density led to the leading corrections to the kinetic energy.  Thus for 1D box problems spectacular improvements in the accuracy of $E\x$ are possible.  Fig. \ref{fig:Exper2} shows total energy errors in an X-only calculation, for a model box problem, while keeping the potential fixed, increasing the particle number, and doubly occupying the orbitals.  Because this is 1D, the electron-electron repulsion is chosen to be non-Coulombic.  For LDAX, the error increases with $N$, but $E\x$ increases much faster, so the fractional error is vanishing for large $N$, consistent with $E\x$ becoming local in that limit.  The blue line shows results when a KS calculation is performed and the semiclassical correction beyond LDA is applied.  The errors are undetectable by eye for $N > 2$.  Moreover, even if we treat the entire energy semiclassically, using the kinetic energy approximations of Sec. \ref{sec:box}, we still achieve extremely high accuracy, far beyond that of any existing 3D approximations.  These results suggest that much more accurate approximations to $E\x$ are possible, but require derivations (or at least insight) beyond 1D.

\sec{Correlation}
\label{sec:C}
In Secs. \ref{sec:Atoms} and \ref{sec:X} respectively we have shown that local approximations for both the KS kinetic and exchange energies become relatively exact in the semiclassical limit, at least for atoms.  As correlation is negligible relative to exchange in this limit, this guarantees that LDA becomes relatively exact for XC in the limit.

But does correlation alone become relatively exact?  This would provide a much stricter condition.   The answer is that in fact it does for atoms.  This follows from a very detailed analysis by Kunz and Rueedi \cite{KR10}.  To see this, start with 
\begin{equation}
\label{LDAC}
E\c\LDA[n] = \int d^3 r\, n(\mbf{r}) \eps\c\unif[n(\mathbf{r})],
\end{equation}
where $\eps\c\unif(n)$ is the correlation energy per electron of a uniform electron gas with density $n$.  This is now well-known to within about 1\% from quantum Monte Carlo calculations \cite{CA80} and exact constraints \cite{SPS10}, including resummed perturbation theory for the high density limit.  As the density is $\z$-scaled to large $\z$, it becomes large.  In this high density limit, Gell-Mann and Brueckner \cite{GB57} showed that
\begin{equation}
\label{CDenLim}
\eps\c\unif = \g\, \ln r_s + \eta + ...,~~~~r\s\to 0,
\end{equation}
where $r_s = (3/4\pi n)^{1/3}$ is the Wigner-Seitz radius, a measure of density, and 
\begin{equation}
\g = \frac{1 - \ln 2}{\pi^2} \approx 0.03109069,
\end{equation}
and $\eta \approx 0.04692032$ \cite{GB57,OMS66,H92} is written in Eq. (21) of Ref. \cite{H92} (up to a factor of -2) as
\begin{equation}
\eta = \frac{3\z(3) + 10}{4\pi^2} - \frac{5}{12} + \frac{\g}{6} [(4\g + 1)\pi^2 + 4\ln(3\pi^2) - 5 - 6Q ],
\end{equation}
where $\z(x)$ is the Riemann zeta function (Sec. 25 of Ref. \cite{DLMF}) and
\begin{equation}
Q = \frac{\int_{-\infty}^{\infty} du\, q^2(u) \ln q(u)}{\int_{-\infty}^{\infty} du\, q^2(u)},
\end{equation}
where $q(u) = 1 - u \arctan(1/u)$ (Eqs. (10) \& (12) of Ref. \cite{H92}).  To 40 digits, $Q$ is
$$-0.5506550741801572697652243519352338115111,$$
which appears not to have been precisely calculated before.
Combining Eqs. (\ref{LDAC}) and (\ref{CDenLim}) yields (independent of the details of the density):
\begin{equation}
\label{ECZasy}
E\c = - (A\c \ln Z - B\c)\, Z + ...,
\end{equation}
where $A\c = 2\g/3 \approx 0.0207271$ and \cite{BCGP16}
\begin{equation}
B\c\LDA = \frac{\g}{3} [\ln(3b^3) - I_2]- \eta \approx -0.00479524,
\end{equation}
where $b = (3\pi/4)^{2/3}/2$ and $I_2 \approx -3.331462$ \cite{BCGP16} is an integral over the TF density:
\begin{equation}
I_2 = \int_{0}^{\infty} dx\, f(x) \ln f(x), \qquad f(x) = \left[ \frac{\Phi(x)}{x} \right]^{3/2}, 
\end{equation}
where $\Phi(x)$ is defined in Sec. \ref{sec:TF}.  The derivation of Kunz and Rueedi produces the first term as the leading contribution to atomic correlation in this limit.  Thus it implies that LDA correlation becomes relatively exact here.  However the LDA value for $B\c$ is highly inaccurate for atoms (about ten times too small and of the wrong sign).  Refs. \cite{BCGP16,PCSB06} used an older estimate for $\eta$ (0.04664) yielding a different value for $B\c\LDA$ (-0.00451).

A striking feature of Eq. (\ref{ECZasy}) is how slowly the large $Z$ limit is approached.  For $T\s$ and $E\x$, the leading corrections differ from the leading terms by factors of $Z^{-1/3}$ and $Z^{-2/3}$ respectively, and even this is annoyingly slow (only reaching 0.2 at the end of the usual periodic table).  But correlation is far far slower, because the dominant term grows only logarithmically.  Thus if $A\c \approx B\c$, then $Z$ must be greater than 20,000 before the second term is reduced to 10\% of the first, and 10$^{43}$ before it is 1\%.  This has important consequences for the role of LDA in functional construction for correlation.  At any earthly value of $Z$, LDA correlation is not close to true correlation, because of its error in $B\c$.  The most popular correlation functional in chemistry, LYP, does not reduce to LDA in the uniform limit for this reason, i.e., that limit is irrelevant, especially for the lighter elements that are crucial to many applications, such as organic chemistry.

On the other hand, the correlation of PBE was designed to respect certain exact conditions, such as recovering the uniform gas correlation energy when the density is constant, but also producing a finite result when coordinate scaling a finite system to its high density limit \cite{PBE96}.  Because of those conditions, its high-density expansion matches that of Eq. (\ref{ECZasy}), with a value of $B\c$ within a few percent of the numerically extracted value for atoms.

\begin{figure}[!htb]
\includegraphics[width=0.9\columnwidth]{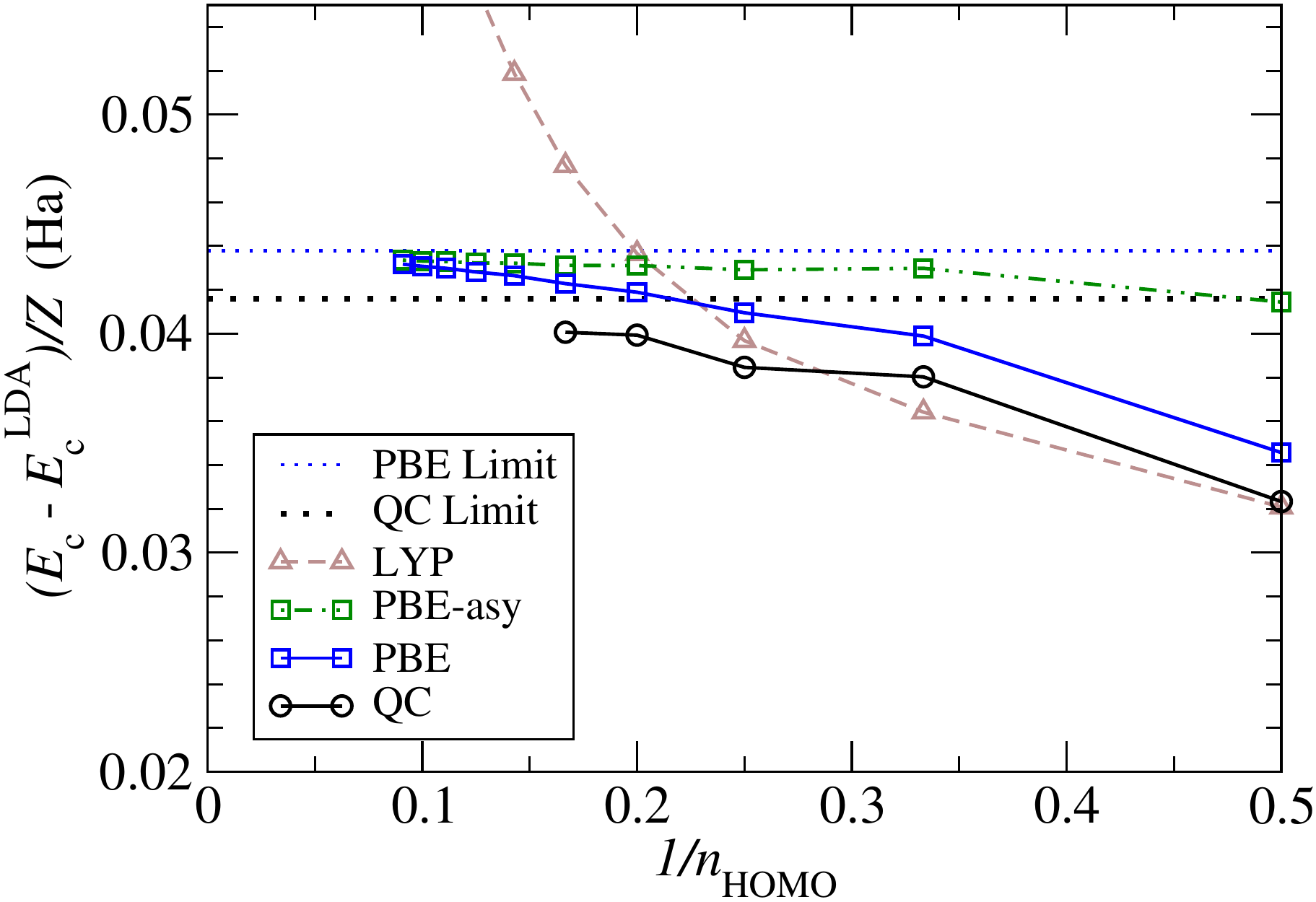}
\caption{The non-local correlation energy per electron within PBE, LYP, and accurate QC calculations as a function of inverse highest occupied shell for noble gases (points).  PBE-asy is Eq. (\ref{CDenLim}) evaluated on the exact density.  Reproduced from Ref. \cite{CCKB18}.}
\label{fig:BydLDAC}
\end{figure}

Fig. \ref{fig:BydLDAC} shows the asymptotic behaviors of various approximate functionals.  The figure does not include LDA, as LDA is too inaccurate to appear on this scale.  We see that even PBE correlation is not quite asymptotically correct, which led to the development of acGGA, where ac denotes 'asymptotically corrected' \cite{CCKB18}.  The acGGA yields the most accurate correlation energies for large-$Z$ atoms of any GGA in existence, precisely because it recovers the asymptotic expansion.

Very recently, it was shown that \cite{KSBW20}, for both exchange and correlation, one can also approach these limits from another direction, by studying the large-$Z$ behavior at fixed $N$ (where $N = 1$ corresponds to the Bohr atom of Sec. \ref{sec:3D}) and then considering $N \ra \infty$.  This might allow the extraction of even the neutral asymptotic coefficients more easily, as large-$N$ interacting calculations may not be necessary.  For more information on correlation, we refer the reader to Refs. \cite{BCGP16,CCKB18}.

\sec{Ionization energies}
\label{sec:IE}
While total energies can be useful as tests of approximations, all practical calculations are of energy differences, such as atomization energies or ionization potentials.  It is entirely conceivable that the semiclassical limit for the total energy is irrelevant to such calculations.  In this section, we explore the simplest possible energy difference, the difference between the $N$ and $N-1$ particle systems, to see how accurate our approximations are for this quantity.

Because our individual eigenvalues are positive in some cases and negative in others, we define the ionization potential of any system as
\begin{equation}
I(N) = | E(N) - E(N-1) |.
\end{equation}
For non-interacting fermions, this is the magnitude of the $N$-th eigenvalue.  For TF, as $N \ra \infty$, $I\TF(N) \ra |\mu|$.  For the flat box
\begin{equation}
I_{\ss L}\TF(N) = \frac{\pi^2 N^2}{2 L^2}\left( 1-\frac{1}{N}+\frac{1}{3N^2} \right),
\end{equation}
to be compared with the exact answer of $\pi^2 N^2/2 L^2$.  Thus, once again, the fractional error in the TF estimate vanishes for large $N$.  For the harmonic oscillator, this difference is exact, as TF yields exact energies.  The exact PT ionization energy is
\begin{equation}
I_{\ss D}(N) = (1 + \lam) \left( N - \half \right) - \frac{N^2}{2}.
\end{equation}
Its TF approximation is
\begin{equation}
I_{\ss D}\TF(N) = \sqrt{2D} \left( N - \half \right) + \half \left(N - \frac{1}{3} \right) - \frac{N^2}{2}.  
\end{equation}
Thus for the PT well the exact ionization energy and its TF approximation agree as $N \ra \infty$ with $N \propto \sqrt{D}$, each matching $|\mu|$ of Eq. (\ref{PTTFDen}).

The LHW does not have a closed form exact expression for $I(N)$, but its TF and next order corrections are
\begin{align}
\label{LHWI}
\begin{split}
I_{\ss F}\TF(N) &= \frac{3}{10} (3 \pi F)^{2/3} [N^{5/3} - (N-1)^{5/3}],\\
\D I^{(1)}_{\ss F}(N) &= \frac{1}{8} (3 \pi F)^{2/3} [N^{2/3} - (N - 1)^{2/3}].\\
\end{split}
\end{align}
For large $N$, $I_{\ss F}\TF(N) \ra (3 \pi F N)^{2/3}/2$ agreeing both with $|\mu|$ from Eq. (\ref{LHWTF}), and the leading order WKB approximation, Eq. (\ref{LHWWKB}).  Since we do not have an exact analytic expression to compare with Eq. (\ref{LHWI}), we show that Eq. (\ref{LHWI}) becomes relatively exact as $N \ra \infty$ in Table \ref{tab:LHWI}.

\begin{table}
$\begin{array}{|c|c|c|c|}
\hline
\multicolumn{2}{|c|}{} & \multicolumn{2}{c|}{\text{Error}}\\
\hline
N & I & \text{TF} & 1^\text{st}\text{ corr. (mH)} \\
\hline
2 & 3.24461 & -0.33 & -5.9 \\
3 & 4.38167 & -0.28 & -3.6 \\
4 & 5.38661 & -0.25 & -2.5 \\
5 & 6.30526 & -0.23 & -1.9 \\
6 & 7.16128 & -0.21 & -1.5 \\
7 & 7.96889 & -0.20 & -1.2 \\
\hline
\end{array}$
\caption{The exact, TF, and leading correction to the LHW ionization energy with $F = 1$.}
\label{tab:LHWI}
\end{table}

For our 3D model systems, we have given answers only for closed shells, so here we define
\begin{equation}
I(s) = | E(N_s)-E(N_{s-1}) |,
\end{equation}
as the energy of the last shell, where $s$ is the number of closed shells.  For the Bohr atom $I(s) = Z^2$, and its TF approximation is
\begin{equation}
\label{ITFBA}
I_{\ss Z}\TF(s) = Z^2 s^{1/3} (a_s - a_{-s}), \qquad a_s^3 = (s+1)\left(s + \half \right).
\end{equation} 
Taking the large $s$ limit yields
\begin{equation}
I_{\ss Z}\TF(s) \sim Z^2\left( 1 + \frac{1}{12 s^2} \right)  , \qquad s \ra \infty.
\end{equation}
Thus the TF results become exact as $s$ becomes large.  The exact ionization energy of the 3D harmonic oscillator can be calculated from Eqs. (\ref{3DHOclosedshellN}) and (\ref{3DHOCSEN}), although the expression is cumbersome.  The TF approximation to the ionization energy is
\begin{equation}
\label{ITF3DHO}
I_\om\TF(s) = \frac{\om}{4} [s(s+1)]^{4/3} [(s+2)^{4/3}-(s-1)^{4/3}],
\end{equation}
and $I\TF(s) \ra \om s^3$ as $s \ra \infty$.  In Table \ref{tab:3DHOI} we show that the relative error of the TF approximation to the 3D harmonic oscillator ionization energy again goes to zero, i.e., the TF result becomes exact as the number of electrons becomes large.
\begin{table}
$\begin{array}{|c|c|}
\hline
s & (I\TF-I)/I \\
\hline
2 & -0.0279 \\
3 & -0.0139 \\
4 & -0.0083 \\
5 & -0.0056 \\
6 & -0.0040 \\
7 & -0.0030 \\
\hline
\end{array}$
\caption{The relative error of the TF shell ionization energy for the 3D harmonic oscillator.}
\label{tab:3DHOI}
\end{table}

So, in all our non-interacting examples, TF yields the exact ionization energy as $N \ra \infty$.  This begs the question: Does TF theory yield $I$ correctly in the large $Z$ limit of real atoms?  Mathematical physicists \cite{S03} have long pondered this question.

A complication arises due to the periodic table.   Each row corresponds to the filling of a shell, and ionization potentials vary across rows, as well as down columns.   We use the Madelung rule \cite{M36} to populate the shells, and consider large $Z$ in the non-relativistic limit.  Then, one can ask if the shell structure survives and if so, is there a well-defined limit?

\begin{figure}[!htb]
\includegraphics[width=0.9\columnwidth]{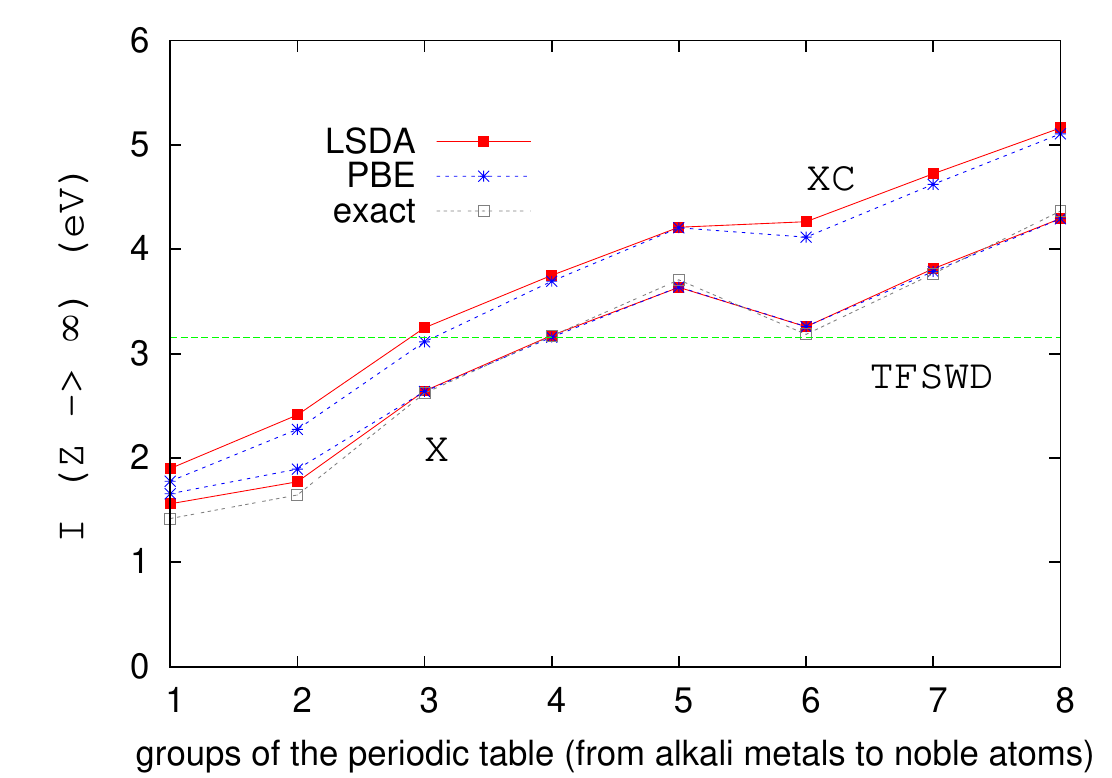}
\caption{Ionization energies extrapolated to infinite $Z$, exactly and with two XC approximations.  The green horizontal line is from extended TF theory.  Exact results are HF calculations, extrapolated to the infiniteth row, and agreeing to within numerical  accuracy with the exchange contribution to both functionals.  Accurate results for XC are still beyond the capabilities of current quantum chemistry.  Reproduced from Ref. \cite{CSPB10}.}
\label{fig:IP}
\end{figure}

The answer appears to be yes, as shown in Fig. \ref{fig:IP}.  By performing Hartree-Fock calculations for up to 3000 non-relativistic electrons, it was possible \cite{CSPB10} to extrapolate down each column to the infinite-$Z$ limit and find a weaker, but still quite distinct, variation across rows.  Moreover, both LDA exchange and PBEX reproduce this curve, within numerical error of the extrapolation.  Thus local approximations yield the correct answer in this limit.  It was also found that, even with correlation turned on, LDA and PBE almost coincided, suggesting that LDA may become exact for the exact ionization potential, even if $Z$ is still very far from the limit where the total energy is accurately given by LDA.

Incredibly, if one averages HF over rows, the resulting single number (3.02 eV)  matches almost exactly the result of an extended-TF calculation on TF densities, a number predicted more than 30 years earlier by Englert \cite{E88,E87}, $I = 3.15$ eV from simple integrals over the TF density, without needing to perform any modern electronic structure calculation.

\sec{Practical functionals}
\label{sec:PracFunc}
Atomic studies \cite{EB09,BCGP16,CCKB18,PRCV08} have estimated the error of the gradient expansion for exchange to the true leading correction to LDA in the semiclassical limit.  Roughly speaking, the gradient expansion was found to underestimate this correction by close to exactly a factor of two.  Because the asymptotic expansion works remarkably well, even down to $Z=1$, this error produces unacceptably inaccurate GEA exchange energies of atoms.  On the other hand, the two most popular GGA's, B88 and PBEX, and a popular meta-GGA at that time, TPSS, recover this coefficient fairly accurately (and very accurately in the case of B88).  This somewhat explains their successes in chemistry and materials, and demonstrates their asymptotic correctness to this order, at least for neutral atoms.  One could hope that this remains true for the energy differences between molecules and atoms, i.e., atomization energies, so that this explanation partially explains their success for bonding, i.e., if they failed to satisfy this condition, they would not be usefully accurate for bond energies.  

On the other hand, this leads also to a kind of reverse engineering as far as materials applications are concerned.  For bulk metals, there are no classical turning points at the Fermi surface, so the gradient expansion approximation should be more appropriate, and B88 and PBEX less so.  Combined with a condition from the surface energy of a uniform gas for correlation, this led to the construction of PBEsol \cite{PRCV08}, a revision of PBE targeted at solids, which restores the correct gradient expansion.  PBEsol proved to yield more accurate lattice parameters and bulk moduli for solids, but sacrifices accurate cohesive energies to attain this.   (It may be impossible to find a single GGA that does both.)  This has since proven very useful in many solid-state applications \cite{PRCV08}.  On the other hand, it further highlights the difficulty of capturing accurate geometries and energetics for both molecules and metals with any GGA.  Since the appearance of PBEsol, the numerically-found leading corrections to LDA energetics of neutral atoms have also been built into several popular functionals, including the popular SCAN meta-GGA and its variants \cite{SRP15}, as well as several from Constantin et al. \cite{CFLD11}, and the acGGA discussed in Sec. \ref{sec:C}.

Just recently \cite{KCBP21}, the question of whether or not turning surfaces exist in solids has been explored using state-of-the-art KS-DFT calculations.  It was found that even for semiconductors with moderate gaps, there are no turning points at the energy of the highest occupied KS eigenvalue, and so no classically forbidden regions in space.  However, usually a moderate expansion of the lattice parameter for the semiconductors does lead to forbidden regions, just as a defect does in a metal.  For insulators, there are typically forbidden regions, even at equilibrium.  Thus, the appearance of such regions can be roughly correlated with conduction properties.  This is very different from molecular calculations, where almost all space is classically forbidden and only a region around the molecule is classically allowed.  The lack of classically forbidden regions in semiconductors and metals explains improved geometries with PBEsol relative to PBE \cite{PRCV08}.

\sec{Summing up}
\label{sec:sums}
In Sec. \ref{sec:box}, we saw that, by creating a uniform approximation for the kinetic energy density in a box problem, we automatically found the leading correction to the local approximation for the kinetic energy.  However, as we discussed above, the uniform approximations described in Sec. \ref{sec:RTP} for open boundaries yield highly accurate densities and kinetic energy densities, but this improved pointwise accuracy over TF does not translate into better energies.  To find corrections to the energy from uniform densities we would have to calculate our approximate density to two more orders, which appears to be an exhausting task \cite{B20b}.  Moreover, the TF approximation is exact for a harmonic oscillator.  Thus any corrections only show up in the difference between the kinetic energy of a well and its harmonic approximation.  This seems likely to explain the extraordinary accuracy of kinetic energy densities in 3D yielding little improvement in overall energies \cite{OP79}.

This would seem to signify the end of the road for this approach to finding leading corrections, and indeed, we need to make a diversion.  That entire approach was based on finding expressions for the density as functionals of the potential, deriving approximations that are uniform in real space.  But the true issue of interest is not the density, but the total energy of occupied orbitals.  So, the heart of the matter is:  Can we find the asymptotic expansion for the energy under semiclassical scaling, regardless of the density?  If we cannot do this, we cannot solve this problem.

Looking back on the results of Ref. \cite{RB17}, we see that in many specific cases with analytic solutions, the semiclassical expansion was extracted for both the eigenvalues and their sums, because in these model cases, exact analytic formulas were available.  Thus the question becomes: Can one (for a general 1D problem) directly sum the energies of $N$ orbitals, and find the semiclassical expansion to arbitrary order for that quantity?  In general, one would expect this to be an asymptotic expansion, with coefficients that are finite only if the potential is infinitely differentiable.  In non-trivial cases, this means modern tricks of asymptotics, including superasymptotics and hyperasymptotics \cite{B99,BH90,B91}, are needed, to be certain that one has indeed found the correct general terms in such an expansion.  This will be the subject of this section.

These corrections can be found by taking the semiclassical limit of eigenvalue sums, to which we now turn our attention.  In this section, we summarize relatively recent work employing techniques of asymptotic analysis to find expressions directly for the expansion of sums of eigenvalues from the expansion of the individual eigenvalues (WKB expansion), without explicit construction of densities, as in Secs. \ref{sec:box} and \ref{sec:RTP} (or subsequent inversion to make density functionals, as in the gradient expansion of Sec. \ref{sec:GEA}).

Here, we will go beyond the leading terms in the semiclassical expansion.  We generalize Eqs. (\ref{PhaseWKB}) and (\ref{BoxQuantCond}) to include the next correction in the WKB expansion, applied to any problem (box boundaries or not) \cite{BO99,KR67}:
\begin{equation}
\label{AllOrdersWKB}
\theta^{(0)} (\eps) + \Delta \theta^{(2)}(\eps) + ... = z\pi.
\end{equation}
Now the superscript 0 denotes the original WKB contribution, and 2 denotes the leading (second-order) correction, which depends on derivatives of the potential.  Here we have generalized the index in Eq. (\ref{AllOrdersWKB}) to a continuous real number $z$.  The value of $z$ is an integer for box boundaries, half-integer for two real turning points, and $j+3/4$, $j=0,1,2...$ for a half-space.  Inversion of the above, power by power, yields $\eps^{(0)}(z) + \Delta \eps^{(2)}(z) + ...$ etc. Eqs. (\ref{PTWKB}) and (\ref{LHWWKB}) are examples of this expansion.

The zero-order case is elementary, as the sum over such levels becomes an integral in the limit:
\begin{equation}
E\TF(N) = I_N[\eps^{(0)}] = \int_{0}^{N} dz\, \eps^{(0)}(z),
\end{equation}
i.e., the semiclassical limit of the sum of WKB levels {\em is} the TF solution \cite{CLEB10}.  One can consider TF as the natural generalization of WKB to three dimensions, and even deduce individual eigenvalues from derivatives of TF energies with respect to $N$, i.e., the TF chemical potential.  In the specific case of a smooth potential with a parabolic minimum, the energy to next order is \cite{B20}
\begin{equation}
\label{SU2}
E(N) = I_N[\eps^{(2)}] - \frac{1}{24} \frac{d\eps^{(0)}}{dz} \bigg|^N_0,
\end{equation}
where $\eps^{(2)} = \eps^{(0)} + \D \eps^{(2)}$ is the WKB solution to 2nd order.  The first term on the right hand side of Eq. (\ref{SU2}) is the universal leading order TF term $I_N[\eps^{(0)}]$ and the analog ($I_N[\D \eps^{(2)}]$) of the term that is kept in the gradient expansion approximation--Eq. (\ref{SPnv}), i.e., the leading correction for a slowly varying gas, where the eigenvalues are continuous.  The second term is the crucial missing contribution for a finite system, where the spectrum is discrete and discrete sums yield corrections depending on the end-points, just as in Secs. \ref{sec:box} and \ref{sec:RTP}.  The significance of Eq. (\ref{SU2}) is that it is a functional of $v(x)$, unlike all specific cases in Sec. 3. 

We illustrate the importance of the correction on the PT well (described in Sec. \ref{sec:ill}).  In this case, the expansion yields an infinite series but, unlike more general cases, one that is absolutely convergent, unless $D$ is very small.  Table \ref{tab:PTS} reports results for a well that binds 6 states.  The TF results become relatively exact in line with the LS theorem as $\z \ra \infty$.  We refer to ignoring the end-point contribution to the integral in Eq. (\ref{SU2}) as the GEA, not because this is exactly the same as using the gradient expansion approximation in the density, but because the same terms have been included (here as a functional of the potential).  We see that, without the end-point contributions, the GEA over-corrects TF, yielding results that are sometimes better, and sometimes substantially worse.  When we include the end-point correction, we find errors are never larger than a milliHartree, i.e., below the threshold for chemical accuracy.  

\begin{table}
$\begin{array}{|c|r|r|r|c|}
\hline
\multicolumn{2}{|c|}{} & \multicolumn{3}{c|}{\text{Error}}\\
\hline
N & \multicolumn{1}{c|}{E(N)} & \multicolumn{1}{c|}{\text{TF}} & \multicolumn{1}{c|}{I_n[\eps^{(2)}]} & 1^\text{st}\text{ corr.} \\
\hline
1 &  2.9221 & 0.07 & -0.04 & 1.5 \times 10^{-5} \\
2 & 11.1886 & 0.13 & -0.08 & 6.2 \times 10^{-5} \\
3 & 23.7993 & 0.16 & -0.12 & 1.4 \times 10^{-4} \\
4 & 39.7543 & 0.18 & -0.17 & 2.5 \times 10^{-4} \\
5 & 58.0536 & 0.17 & -0.21 & 3.8 \times 10^{-4} \\
6 & 77.6972 & 0.14 & -0.25 & 5.5 \times 10^{-4} \\
\hline
\end{array}$
\caption{The exact, TF, TF + $I_n[\eps^{(2)}]$, and true leading TF correction (TF + $I_n[\eps^{(2)}]$ + boundary term) for the PT well of depth $D = 20$.}
\label{tab:PTS}
\end{table}

\begin{table}
$\begin{array}{|c|r|r|r|r|c|}
\hline
\multicolumn{2}{|c|}{} & \multicolumn{4}{c|}{\text{Error}}\\
\hline
\text{j} & \multicolumn{1}{c|}{\eps_j} & \multicolumn{1}{c|}{\text{WKB}} & \multicolumn{1}{c|}{\text{TF}} & \multicolumn{1}{c|}{I_n[\eps^{(2)}]} & 1^\text{st}\text{ corr.} \\
\hline
0 &  2.92214 & 0.12  &  0.07 & -0.04 & 1.5 \times 10^{-5} \\
1 &  8.26643 & 0.10  &  0.05 & -0.04 & 4.6 \times 10^{-5} \\
2 & 12.61072 & 0.08  &  0.03 & -0.04 & 7.7 \times 10^{-5} \\
3 & 15.95501 & 0.06  &  0.01 & -0.04 & 1.1 \times 10^{-4} \\
4 & 18.29930 & 0.04  & -0.01 & -0.04 & 1.4 \times 10^{-4} \\
5 & 19.64359 & 0.02  & -0.03 & -0.04 & 1.7 \times 10^{-4} \\
\hline
\end{array}$
\caption{Same as Table \ref{tab:PTS} but for eigenvalues calculated from WKB and $\eps_j = E(j+1) - E(j)$.  The first correction to TF is identical to the leading correction to WKB, $\eps^{(2)}$ from Eq. (\ref{SU2}).}
\label{tab:PTe}
\end{table}

\begin{table}
$\begin{array}{|c|r|c|c|c|}
\hline
\multicolumn{2}{|c|}{} & \multicolumn{3}{c|}{\text{Error}}\\
\hline
N & \multicolumn{1}{c|}{E(N)} & \text{TF} & 1^\text{st}\text{ corr. (mH)} & 2^\text{nd}\text{ corr. (mH)} \\
\hline
1 &  1.8558 & -0.5172 & 40.5 & -6.4 \\
2 &  5.1004 & -0.8507 & 34.6 & -2.6 \\
3 &  9.4820 & -1.1291 & 31.0 & -1.5 \\
4 & 14.8686 & -1.3769 & 28.5 & -1.0 \\
5 & 21.1739 & -1.6042 & 26.7 & -0.8 \\
6 & 28.3352 & -1.8164 & 25.2 & -0.6 \\
\hline
\end{array}$
\caption{Same as Table \ref{tab:PTS} but for the LHW with $F = 1$.  Boundary corrections to $E(N)$ do not appear until the second order.}
\label{tab:LHWS}
\end{table}

But it is entirely possible that we are being misled by potentials for which we have analytic formulas for their energies (box, harmonic oscillator, PT, and Morse).  Because of this simplicity, the semiclassical expansion is finite or at least convergent in these cases.  But we expect that in general this expansion is asymptotic and we need to know if these methods work as well in such cases.

To study this, the simplest case is the linear half-well of Sec. \ref{sec:ill}.  The eigenvalues are given by zeroes of the Airy function, and their expansion is indeed asymptotic.  To deal with such expansions, we introduce technology that may be unfamiliar to the typical reader \cite{BH90,B91,BH93,B99}.  If we fix the order in the expansion, just as in the PT case (two terms), errors decrease as the level increases, implying that the worst case is the ground-state.  Because the expansion is asymptotic, at some (not very high) order, adding terms increases the error.  But one can look at the magnitude of the contribution to each order, and identify at which order the magnitude of the addition is smallest.  For any given level, truncation at the smallest addition is called optimal truncation, yielding the optimal choice for that level.  Thus optimal truncation at each level yields overall results that are typically better than those of truncation at any fixed order, and so are called superasymptotic.

But one can go considerably further.  By finding the leading behavior of the coefficients in the expansion as the order grows large (usually an asymptotic expansion, this time in the order), one can re-sum the series to all orders, picking up even more of the subdominant terms, achieving still higher accuracy.  One can also analyze optimal truncation, and derive a better choice of how much of the next term should be included.  Incredibly tiny errors can be achieved with such methods, but they require knowing the expansion to many orders \cite{B91,BH90,B99}.

\begin{figure}[!htb]
\includegraphics[width=0.9\columnwidth]{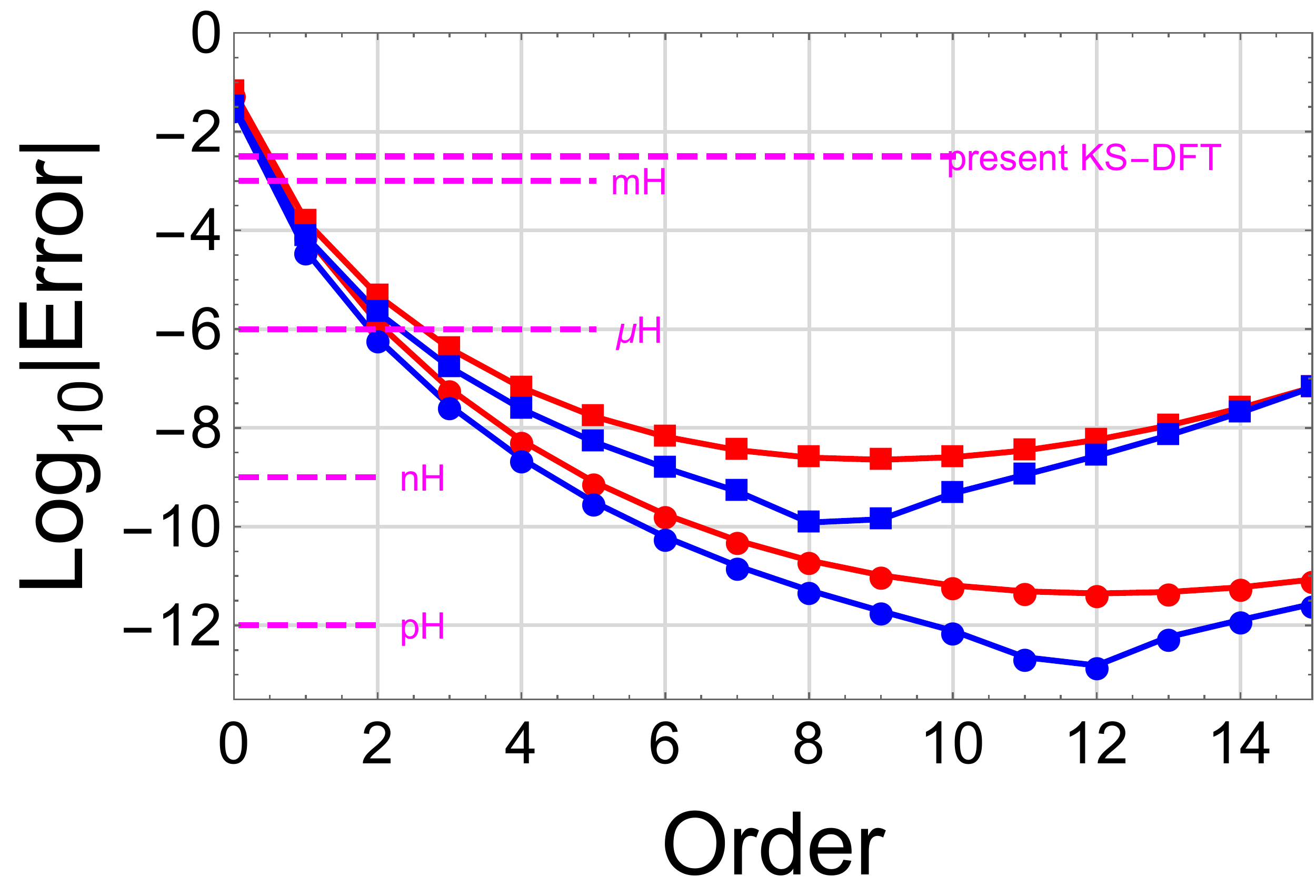}
\caption{The summation approximation for the linear half well with $N = 2$ (squares) and $N = 3$ (circles), with the regular asymptotic approximation (red) and the hyperasymptotic approximation (blue).}
\label{fig:LHW}
\end{figure}
So we apply our summation procedures to the asymptotic expansion of the eigenvalues, at least in some specific cases.   If the problem is one with an infinite discrete spectrum, we can find an extremely cool result, using the Euler-Maclaurin summation formula \cite{BB20}:
\begin{equation}
\label{SUForm}
E^{(M)}(N) = E_\infty - \int_{N}^{\infty} dj\, \tilde{\eps}(j) - \frac{\eps_N}{2}+ \sum_{m = 1}^{M} b_m\, \tilde{\eps}^{(2m-1)}(N),
\end{equation}
where $\tilde{\eps}(j)$ is a smooth monotonic interpolation of $\eps_j$ for non-integer values of $j$, $M$ is the order of the asymptotic expansion, and $b_m = B_{2m}/(2m)!$ where the $B_{2m}$ are Bernoulli numbers \cite{W21}.  Here $E_\infty$ is the (very carefully) regularized infinite sum over eigenvalues \cite{T10}.  In the right hand sides of Eqs. (\ref{SUForm})- (\ref{Deltap}) $f^{(n)}(x)$ always denotes the n-th derivative, not the n-th order of $f(x)$.  In Fig. \ref{fig:LHW}, we show just how small the errors can get with asymptotics by showing the asymptotic and hyperasymptotic approximation to the sum of the first two and three linear half-well levels.  By performing the expansion up to 12 orders, we can optimally truncate, yielding the red curve.  The error is about $0.1$ nanoHartrees!  Moreover, we can {\em reduce} that error by almost two orders of magnitude, using hyperasymptotics.

Why is this so important?  This shows that, in an extremely simple case, in principle it is possible to achieve this ridiculous level of accuracy, once the asymptotic expansion can be calculated to many orders.  Since we have shown that the gradient expansion of DFT is simply the semiclassical expansion in the specific case of a slowly-varying gas, this strongly suggests that similar accuracy is in principle possible, which would increase the accuracy of DFT by many orders of magnitude, moving it from moderate accuracy to far beyond chemical accuracy.

Is getting errors of order picoHartrees not something of overkill?  No.  In this simplest of all possible cases, the point is to show how far one can go with these summation formulas.  Moreover, for the individual eigenvalues the generation of arbitrarily high orders and the asymptotic behavior of the coefficients in the asymptotic expansion is essentially the equivalent of finding the exact answer.  The same applies to the sum over eigenvalues.  By achieving such accuracy, and more importantly by understanding it, we can have complete confidence in such expansions.

Why can't this be done immediately for DFT, at least for the KS kinetic energy?   The answer is that we have only done this for a few simple potentials, not as a general {\em functional} of the potential.  In one dimension, we do know the semiclassical expansion of the individual eigenvalues as functionals, as well as the gradient expansion.  Thus we should be able to derive the  correction terms, as functionals of the potential.  A small step needed toward this goal is to better understand the role of the boundaries.  These show up in the Maslov index \cite{MF81} in the semiclassical quantization condition.  If these are explicitly included, one can find a general result that explicitly accounts for the boundary conditions (hard walls, true turning points, or periodic) \cite{B20b}.  

To understand this general expression, begin from the Euler-Maclaurin formula given by Hua \cite{H12}, and integrate by parts $p$ times:
\begin{equation}
\sum_{a\leq j \le b} f(j) = \sum_{k=0}^p \frac{(-1)^k}{k!} D_k + \frac{(-1)^{p+1}}{p!} R_p,
\label{EEM}
\end{equation}
where the end-point contributions are
% $h_k(x)$ are $P_k(x) f^{(k-1)}(x)$
\begin{equation}
D_k= \left[ P_k(x) f^{(k-1)}(x) \right]_a^b,
\end{equation}
and the remainder is
\begin{equation}
R_p= \int_a^b dx\, P_p(x)\, f^{(p)}(x).
\end{equation}
Eq. (\ref{EEM}) is true for any $p \geq 1$, so long as the $p$-th derivative of $f(x)$ is continuous.  Here, the term $f^{(-1)}(x)$ is simply the anti-derivative, so the integral is $D_0$.   The $P_k$ are periodized Bernouilli polynomials (Sec. 24 of Ref. \cite{DLMF}), and $P_k(1)=B_k$.  This leads to the following exact formula for the sum of eigenvalues
\begin{equation}
E(N) = \int_a^b dz\, \eps(z) - \sum_{k=1}^{\floor{p/2}} c_k\, \tilde{\eps}^{(2k-1)}(b) + \Delta_p,
\label{ENfin}
\end{equation}
where $b=N+1/2-\nu$, $a=1-\nu$, $\nu$ is the Maslov index, 
$$c_k = \frac{B_{2k}}{(2k)!} \left(1-\frac{2}{4^k}\right),$$
and $\Delta_p$ is of order $\tilde{\eps}^{(2p)}(N)$ and is given exactly by
\begin{equation}
\label{Deltap}
\Delta_p=-\sum_{k=1}^{p}\frac{B_k}{k!}\, \tilde{\eps}^{(k-1)}(a) + \frac{(-1)^{p+1}}{p!} R_p.
\end{equation}
(Since the integration interval is no longer an integer, $R_p$ does not vanish beyond a maximum $p$ for simple powers.)  This is an exact formula for all potentials that are sufficiently smooth (the $p$-th derivative must be continuous), and can be applied with any $p \geq 1$, and to any boundary conditions.  It recovers the exact result for the box $(\nu=1)$, the harmonic oscillator and PT wells with two real turning points $(\nu=1/2)$, and the asymptotic expansion for the linear half well ($\nu=3/4$) with one real turning point and one hard wall.  Moreover, when the WKB expansion is asymptotic, Eq. (\ref{ENfin}) recovers the asymptotic expansion of the sum of the eigenvalues, Eq (\ref{SUForm}).

All the applications to model systems described here use potentials that are smooth, and that can be differentiated to all orders.  An interesting question is to consider less smooth potentials, such as a truncated linear well or harmonic oscillator.  Ref. \cite{BB19} shows the regular asymptotic expansion misses the contributions due to truncation in every order, but also how to use asymptotics to recover these exponentially small contributions.

There is also the question of dimensionality \cite{B83}.  The WKB formulas apply only in one dimension.  These summation procedures can be applied in any dimension, and the particle in a two-dimensional incommensurate box was examined in Ref. \cite{BB20}, showing that the summation formulas did not yield worse results than standard semiclassical methods.

We are also studying the general quartic oscillator, $v(x) = a x^2 + b x^4$, with an emphasis on the double-well potential, in order to better understand these asymptotic expansions.  Toward this goal, we recently published benchmark results for these systems \cite{OB21}, which should prove useful in various contexts in chemical physics and beyond.  Unlike the other potentials we have discussed, the general quartic oscillator has no simple analytic solutions and its potential is not scale invariant.

\sec{Conclusions}
\label{sec:Conc}
So, what has been learned in this romp through the connection between semiclassics and DFT?  Probably the single most important thing is that functionals become local in the semiclassical limit discussed here.  Thus local density approximations are a universal limit of all electronic systems, be they atoms, molecules, or solids.  One can then use a uniform gas calculation to deduce the exact form of a local approximation, or even fit it with an inhomogeneous system by scaling to the semiclassical limit.  We claim that this is the key to understanding approximations in DFT, and the success of semilocal approximations.

We have followed a thread from the land of TF theory for atoms all the way to modern XC approximations for use in KS-DFT.  Along the way, we have considered mostly the kinetic energy of non-interacting electrons in 1D, where we can derive many results from the WKB approximation, either for the eigenfunctions (to approximate densities) or eigenvalues.  We can bootstrap the functional forms that can be derived in 1D to their 3D counterparts for XC.

\begin{table}
\noindent\(\begin{array}{|c|c|r|r|r|r|}
\hline
\multicolumn{3}{|c|}{} & \multicolumn{3}{c|}{\text{Error}} \\
\hline
\text{Atom} & \text{Z} & \multicolumn{1}{c|}{\text{Exact}} & \multicolumn{1}{c|}{\text{LSD}} & \multicolumn{1}{c|}{\text{PBE}} & \multicolumn{1}{c|}{3^{\text{rd}} \text{ corr.}} \\
\hline
\text{H}  &  1 &     -0.500      &  0.021      & -0.008      & -0.097       \\
\text{He} &  2 &     -2.904      &  0.069      & -0.072      &  0.027       \\
\text{Ne} & 10 &   -128.937      &  0.704      &  0.036      & -0.461       \\
\text{Ar} & 18 &   -527.539      &  1.593      &  0.144      &  0.679       \\
\text{Kr} & 36 &  -2753.94\cw{0} &  3.79\cw{0} &  0.45\cw{0} & -0.58\cw{0}  \\
\text{Xe} & 54 &  -7235.23\cw{0} &  6.37\cw{0} &  0.91\cw{0} &  0.22\cw{0}  \\
\text{Rn} & 86 & -21872.5\cw{00} & 11.2\cw{00} &  1.8\cw{00} & -0.2\cw{00}  \\
\hline
\end{array}\)
\caption{Same as Table \ref{tab:AE} but comparing LDA, PBE, and a third order correction extracted numerically.}
\label{tab:AEConc}
\end{table}

To illustrate these connections, we apply some of the methods from the middle of this article to connect our starting point (TF theory in Sec. \ref{sec:TF}) to our end point (asymptotics in Sec. \ref{sec:sums}).  We began with TF theory, and showed it yields the leading term in the asymptotic expansion of the energies of atoms.   We also showed how the next two corrections are given by simple integrals over the TF density, and we found results that were typically within a factor of 4 of HF results.  But we can also make a naive guess at the next term.   We guess it is proportional to $Z^{4/3}$, and fit the constant to the numerical data, crudely, to find $c_3 \approx -0.058$.  Addition of this term yields the last column of Table \ref{tab:AEConc}, which now has extraordinarily small errors (which alternate in sign, due to the fitting). With our guessed correction, our four simple terms yield energies competitive with a modern GGA, as we show in Table \ref{tab:AEConc}.  Naturally, it beats these calculations for the largest $Z$ values, which are most dominated by the asymptotic expansion.  If we only knew how to calculate this series for molecules and solids, we would not need the KS scheme at all!

\begin{figure}[!htb]
\includegraphics[width=0.9\columnwidth]{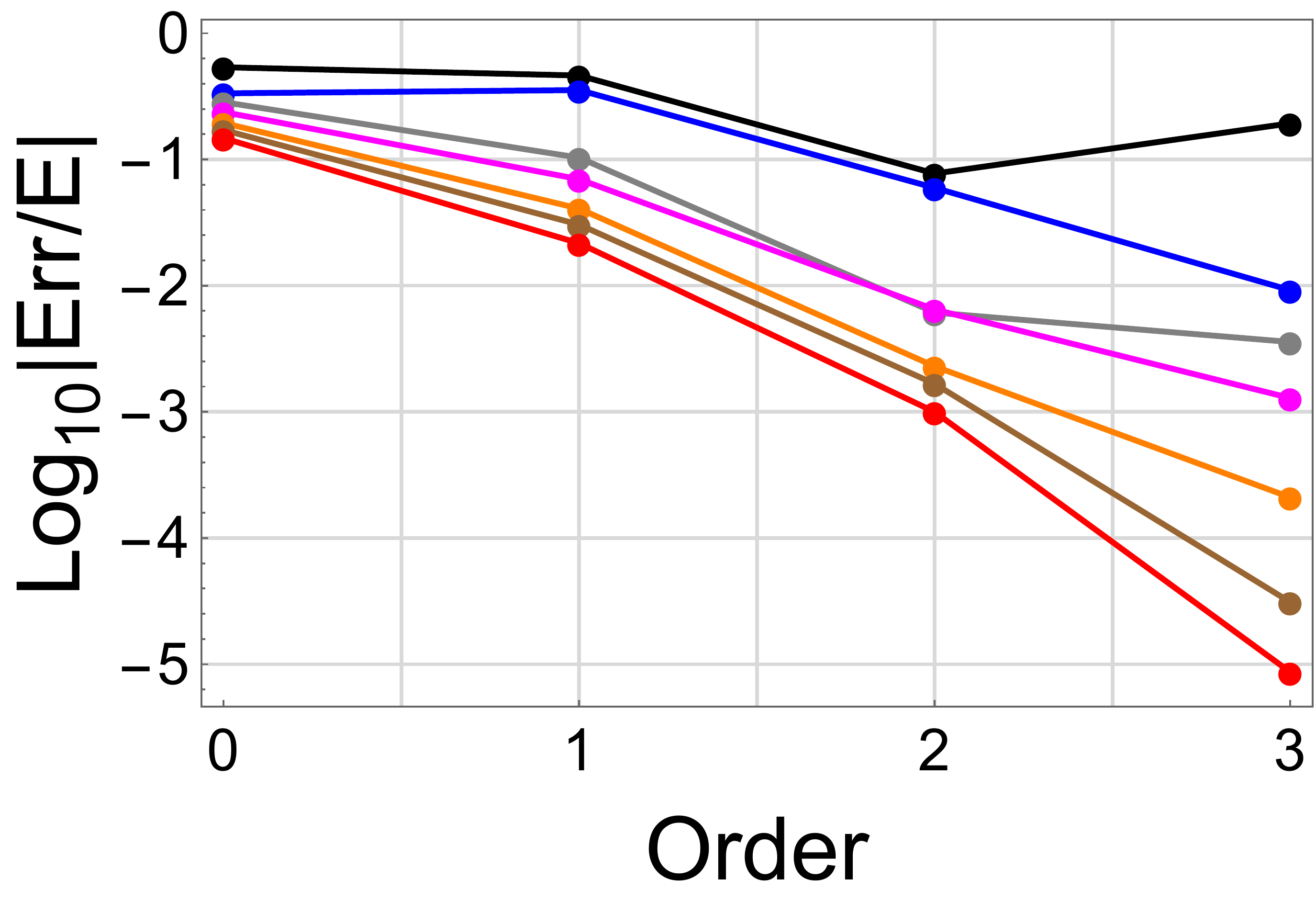}
\caption{The errors of successive orders of the large Z asymptotic expansion of neutral atomic energy from zeroth (TF) to third order, for H (black), He (blue), Ne (gray), Ar (magenta), Kr (orange), Xe (brown), Ra (red).  Here the 3rd order is the assumed form with fitted coefficient $-0.058 Z^{4/3}$.}
\label{fig:AtmEnErr}
\end{figure}

To make the connection between the  expansion we began from (Eq. (\ref{EZasy})) and the asymptotic expansions of Sec. \ref{sec:sums}, in Fig. \ref{fig:AtmEnErr} we plot the fractional errors for the energy of successively heavier neutrals (different colors) as a function of the number of terms included.  We see many trends similar to those of Fig. \ref{fig:LHW}, although little sign of increasing error (and the magnitude of successive terms always decreases, even for $Z=1$).  

Thus 1D examples can give insight into the 3D world.  The end-point corrections to the integral forms, arising from the boundaries of finite systems, such as Eq. (\ref{SU2}), mean that the gradient expansion does not yield all contributions to the semiclassical expansion for such systems.  We saw in Sec. \ref{sec:sums} that it is possible to derive corrections for simple systems as functionals of the potential.  We do not yet know how to turn these expressions into density functionals.  The null result of Sec. \ref{sec:RTP}, showing how improvements in kinetic energy densities pointwise can produce little or no improvement in the kinetic energy, appears very relevant to studies of the kinetic energy density in DFT \cite{PC07,LMA14}.  Moreover, the different cases studied in Ref. \cite{LMA14} can be classified by their Maslov index.  Radial problems have a Maslov index of 3/4 just like the LHW, because the origin acts as a hard wall, at least if the TF density has finite measure.  Possibly one should classify Coulomb potentials differently from those lacking a cusp at the nucleus.  The same analysis can also be applied to the older literature on surface energies \cite{SGP82}, jellium spheres \cite{EP91,ELD94}, and defects in solids \cite{YPKZ97}, as well as yielding a connection between the traditional gradient expansion and the Airy gas \cite{KM98} and subsystem functionals \cite{AM02,MA10,HAM10}.

In the meantime, we can use insight from these studies to improve understanding of the behavior of the XC energy used in KS-DFT calculations.  We want to understand the leading corrections to local density approximations for any component of the energy functional.  We see that while exchange rapidly approaches its local limit with increasing $Z$, this is not true for correlation, where the LDA contribution scales only as $Z\ln Z$, and thus requires unphysically large values of $Z$ to dominate.  This explains why popular molecular approximations for exchange usually respect the uniform limit, but those for correlation do not.  If they did not do so, they would be highly inaccurate for large $Z$ atoms.

Moreover, standard GGA's for exchange are highly accurate for the leading correction to LDA in this limit, so they have very small relative errors for any $Z$.  On the other hand, they cannot then recover the usual gradient expansion, which is the correct expansion if the HOMO is everywhere above the KS potential, as it is in bulk metals and many small and moderate gap insulators.  Finally, when a bond is stretched, eventually a single classically allowed region must bifurcate into two distinct such regions, Fig. 1 of Ref. \cite{ELCB08}.  The asymptotic expansions at equilibrium and stretched bond lengths must be distinct, because the topology of their turning surfaces differs.  At the bond length at which that occurs, the semilocal approximations that most functionals use must fail, and in fact they do.  The KS equations then yield a broken-symmetry solution which has lower energy \cite{KCBP21,MZ20,PRSN21}.

There are dozens if not hundreds of further questions to be explored, each of which should shed further light on the connection between semiclassics and density functionals.  The landscape stretches from mathematical proofs to semiclassical approximations to density functional construction; from simple one dimensional potentials, to atoms and ions, molecules, clusters, surfaces, and bulk solids; from non-interacting to weakly correlated and strongly correlated systems; and from mathematical physics to computational chemistry and materials science and beyond.

KB thanks the organizers of the Singapore workshop on Density Functionals for Many-Particle Systems: Mathematical Theory and Physical Applications of Effective Equations, and the Institute for Mathematical Sciences.  We thank John Snyder, Jeremy Ovadia, and Krishanu Ray, for their unpublished notes on the Bohr atom.  We thank Attila Cangi, Antonio Cancio, John Perdew, and Berthold-Georg Englert for proofreading our manuscript and offering helpful corrections.  Above and beyond we thank Nathan Argaman for his particularly detailed suggestions, insightful questions, and corrections which have significantly altered our text.  Work supported by NSF grant number CHE-1856165.

\bibliography{Bib}

\end{document}